%% file: susysurfaces.tex
\RequirePackage{pdf14}

\documentclass[11pt]{article}
\usepackage[usenames,dvipsnames,table]{xcolor}
\usepackage[utf8]{inputenc}
\usepackage{graphicx}
\usepackage{caption}
\usepackage{subcaption}
\usepackage{bbold,array,setspace,mathrsfs,amsthm,amsfonts,amsmath,yfonts,dsfont,bbm,colonequals,mathtools,tikz}
\usepackage{./aux/jheppub}
\usepackage{afterpage}
\usepackage{xspace}
\usepackage{braket}
\usepackage[normalem]{ulem}
\input{./aux/topmatter}

\title{Superconformal surfaces in four dimensions}

\author[a]{Lorenzo Bianchi,}
\author[b]{Madalena Lemos}

\affiliation[a]{Center for Research in String Theory - School of Physics and Astronomy Queen Mary University of London, Mile End Road, London E1 4NS, UK}
\affiliation[b]{Theoretical Physics Department, CERN, 1211 Geneva 23, Switzerland}

\emailAdd{lorenzo.bianchi@qmul.ac.uk}
\emailAdd{madalena.lemos@cern.ch}

\preprint{CERN-TH-2019-190}

\bigskip
\abstract{
We study the constraints of superconformal symmetry on codimension two defects in four-dimensional superconformal field theories.
We show that the one-point function of the stress tensor and the two-point function of the displacement operator are related, and we discuss the consequences of this relation for the Weyl anomaly coefficients as well as in a few examples, including the supersymmetric R\'enyi entropy. Imposing consistency with existing results, we  propose a general relation that could hold for sufficiently supersymmetric defects of arbitrary dimension and codimension.
Turning to $\NN=(2,2)$ surface defects in $\NN \geqslant 2$ superconformal field theories, we study the associated chiral algebra.
We work out various properties of the modules introduced by the defect in the original chiral algebra. In particular, we find that the one-point function of the stress tensor controls the dimension of the defect identity in chiral algebra, providing a novel way to compute it, once the defect identity is identified. Studying a few examples, we show explicitly how these properties are realized. 
}


\begin{document}
\setcounter{tocdepth}{2}
\maketitle
\setcounter{page}{1}

\input{sections/1_introduction}
\input{sections/2_kinematics}

\input{sections/2_1_Neq1}

\input{sections/2_2_Neq2}
\input{sections/3_consequences}

\input{sections/4_chiralalgebra}

\input{sections/acknowledgments}

\appendix
\input{sections/A_conventions}
\input{sections/B_index}
\input{sections/C_correlators}
\input{sections/D_monodromy}

\bibliography{./aux/biblio}
\bibliographystyle{./aux/JHEP}

\end{document}

%% file: aux/topmatter.tex
\newtheorem*{result*}{Result}

\newtheorem*{conj*}{Conjecture}

\def\deft{\mathrm{def.}}
\def\d{\mathrm{d}}

\def\Lchi{L^{\chi}}
\def\Lbchi{\Lb^{\chi}}
\def\Lhchi{\Lh^{\chi}}
\def\LT{L^{T}}

\def\pd{\dot{\bf{1}}}
\def\md{\dot{\bf{2}}}

\def\wb{\bar{w}}

\def\log{\mathrm{log}}

\def\sl{sl}

\def\BB{{\mathcal{B}}}
\def\CC{{\mathcal{C}}}
\def\DD{{\mathcal{D}}}

\def\OO{{\mathcal{O}}}
\def\NN{{\mathcal{N}}}
\def\LL{{\mathcal{L}}}

\def\II{{\mathcal{I}}}
\def\DD{{\mathcal{D}}}
\def\QQ{Q}
\def\MM{{\mathcal{M}}}
\def\HH{D}
\def\JJ{{\mathcal{J}}}
\def\TT{{\mathcal{T}}}
\def\PP{P}
\def\KK{K}
\def\SS{S}
\def\RR{{\mathcal{R}}}

\def\Qt{\tilde{Q}}
\def\St{\tilde{S}}

\def\Lb{\bar{L}}

\def\Od{\mathbb{O}}
\def\Ld{\mathbb{\Lambda}}
\def\Dd{\mathbb{D}}

\def\aa{\alpha}
\def\aad{{\dot{\alpha}}}
\def\bb{\beta}
\def\bbd{{\dot{\beta}}}
\def\ggd{{\dot{\gamma}}}

\def\ddd{\dot{\delta}}
\def\Gb{\bar{G}}
\def\Lb{\bar{L}}
\def\Lh{\hat{L}}
\def\Dh{\hat{\Delta}}

\def\Qt{\tilde{Q}}
\def\St{\tilde{S}}
\def\ii{i}
\def\log{\mathrm{log}}

\def\sl{sl}

\def\ad{\dot{\alpha}}

\renewcommand{\a}{\alpha}

\newcommand{\pa}{\partial}
\newcommand{\g}{\gamma}

\newcommand{\D}{\Delta}
\newcommand{\e}{\epsilon}

\renewcommand{\L}{\Lambda}

\newcommand{\n}{\nu}

\newcommand{\s}{\sigma}
\renewcommand{\S}{\Sigma}

\renewcommand{\O}{\Omega}

\newcommand{\zb}{\bar{z}}

\newcommand{\Tr}{\textup{Tr}}

\newcommand{\be}{\begin{equation}}
\newcommand{\ee}{\end{equation}}
\newcommand{\bea}{\begin{eqnarray}}
\newcommand{\eea}{\end{eqnarray}}
\newcommand{\nn}{\nonumber}
\newcommand\qq{\mathbbmtt{Q}}

\newcommand{\beq}{\begin{equation}}
\newcommand{\eeq}{\end{equation}}

\newcommand\eg{{\it e.g.}}
\newcommand\ie{{\it i.e. }}

%% file: sections/1_introduction.tex
%!TEX root = ../susysurfaces.tex
%%%%%%%%%%%%%%%%%%%%%%%%%%%%%%%%%%%%%%%%%%%%%
\pagebreak
\section{Introduction and summary}
\label{sec:intro}
%%%%%%%%%%%%%%%%%%%%%%%%%%%%%%%%%%%%%%%%%%%%%
Our modern understanding of Quantum Field Theories (QFT) suggests that symmetries and dualities are the correct paradigm to unveil non-perturbative features that are not accessible to a Lagrangian description. This is especially true in the presence of conformal invariance, when we have the concrete hope that symmetries and internal consistency may suffice to completely fix the dynamics of a Conformal Field Theory (CFT). The (super-)conformal bootstrap program, based on this philosophy, has provided a large wealth of results on correlation functions of strongly coupled (super-)conformal field theories, see \cite{Poland:2018epd} for a recent review. While restricting to local operators is a consistent truncation of the CFT operator algebra that allows to study a more tractable problem, the goal is to move beyond this restriction and enlarge our set of observables to include correlation functions in the presence of  non-local operators, or defects. This is especially important if we take into account that extended excitations probe aspects of a CFT that are not accessible to correlation functions of local operators only. Even more surprisingly, it is now clear that CFTs with the same spectrum of local operators may support different and incompatible spectra of defects, resulting in different low-energy dynamics and interesting phase transitions \cite{Gaiotto:2010be,Aharony:2013hda,Gaiotto:2011tf}. 

A conformal defect generically preserves conformal invariance along its profile and rotations in the orthogonal directions. The spectrum splits into defect and bulk operators. Correlation functions involving only the former are constrained by the residual symmetry in the same way as for a lower dimensional CFT (notice, however, that the exchange of energy with the bulk prevents the presence of a conserved stress energy tensor). In particular, one can take the OPE of defect operators inside correlation functions until reaching the only defect operator with non-vanishing one-point function, i.e. the defect identity. The latter corresponds to the empty defect and its one-point function is given by the defect expectation value. To fully characterize a conformal defect, however, one needs to include interactions with the bulk degrees of freedom. One-point functions of bulk operators and bulk to defect couplings are then added to the defect spectrum and OPE coefficients to describe the full set of defect CFT data \cite{Billo:2016cpy}. Their allowed values are further constrained by crossing relations involving bulk, defect, and mixed correlators and the long term goal of the defect bootstrap program is to put stringent bounds on the space of consistent defects. In this context, numerical techniques can be directly applied to correlation functions of defect operators \cite{Gaiotto:2013nva,Liendo:2018ukf,Gimenez-Grau:2019hez}, however, the naive application fails if one wishes to study correlation functions that probe the bulk to defect couplings. In this case, one of the OPE channels lacks the positivity required for the numerical tools to apply.\footnote{An alternative approach to study the crossing equations that does not rely on positivity has been applied to the case of defect CFTs in \cite{Gliozzi:2015qsa,Gliozzi:2016cmg}. In this approach one does an extreme truncation of the CFT spectrum to find approximate solutions to the crossing equations. By contrast with the numerical bootstrap one does not get rigorous bounds on the CFT data, but rather estimates with unknown errors. See also \cite{Hogervorst:2017kbj,Rastelli:2017ecj,Lemos:2017vnx,Mazac:2018biw,Bissi:2018mcq,Liendo:2019jpu} for progress in analytical approaches to defect CFTs.} As such, the task of obtaining non-perturbative information on these couplings is harder than in the case of CFTs without defects, and has only been studied in the case of boundaries where positivity was assumed \cite{Liendo:2012hy}. Supersymmetry gives us additional tools for constraining the dynamics of defect CFTs and this is the approach we will use in this paper.

Generically, a \emph{conformal defect} is characterized by an infinite number of defect CFT data. Nevertheless, it is interesting to isolate a subset which is both physically interesting and universal. For the case of homogeneous CFTs in four dimensions, the Weyl anomaly coefficients $a$ and $c$ match these requirements. On the one hand, they appear in the two- and three-point functions of the stress tensor operator, implying that they must be present in any local CFT. On the other hand, they feature in the energy flux measured in ``conformal collider experiments'' \cite{Hofman:2008ar}. Requiring that the integrated energy flux is positive provides important bounds on their allowed values \cite{Hofman:2008ar}. For the case of conformal defects, the set of physically interesting operators is enlarged by defect excitations. Among them, a distinguished role is played by the \emph{displacement operator} that is related to the broken invariance under translations in the orthogonal directions and, as such, it is present for any extended excitation inserted in a local CFT. Its two-point function is an important piece of defect CFT data and, together with the one-point function of the stress tensor, they determine two of the three defect anomaly coefficients featured by a two-dimensional defect \cite{Lewkowycz:2014jia,Bianchi:2015liz}. Their relation with deformations in the shape of the defect, or in the background geometry \cite{Lewkowycz:2013laa,Fiol:2015spa,Bianchi:2019dlw}, as well as their role in the computation of the emitted radiation \cite{Correa:2012at,Fiol:2015spa,Bianchi:2018zpb}, make these two parameters a good starting point for the full characterization of an extended excitation. One of the main results of this paper is to show that for any superconformal surface defect in four dimensions these two quantities are related by a simple, theory independent, numerical factor. 

The interest in surface defects in four-dimensional superconformal theories (SCFTs) has taken different directions. The initial attention for defects in $\mathcal{N}=4$ Super Yang-Mills was triggered by the AdS/CFT correspondence and it led to the discovery of systems of intersecting branes corresponding to supersymmetry preserving surface defects \cite{Constable:2002xt}. This holographic description received a field theoretical counterpart in the work of \cite{Gukov:2006jk}, which was followed by several generalizations and explicit computations \cite{Gomis:2007fi,Buchbinder:2007ar,Gukov:2008sn,Drukker:2008wr,Koh:2008kt}. For lower supersymmetry, the most studied examples are surely surface defects preserving a two-dimensional $\mathcal{N}=(2,2)$ superconformal algebra inside a four-dimensional $\mathcal{N}=2$ SCFT \cite{Alday:2009fs,Gaiotto:2009fs,Dimofte:2010tz,Gaiotto:2011tf,Kanno:2011fw,Gaiotto:2012xa,Gaiotto:2013sma,Gadde:2013ftv,Gomis:2014eya,Cordova:2016uwk,Cordova:2017mhb,Cordova:2017ohl,Gorsky:2017hro,Ashok:2017lko,Pan:2017zie,Ashok:2018zxp,Nishinaka:2018zwq}. A protected subsector of these defects is also captured by a two-dimensional chiral algebra \cite{defectLCW,Cordova:2017mhb}, and its study will be one of the main focuses of this work.
Finally, supersymmetric surface defects in $\NN=1$ SCFTs preserve an $\NN=(2,0)$ superconformal algebra, and have been studied in \cite{Koh:2009cj,Drukker:2017dgn,Razamat:2018zel}.

In \cite{Beem:2013sza} it was shown that any $\NN\geqslant 2$ SCFT possesses a subsector of protected operators isomorphic to a two-dimensional chiral algebra.\footnote{A similar construction holds for $6d$ SCFTs with $\NN=(2,0)$, and $2d$ SCFTs with at least $\NN=(4,0)$ \cite{Beem:2014kka}.}  This subsector is obtained by restricting the operators to a plane, and passing to the cohomology of a certain nilpotent supercharge $\qq$, with the cohomology classes of $\qq$ having the structure of a two-dimensional chiral algebra.
This provides a powerful tool to obtain non-perturbative dynamical information on interacting $\NN=2$ SCFTs, by knowledge of their associated chiral algebras, independently of whether they admit a Lagrangian description or not. 
The construction can be further enriched by adding surface defects as anticipated in \cite{Beem:2013sza}, and made precise in \cite{defectLCW,Cordova:2017mhb}.\footnote{In a similar way, for $\NN=4$ SCFTs one can obtain a subsector captured by a topological theory that can be enriched by adding half-BPS defects \cite{Liendo:2016ymz}.}
Specifically, an $\NN=(2,2)$ surface defect intersecting the chiral algebra plane at a point preserves $\qq$, and in chiral algebra it appears as the insertion of a local operator. In \cite{defectLCW,Cordova:2017mhb} it was shown that the defect gives rise to a module over the original chiral algebra of the bulk SCFT, and the (graded) partition function of the module is obtained by computing the four-dimensional Schur index.

\section*{Summary of results}

\paragraph{Constraints on superconformal surfaces.}

In the first part of this work, we consider uncharged codimension two superconformal defects in four-dimensional SCFTs. In particular, our results are valid for $\NN\geqslant 1$ SCFTs in the presence of surface defects that preserve at least an $\NN=(2,0)$ subalgebra. We are interested in the correlation functions of the most universal multiplets in these theories, namely the \emph{stress tensor} of the bulk SCFT, present in any local theory, and the \emph{displacement operator}, associated with the breaking of translation invariance in the orthogonal directions. In a SCFT these two operators belong in superconformal multiplets, and the multiplets' correlation functions are the subject of our work. Following the bootstrap approach, the first task is to  fix all that is dictated by symmetry, \ie fixing the kinematics of the correlation functions.
The lowest non-trivial $n-$point functions involving these operators are the \emph{bulk one-point functions} of operators in the stress tensor multiplet, the \emph{defect two-point functions} of those in the displacement multiplet, and the \emph{bulk to defect two-point functions} between operators in each of these multiplets. We find in section~\ref{sec:kin} that superconformal symmetry fixes all of these correlators in terms of a \emph{single} dynamical number. This follows from the following universal relation:
\begin{result*}
For supersymmetric surface defects in $4d$ $\NN\geqslant1 $ SCFTs, the one-point function of the stress tensor, $h$, and the two-point function of the displacement operator, $C_D$, are related by supersymmetry as
\be
 C_D=48 h\,, 
 \nn
\ee
\end{result*}
\noindent
where the precise definitions of $C_D$ and $h$ are given in \eqref{eq:displ2pt} and \eqref{Tonept} respectively.
Following \cite{Bianchi:2018zpb}, where a similar relation was obtained for half-BPS line defects in $\NN=2$ SCFTs, this relation is obtained by studying the bulk to defect coupling of the full stress tensor and displacement superconformal multiplets. By imposing supersymmetric Ward identities for the preserved and broken supersymmetries we find that this coupling is determined by a single parameter and this yields to the relation quoted above. 
The latter implies two of the \emph{Weyl anomaly coefficients} are equal, as described in section~\ref{sec:consequences}, along with a few examples and consequences for the stress tensor defect OPE. Finally, based on cases where a relation between  $C_D$ and $h$ is known, or conjectured, we also put forward  the proposal that
\begin{conj*}
For a supersymmetric defect of dimension $p$ and codimension $q$, the one-point function of the stress tensor and the two-point function of the displacement operator are related as
\be 
 C_D=\frac{2^{p+1} (q+p-1)(p+2)}{q-1} \frac{\Gamma(\frac{p+1}{2})}{\pi^{\frac{p+1}{2}}} \frac{\pi^{\frac{q}{2}}}{\Gamma(\frac{q}{2})} h\,.
 \nn
\ee
\end{conj*}
\noindent
While we cannot say what amount of supersymmetry is needed for such a relation to hold in dimensions $d=p+q \neq 4$, if it exists, consistency with known results fixes the proposed relation.

\paragraph{$\NN=(2,2)$ surfaces and chiral algebras.} 

The rest of this work concerns  $\NN \geqslant 2$ SCFTs in the presence of uncharged two-dimensional $\NN=(2,2)$ defects.
In section~\ref{sec:Neq2} we identify the superconformal multiplet that accommodates the displacement supermultiplet for these defects \cite{Gaiotto:2013sma}, and fix the one-point function of the stress tensor supermultiplet, the two-point function of the displacement supermultiplet, and the two-point function between the displacement and stress tensors supermultiplets, in terms of $h$.  Turning to \emph{dynamics}, we study the chiral algebras of these defects in section~\ref{sec:chiralalgebra}. We show that the scaling dimension, in chiral algebra, of the operator inserted by the defect identity is given in terms of $h$, thus providing a new way to compute it in SCFTs.\footnote{Here and in the following we use the expression defect identity to denote the empty defect, \ie the vacuum of the defect CFT. \label{defectid}}
Apart from the defect identity, the superprimaries of certain short defect supermultiplets are captured by the chiral algebra, and we describe a few noteworthy cases. In particular, we notice that not all defect operators in cohomology can be obtained as descendants of the defect identity under the action of the chiral algebra generators. In other words, the defect can insert a reducible module over the original chiral algebra.  Among the defect operators in cohomology, one finds the superprimary of the \emph{displacement} supermultiplet, which is the defect operator associated with the breaking of the $su(2)_R$ symmetry.
This allows to compute correlation functions in the presence of the defect if one can identify the defect
identity in chiral algebra. To this end, we determine how the bulk chiral algebra modes act on the defect identity, from defect OPE selection rules in four dimensions,
\begin{result*}
The defect identity introduces in chiral algebra a state $|\sigma\rangle$ that obeys
\be 
\LT_{n >0} |\sigma \rangle = 0\,, \qquad \LT_0 | \sigma\rangle = -3\pi^2 h |\sigma\rangle + (\substack{\text{defect marginal} \\ \text{operators}})\,, \qquad \LT_{-1} |\sigma\rangle \sim |\mathbb{O}_\uparrow\rangle \,, \qquad J_{n \geqslant 1} |\sigma\rangle =0\,,
\nn
\ee
where $\LT_n$ are modes of the two-dimensional stress tensor, $J_n$ those of affine Kac Moody currents associated to possible bulk flavor symmetries of the bulk theory, and $\mathbb{O}_\uparrow$ is the superprimary of the displacement operator.
\end{result*}
By studying the form of correlation functions involving bulk and defect operators in chiral algebra, we make a proposal for the two-dimensional scaling weight of defect operators. Finally, we see how these results are realized in a few examples. We also test the proposal of \cite{Cordova:2017mhb}, that monodromy defects are obtained in chiral algebra by the spectral flow, in the case of a single free hypermultiplet, by explicitly computing the one-point function of the stress tensor and of the flavor currents.

%% file: sections/2_kinematics.tex
%!TEX root = ../susysurfaces.tex
%%%%%%%%%%%%%%%%%%%%%%%%%%%%%%%%%%%%%%%%%%%%%
\section{Kinematics of supersymmetric correlation functions}
\label{sec:kin}
%%%%%%%%%%%%%%%%%%%%%%%%%%%%%%%%%%%%%%%%%%%%%
In the first part of this paper, our considerations are purely algebraic and we do not need to specify any microscopic detail of the defect. We will use the preserved and broken superconformal symmetries to constrain the kinematics of defect correlation functions, and obtain a relation between the one-point function of the stress tensor and the two-point function of the displacement operator, valid for any four-dimensional supersymmetric defect.  Half-BPS surface defects in four dimensions preserve a superalgebra
\begin{align}\label{subalgebra}
 su(1,1|\mathcal{N}_1)\oplus su(1,1|\mathcal{N}_2) \subset su(2,2|\mathcal{N})\,,
\end{align}
for non-negative integers $\mathcal{N}_1$ and $\mathcal{N}_2$ such that $\mathcal{N}_1+\mathcal{N}_2=\mathcal{N}$. In \eqref{subalgebra} it is understood that $su(1,1|0)\equiv sl(2)$, and there will often be a commutant of the defect superalgebra inside the four-dimensional one leading to an extra $u(1)$ factor. One may wonder why a surface defect could not preserve a $osp(\mathcal{N}|2)$ subalgebra. It is a straightforward exercise to verify that the embedding of such an algebra inside $su(2,2|\mathcal{N})$ must involve a linear combination of $Q$s and $\tilde Q$s which breaks invariance under rotations in the directions orthogonal to the defect.\footnote{A related result was obtained in \cite{Bobev:2015jxa} where the authors wrote a superconformal algebra with four supercharges in dimensions $2\leqslant d \leqslant 4$. The authors start from a four-dimensional $\NN=1$ superconformal algebra and reduce to lower dimensions by restricting to the conformal algebra of a lower dimensional theory. This could be thought of as placing a codimension one or two defect in the four-dimensional theory. In their construction invariance under rotations in the orthogonal directions is automatically preserved, ending up as an $R-$symmetry in the lower dimensional theory. The two-dimensional superconformal algebra they obtain inside the $4d$ $\NN=1$ one is precisely the $su(1,1|1)\oplus su(1,1)$ we consider in this work. Similar results were obtained with eight supercharges in \cite{Bobev:2017jhk} starting from six dimensions, thus relevant for defects in $6d$ $(1,0)$ theories. We thank N.~Bobev for very useful discussions on these points.}
In this paper, we only consider superconformal defects preserving rotations in the orthogonal directions. Given this restriction, defects preserving less than half supersymmetry can be viewed as half-BPS defects in a bulk theory with less supersymmetry. For example, one quarter-BPS defects in $\mathcal{N}=2$ theories would preserve $su(1,1|1)\oplus sl(2)\oplus u(1)$ and can be seen as a half-BPS defects in $\mathcal{N}=1$, the only difference being that the extended R-symmetry may produce some additional global symmetry commuting with all fermionic generators. Nevertheless, it should be clear from this reasoning that every constraint that is found for half-BPS defects in $\mathcal{N}=1$ applies to any BPS defect with extended supersymmetry. 

Even though eq.~\eqref{subalgebra} is in Lorentzian signature, in what follows we will study defects in Euclidean four-dimensional space. Let us consider a flat conformal surface defect stretched along the directions $x_1$ and $x_2$, where we introduce complex coordinates
\begin{align}
  w&=x_1+i x_2\,, & \bar{w}&=x_1-i x_2\,.
\end{align}
We require that the preserved superalgebra includes the global part of the two-dimensional conformal algebra as well as the $u(1)$ generator $\mathcal{M}=M_\mathbf{1}{}^\mathbf{1}+ M^{\dot{\mathbf{1}}}{}_{\dot{\mathbf{1}}}$ of rotations in the orthogonal directions parameterized by\footnote{Note that we are using $\alpha=\mathbf{1}, \mathbf{2}$ for the $\pm$ spinor indices, and similarly for the dotted ones. This is to avoid confusion with the $\pm$ appearing later in the two-dimensional supercharges.}
\begin{align}
  z&=x_3-i x_4\,, & \bar{z}&=x_3+i x_4\,.
  \label{eq:zzb}
\end{align}
In our conventions, summarized in appendix \ref{sec:conventions}, the preserved two-dimensional conformal algebra is generated by
\begin{align}\label{2dconf}
 L_{-1}&=\frac12 \left(P_1 - i P_2\right)\,, & L_0&=\frac12(D+\mathcal{M}_{\parallel})\,, & L_1&=\frac12 \left(K_1+i K_2\right) \,,\\
 \bar{L}_{-1}&=\frac12 \left(P_1+i P_2\right)\,, & \bar L_0&=\frac12(D-\mathcal{M}_{\parallel})\,, & \bar L_1&=\frac12 \left(K_1-i K_2\right) \,,
\end{align}
with $\mathcal{M}_{\parallel}=M_\mathbf{1}{}^\mathbf{1}- M^{\dot{\mathbf{1}}}{}_{\dot{\mathbf{1}}}$ generating rotations along the defect plane.

In the following, we will consider a set of correlation functions involving the stress tensor $T_{\mu \nu}$ and the displacement operator $\mathbb{D}_{\uparrow,\downarrow}$. The latter is a defect degree of freedom defined by the Ward identity associated with the breaking of translational invariance in the directions orthogonal to the defect
\begin{align}\label{WardDbarD}
 \pa^{\mu}T_{\mu z}&=-\delta^2(z) \mathbb{D}_{\uparrow}\,, & \pa^{\mu}T_{\mu \bar{z}}=-\delta^2(z) \mathbb{D}_{\downarrow}\,,
\end{align}
with  $T_{\mu z}=\frac12(T_{\mu 3}+iT_{\mu 4})$ and $T_{\mu \bar{z}}=\frac12(T_{\mu 3}-iT_{\mu 4})$, and $\mu$ is a $4d$ bulk index. As such, it is associated to deformations in the shape of the surface. In particular, we can consider an arbitrary correlation function of local operators in the presence of the defect $\S$ defined by
\begin{align}
\braket{\mathcal{X}}_{\S} \colonequals \braket{ O(x_1)\dots O(x_n)\hat{O}(w_1)\dots \hat{O}(w_m)}_{\S} \colonequals\frac{\braket{ O(x_1)\dots O(x_n) \hat{O}(w_1)\dots \hat{O}(w_m) \S}}{\braket{\S}}\,.
\label{eq:defcorrfn}
\end{align}
Here $O(x_i)$ are operators living in the bulk $4d$ SCFT, while $\hat{O}(w_i)$ are defect operators, \ie operators of the two-dimensional conformal theory on the defect. Here and in the following we will add a hat to distinguish defect operators and their quantum numbers from bulk ones. The displacement operator accounts for the variation of this correlation function after a small deformation of the defect, $\delta z(w)$,
\begin{align}\label{dispins}
 \braket{\mathcal{X}}_{\delta \S}\sim\int \d^2w \braket{\mathcal{X}\mathbb{D}_{\uparrow}(w)}_{\S} \delta z (w) + \int \d^2w \braket{\mathcal{X}\mathbb{D}_{\downarrow}(w)}_{\S} \delta \bar{z} (w)\,.
\end{align}

Alternatively, one can consider the insertion of the displacement operator as the action of the broken translation generators $\mathsf{P}_{\uparrow}$ and $\mathsf{P}_{\downarrow}$ on the non-local operator $\S$. To make this precise, we consider for a moment a spherical defect and we define the charges in radial quantization
\begin{align}
 \mathsf{P}_{\uparrow}&=-\int_{\s} \d \O^{\mu} T_{\mu z}\,, & \mathsf{P}_{\downarrow}&=-\int_{\s} \d \O^{\mu} T_{\mu \bar z}\,,
\end{align}
where the integral is performed over a sphere $\s$. As usual, one can compute the action of the generator by considering the commutator $[\mathsf{P}_{\uparrow},\S]$ and, using the fact that $\mathsf{P}$ is topological, deform the contour to a shell surrounding the defect. Then, using \eqref{WardDbarD}, we get
\begin{align}\label{brokentransl}
 [\mathsf{P}_{\uparrow},\S]&=\int \d^2w \ \mathbb{D}_{\uparrow} \S\,, & [\mathsf{P}_{\downarrow},\S]&=\int \d^2w \ \mathbb{D}_{\downarrow} \S\,,
\end{align}
where both sides of these equations must be thought of as inserted in a radially ordered correlator. We will see below that a similar derivation applies to global bosonic symmetry as well as for fermionic generators. 

The class of correlation functions we will be interested in includes the displacement two-point function\footnote{The factor of two is included to make contact with the usual definition in terms of orthogonal indices.}
\begin{align}
 \braket{\mathbb{D}_{\uparrow}(w)\mathbb{D}_{\downarrow}(0)}_{\S}=\frac{C_D}{2 w^3\bar w^3}\,,
 \label{eq:displ2pt}
\end{align}
and the stress tensor one-point function, with non-vanishing components\footnote{$h$ defined here is related to $a_T$ of \cite{Billo:2016cpy} by $a_T= -4 h$.}
\be
\label{Tonept}
\begin{aligned}
 \braket{T_{zz}}_{\S}&=-\frac{h}{z^3 \bar z}\,, & \braket{T_{\bar z \bar z}}_{\S}&=-\frac{h}{z \bar z^3}\,,\\
 \braket{T_{z\bar z}}_{\S}=\braket{T_{\bar z z}}_{\S}&=\frac{h}{2 z^2 \bar z^2}\,,\qquad  & \braket{T_{w\bar w}}_{\S}=\braket{T_{\bar w w}}_{\S}&=-\frac{h}{2 z^2 \bar z^2}\,,
\end{aligned}
\ee
where an index $w$ corresponds to $X_w = \frac12 (X_1 - i X_2)$ and $X_{\bar{w}}=\frac12 (X_1 + i X_2)$.
The form of these correlators is fixed by conformal symmetry, see \eg, \cite{Billo:2016cpy}. However, for a general defect CFT, $C_D$ and $h$ are independent pieces of CFT data that depend on the particular theory being studied. Nevertheless, in the presence of supersymmetry we will prove that
\begin{align}\label{relation}
 C_D=48 h\,,
\end{align}
following only from symmetry considerations, and independently of the dynamics of the CFT in question. Note that in particular this implies $h$ is non-negative, due to positivity of the displacement two-point function.
To that end, we will consider a third correlator, namely the bulk to defect two-point function of the stress tensor and the displacement operator. Generically, a correlator of a spin two bulk conformal primary and an orthogonal defect vector, is fixed in terms of three parameters. However, it was shown in \cite{Billo:2016cpy} that this specific two-point function is fully determined by $C_D$ and $h$. The derivation of \cite{Billo:2016cpy} is valid for any dimension and codimension and is based on two sets of Ward identities. We rewrite them here in our notation for a surface defect in $4d$. The first set of identities relates the two-point function to $h$ and it is a direct consequence of \eqref{brokentransl}
\begin{align}\label{Wardbrokentransl}
 \pa_z \braket{T_{\mu\nu}(z,0)}_{\S}&=-\int \d^2w \braket{T_{\mu\nu}(z,0)\mathbb{D}_{\uparrow}(w)}_{\S}\,,
\end{align}
where $\mu$ and $\nu$ run over the two sets of complex coordinates and other inequivalent identities are obtained by complex conjugation.
The second set of identities is realized in terms of distributions and it descends from \eqref{WardDbarD}
\begin{align}\label{WardTcons}
 \pa^{\mu} \braket{T_{\mu z}(z,w)\mathbb{D}_{\downarrow}(0)}_{\S}=-\delta^2 (z) \braket{\mathbb{D}_{\uparrow}(w) \mathbb{D}_{\downarrow}(0)}_{\S}\,.
\end{align}
This equation and its conjugate establish a relation between the bulk to defect correlator and $C_D$. As we already remarked above, if a relation like \eqref{relation} holds for all $\mathcal{N}=1$ surface defects, then it will automatically hold for extended supersymmetry. We thus consider the $\mathcal{N}=1$ case and show explicitly that \eqref{relation} is a consequence of supersymmetric Ward identities. After that, we also describe in some detail the case of surface defects preserving $\mathcal{N}=(2,2)$, in an $\mathcal{N}=2$ four-dimensional SCFT. This analysis, though unnecessary for the sake of proving \eqref{relation}, will be extremely useful in the second part of the paper, where we will explore the two-dimensional chiral algebras in the sense of \cite{Beem:2013sza} associated to this type of defects \cite{defectLCW,Cordova:2017mhb}.

%% file: sections/2_1_Neq1.tex
%!TEX root = ../susysurfaces.tex
%%%%%%%%%%%%%%%%%%%%%%%%%%%%%%%%%%%%%%%%%%%%%
\subsection{Half-BPS surfaces in \texorpdfstring{$\NN=1$}{N=1} SCFTs}
\label{sec:Neq1}
%%%%%%%%%%%%%%%%%%%%%%%%%%%%%%%%%%%%%%%%%%%%%
Following the pattern \eqref{subalgebra}, for $\mathcal{N}=1$ the only possible preserved symmetry is
\begin{align}
 su(1,1|1)\oplus sl(2)\oplus u(1)_{Z}\subset su(2,2|1)\,,
\end{align}
corresponding to an $\mathcal{N}=(2,0)$ surface defect.
The commutation relations for the $\mathcal{N}=1$ generators in four dimensions can be found in appendix \ref{sec:Neq1algebra}. In order to generate the full $\mathcal{N}=(2,0)$ subalgebra the bosonic generators \eqref{2dconf} must be supplemented by the fermionic charges 
\begin{align}
 G_{-\frac12}^+&=Q_\mathbf{1} \,, & G_{-\frac12}^-&=\tilde{Q}_{\dot{\mathbf{2}}}\,, & G_{\frac12}^+&=\tilde{S}^{\dot{\mathbf{2}}}\,, & G_{\frac12}^-&=S^{\mathbf{1}}\,,
 \label{eq:N20charges}
\end{align}
as well as the bosonic generator $J$ and the commutant $Z$, which are linear combinations of the $u(1)_{\hat{r}}$ R-symmetry generator, $\hat{r}$, and the orthogonal rotations, $\MM$
\begin{align}
J&=3 \hat{r} - \mathcal{M}\,, & Z&=-\hat{r}+\mathcal{M}\,.
\end{align}
The resulting $2d$ commutation relations are given in appendix \ref{sec:2d20algebra}. Defect operators can be organized in representations of this preserved subalgebra. Representations of $su(1,1|1)$, and a convenient superspace formalism, have been known for a long time \cite{DiVecchia:1985ief,Mussardo:1988ck,Blumenhagen:1992sa} (see also \cite{Fitzpatrick:2014oza,Cornagliotto:2017dup,Buric:2019rms} for the computation of the superblocks). However, here we are interested in the coupling between bulk and defect degrees of freedom and, in order to fully exploit the symmetries of the problem, we find it more convenient to work in components. We start by determining which $2d$ supermultiplet can accommodate the displacement operator. The exact same question was asked and answered in \cite{Bianchi:2018scb} in the context of line defects in three dimensions. Here we review that argument using a more algebraic approach.

\subsubsection{Displacement supermultiplet}
We start by looking at broken supercharges. The defect breaks two supercharges
\begin{align}
 \mathsf{Q}_{\uparrow}^-&=\tilde Q_{\dot{\mathbf{ 1}}}\,, & \mathsf{Q}^+_{\downarrow}&=- Q_{\mathbf{2}}\,,
\end{align}
and the associated supercurrents are no longer conserved. Analogously to \eqref{WardDbarD}, one can write
\begin{align}
 \pa^{\mu} \tilde{J}_{\mu \dot{\mathbf{ 1}}}&=-\delta^2(z) \mathbb{\L}_{\uparrow}^-\,, & \pa^{\mu} J_{\mu \mathbf{2}}&=\delta^2(z) \mathbb{\Lambda}_{\downarrow}^+\,,
\end{align}
where $\mathbb{\L}_{\uparrow}^-$ and $\mathbb{\L}_{\downarrow}^+$ are fermionic defect operators that are produced by the action of the broken supercharges on the defect
\begin{align}
 [\mathsf{Q}_{\uparrow}^-, \S]&=\int \d^2w\ \mathbb{\L}_{\uparrow}^-(w) \S \,, &  [\mathsf{Q}^+_{\downarrow}, \S]&=\int \d^2w\ \mathbb{\L}_{\downarrow}^+(w) \S\,.
 \label{eq:Neq1susybroken}
\end{align}
The defect operators $\mathbb{\L}_{\uparrow}^-$ and $\mathbb{\L}_{\downarrow}^+$ have $L_0=1$, $\Lb_0=\frac32$, $J=\mp 2$. As such, the former is an anti-chiral operator, and the latter a chiral operator, with respect to the left $\NN=2$ superalgebra and thus they must be superconformal primaries. We can act with the preserved supercharges to build the whole multiplet, and use the commutator of broken and preserved supercharges to identify the displacement supermultiplet \eqref{brokentransl} as the action of preserved supercharges on \eqref{eq:Neq1susybroken}
It is a purely algebraic exercise to show that
\be
\begin{aligned}
 \{G_{-\frac12}^+, \mathbb{\L}_{\uparrow}^-\}&=\mathbb{D}_{\uparrow}\,, & \{G_{-\frac12}^-, \mathbb{\L}_{\uparrow}^-\}&=0\,, & \{G_{-\frac12}^+, \mathbb{\L}_{\downarrow}^+\}&=0\,, & \{G_{-\frac12}^-, \mathbb{\L}_{\downarrow}^+\}&=\mathbb{D}_{\downarrow}\,,\\
 \{G_{-\frac12}^+, \mathbb{D}_{\uparrow}\}&=0\,, & \{G_{-\frac12}^-, \mathbb{D}_{\uparrow}\}&=\pa_w \mathbb{\L}_{\uparrow}^- \,, & \{G_{-\frac12}^+, \mathbb{D}_{\downarrow}\}&=\pa_w \mathbb{\L}_{\downarrow}^+ \,,& \{G_{-\frac12}^-, \mathbb{D}_{\downarrow}\}&=0\,.
\end{aligned}
\ee
Therefore, the displacement supermultiplets have the following structure 
\begin{center}
\begin{tikzpicture}
\node at (-1.5, 1) {$Z=1$};
\node at (1.5, 1) {$Z=-1$};
\node at (-2, 0) {$\mathbb{\L}_{\uparrow}^-$};
\node at (-1, -1) {$\mathbb{D}_{\uparrow}$};
\node at (1, -1) {$\mathbb{D}_{\downarrow}$};  
\node at (2, 0) {$\mathbb{\L}_{\downarrow}^+$};  
\node at (-2,-2) {$-2$};
\node at (-1,-2) {$-1$};
\node at (1,-2) {$1$};
\node at (2,-2) {$2$};
\node at (-3,-1) {$3$};
\draw[->] (-1.6,-0.4)--(-1.4,-0.6);
\draw[->] (1.6,-0.4)--(1.4,-0.6);
\node at (-3,0) {$\frac52$};
\node at (-3,-2) {$\Dh / J$};
\end{tikzpicture}
\end{center}
where $\Dh$ is the eigenvalue of $L_0 + \Lb_0$. These multiplets were also obtained in superspace in \cite{Drukker:2017dgn}.

\subsubsection{Correlation functions}
We start by considering the one-point function of the operators in the stress tensor multiplet. The $\mathcal{N}=1$ supercurrent multiplet contains the stress tensor operator, the supercurrents $J_{\mu \a}$ and $\tilde{J}_{\mu \dot\a}$ and the $R-$symmetry current $j_{\mu}$. Our conventions for the supersymmetry transformations are summarized in appendix \ref{sec:susyvarsNeq1}. Using the Ward identities
\begin{align}
\braket{\{G_{-\frac12}^+,\tilde{J}_{\mu \dot \a}\}}_{\S}&=0\,, & \braket{\{G_{-\frac12}^-,J_{\mu\a}\}}_{\S}&=0 \,,
\end{align}
we find the following non-vanishing components for the R-current one-point function,\footnote{The reason why a spin one operator can acquire a non-vanishing one-point function is related to the non-chiral nature of the $R-$symmetry current. If one allows for parity odd contributions, it is not hard to see that, in the presence of a surface defect in four dimensions, a spin one current $j_{\mu}$ can acquire a non-vanishing one-point function only for the orthogonal directions $i=3,4$
\begin{align}
 \braket{j_i}_{\S}=a\frac{ \e_{ik}x^k}{|x_{\perp}|^{\D+1}}\,,
\end{align}
where $a$ is, in general, some undetermined constant. In our case we saw that this constant is determined in terms of $h$, the one-point function of the stress tensor \eqref{Tonept}.}
\begin{align}\label{u1curr1pt}
 \braket{j_z}_{\S}&=-\frac{h}{2z^2 \bar z}\,, &  \braket{j_{\bar z}}_{\S}&=\frac{h}{2z \bar z^2}\,,
\end{align}
where $j_z=\frac12(j_3+i j_4)$ and $j_{\bar z}=\frac12(j_3-i j_4)$. Notice that the $U(1)$ R-symmetry is preserved by the defect, but this does not imply that the current one-point function is vanishing as one can easily show using Stokes theorem. This statement holds true for any abelian symmetry that is preserved by the defect (the requirement of being abelian is crucial for the current itself to be uncharged under the preserved symmetry).

We now consider the defect two-point function of the operators in the displacement supermultiplet.
Using the results of the previous section one can derive relations between fermionic and bosonic correlators simply by considering the Ward identity
\begin{align}
\braket{\{G_{\frac12}^+,\mathbb{\L}_{\uparrow}^-(w) \mathbb{D}_{\downarrow}(0)\}}_{\S}=0 \,,
\end{align}
which leads to
\begin{align}
 \pa_w \braket{\mathbb{\L}_{\uparrow}^-(w) \mathbb{\L}_{\downarrow}^+(0)}_{\S}=\braket{\mathbb{D}_{\uparrow}(w) \mathbb{D}_{\downarrow}(0)}_{\S}\,,
\end{align}
and, in turn
\begin{align}
 \braket{\mathbb{\L}_{\uparrow}^-(w) \mathbb{\L}_{\downarrow}^+(0)}_{\S}=-\frac{C_D}{w^2 \bar w^3}\,.
\end{align}
We are now ready to consider the bulk to defect coupling.

In this case there are two types of supersymmetric Ward identities one needs to consider. First, we have the ordinary Ward identities with the preserved supercharges
\begin{align}
 \braket{\{G_{\frac12}^+,\tilde{J}_{\mu \dot \a}(z,0)\mathbb{D}_{\downarrow}(w)\}}_{\S}&=0\,, & \braket{\{G_{\frac12}^+,T_{\mu \nu}(z,0)\mathbb{\L}_{\uparrow}^-(w)\}}_{\S}&=0\,,
\end{align}
and analogous relations with other operators and other preserved supercharges. Secondly, we have other Ward identities, along the lines of \eqref{Wardbrokentransl} and \eqref{WardTcons}, generated by broken supercharges. For instance
\begin{align}
\braket{\{\mathsf{Q}_{\uparrow}^-,J_{\mu \a}(z,0)\}}_{\S}&=\int \d^2 w \braket{J_{\mu \a}(z,0)\mathbb{\L}_{\uparrow}^-(w)}_{\S} \,,\\
\pa^{\mu} \braket{\tilde{J}_{\mu \dot{\mathbf{1}}}(z,w)\mathbb{\L}_{\downarrow}^+(0)}_{\S}&=-\delta^2(z) \braket{\mathbb{\L}_{\uparrow}^-(w)\mathbb{\L}_{\downarrow}^+(0)}_{\S}\,.
\end{align}
Implementing all the constraints we find that the only consistent solution requires the validity of \eqref{relation}. For completeness, we report the result of all the correlators in appendix \ref{sec:allcorr}, all of which are fixed in terms of $h$. As we have already stressed, the argument we just outlined is sufficient to prove the validity of \eqref{relation} for any superconformal surface defect in four dimensions. Nevertheless, in section \ref{sec:chiralalgebra} we will be interested in the specific case of $\mathcal{N}=(2,2)$ surfaces in $\mathcal{N}=2$ superconformal theories. For this reason, in the next section we provide some additional details on the $\mathcal{N}=(2,2)$ example.

%% file: sections/2_2_Neq2.tex
%!TEX root = ../susysurfaces.tex
%%%%%%%%%%%%%%%%%%%%%%%%%%%%%%%%%%%%%%%%%%%%%
\subsection{\texorpdfstring{$\NN=(2,2)$ surfaces in $\NN=2$ SCFTs}{N=(2,2) surfaces in N=2 SCFTs}}
\label{sec:Neq2}
%%%%%%%%%%%%%%%%%%%%%%%%%%%%%%%%%%%%%%%%%%%%%

An $\mathcal{N}=(2,2)$ surface defect preserves 
\begin{align}
su(1,1|1)\oplus su(1,1|1) \oplus u(1)_{\mathcal{C}} \subset su(2,2|2)\,.
\label{eq:N22inN2}
\end{align}
Out of the generators of the $4d$  $\NN=2$ superconformal algebra collected in appendix~\ref{sec:superalgebra} the defect 
superalgebra has as fermionic generators the supercharges
\begin{align}
G_{-\tfrac12}^+&=Q_{\mathbf{1}}^2\,, & G_{-\tfrac12}^-&=\Qt_{2,\dot{\mathbf{2}}}\,, & \Gb_{-\tfrac12}^+&=Q^1_\mathbf{2}\,, & \Gb_{-\tfrac12}^-&=\Qt_{1,\dot{\mathbf{1}}}\,,
\label{eq:defQ}
\end{align}
and conjugate conformal supercharges
\begin{align}
G_{+\tfrac12}^- &= S_2^\mathbf{1}\,, & G_{+\tfrac12}^+&=\St^{2 \dot{\mathbf{2}}}\,, &  \Gb_{+\tfrac12}^-&=S_1^\mathbf{2}	\,, &  \Gb_{+\tfrac12}^+&=\St^{1,\dot{\mathbf{1}}}\,,
\label{eq:defS}
\end{align}
with the commutation relations given in appendix~\ref{app:defectalgebra}. The defect also preserves the $u(1)_r$ generator $r$ and the Cartan of the $su(2)_R$ symmetry, $\mathcal{R}=\frac12(\RR^1{}_1-\RR^2{}_2)$, which together with the orthogonal rotations $\mathcal{M}$, also preserved by the defect, can be recombined in the three $u(1)$ generators
\begin{align}
\mathcal{J}&= -2\mathcal{R}-\mathcal{M}+r\,, & \bar{\mathcal{J}}&= 2\mathcal{R}+\mathcal{M}+r\,, & \mathcal{C}&=\mathcal{R}+\mathcal{M}\,.
\label{eq:u1s}
\end{align}
The first two are part of the $2d$ $\NN=(2,2)$ superconformal algebra, and the last is a commutant.

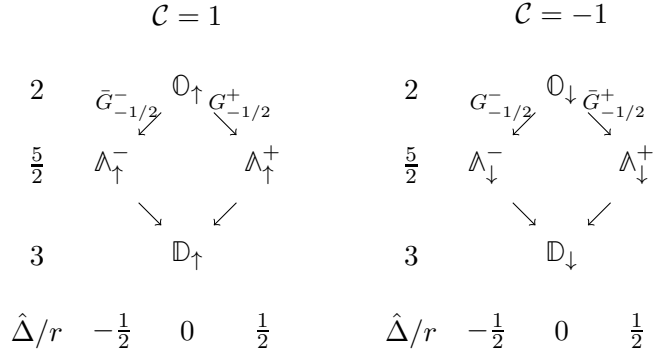
\begin{figure}
\begin{center}
\begin{tikzpicture}
\node at (-1,2.8) {$\Dh/r$};
\node at (-1,3.8) {$3$};
\node at (-1,5) {$\frac52$};
\node at (-1,6) {$2$};
\node at (0,2.8) {$-\frac12$};
\node at (1,2.8) {$0$};
\node at (2,2.8) {$\frac12$};
\node at (1, 7) {$\CC=1$};
    \node at (0,5) {$\mathbb{\L}_{\uparrow}^-$};
  \draw[->]  (0.35,4.5)--(0.65,4.2);
   \draw[->]  (1.65,4.5)--(1.35,4.2);
  \node at (1,3.8) {$\mathbb{D}_{\uparrow}$};
    \node at (2,5) {$\mathbb{\L}_{\uparrow}^+$};
      \draw[<-]  (0.35,5.4)--(0.65,5.7);
   \draw[<-]  (1.65,5.4)--(1.35,5.7);
\node at (1,6) {$\mathbb{O}_{\uparrow}$}; 
\node at (1.7,5.8) {$\scriptstyle{G_{-1/2}^+}$};
\node at (0.2,5.8) {$\scriptstyle{\bar{G}_{-1/2}^-}$};
  \end{tikzpicture}\hspace{1 cm}
\begin{tikzpicture}
\node at (-1,2.8) {$\Dh/r$};
\node at (-1,3.8) {$3$};
\node at (-1,5) {$\frac52$};
\node at (-1,6) {$2$};
\node at (0,2.8) {$-\frac12$};
\node at (1,2.8) {$0$};
\node at (2,2.8) {$\frac12$};
\node at (1, 7) {$\CC=-1$};
    \node at (0,5) {$\mathbb{\L}_{\downarrow}^-$};
  \draw[->]  (0.35,4.5)--(0.65,4.2);
   \draw[->]  (1.65,4.5)--(1.35,4.2);
  \node at (1,3.8) {$\mathbb{D}_{\downarrow}$};
    \node at (2,5) {$\mathbb{\L}_{\downarrow}^+$};
      \draw[<-]  (0.35,5.4)--(0.65,5.7);
   \draw[<-]  (1.65,5.4)--(1.35,5.7);
\node at (1,6) {$\mathbb{O}_{\downarrow}$};   
\node at (1.7,5.8) {$\scriptstyle{\bar{G}_{-1/2}^+}$};
\node at (0.2,5.8) {$\scriptstyle{G_{-1/2}^-}$};
  \end{tikzpicture}
  \end{center}
  \caption{The supermultiplets containing the displacement operator, and the operators appearing from the breaking of $su(2)_R$ and supersymmetry. Both supermultiplets are short, with the one on the left being $(a,c)$ and the one on the right $(c,a)$.}
  \label{fig:displacement}
  \end{figure}

Following same procedure used in section \ref{sec:Neq1}, we now obtain the structure of the displacement supermultiplet, which has been worked out in \cite{Gaiotto:2013sma}. 
We start from the broken currents. In this case, the lowest dimensional conserved currents that are broken are precisely the $su(2)_R$  currents, $t_{\mu \mathcal{I}}{}^{\mathcal{J}}$ with $\mathcal{I}\neq \mathcal{J}$, and, accordingly, two dimension two defect scalar operators are produced by the Ward identities
\begin{align}
 \pa^{\mu}t_{\mu 2}{}^{1}&=-\delta^2(z) \mathbb{O}_{\uparrow}\,, &  \pa^{\mu}t_{\mu 1}{}^{2}&=-\delta^2(z) \mathbb{O}_{\downarrow}\,.
\end{align}
Also in this case, these defect excitations can be interpreted as the result of the action of two broken generators $\mathsf{R}_{\uparrow}=\RR^1_{\phantom{1}2}$ and $\mathsf{R}_{\downarrow}=\RR^2_{\phantom{2}1}$ on the defect
\begin{align}
 [\mathsf{R}_{\uparrow}, \S]&=\int \d^2w\ \mathbb{O}_{\uparrow}(w) \S\,, &  [\mathsf{R}_{\downarrow}, \S]&=\int \d^2w\ \mathbb{O}_{\downarrow}(w) \S\,.
\end{align}
Similarly, the action of the broken supercharges 
 \begin{align}
 \mathsf{Q}_{\uparrow}^{+}&=Q^1_\mathbf{1}\,, & \mathsf{Q}_{\uparrow}^{-}&=\tilde 
 Q_{2\dot{\mathbf{1}}} \,,  & \mathsf{Q}_{\downarrow}^{+}&=-Q^2_\mathbf{2}\,,  &  \mathsf{Q}^{-}_{\downarrow}&=-\tilde{Q}_{1\dot{\mathbf{2}}}\,,
\end{align}
produces a total of four defect fermions
\begin{align}
 [\mathsf{Q}_{\uparrow}^\pm, \S]&=\int \d^2w\ \mathbb{\L}^{\pm}_{\uparrow}(w) \S\,, &  [\mathsf{Q}^{\pm}_{\downarrow}, \S]&=\int \d^2w\ \mathbb{\L}^{\pm}_{\downarrow}(w) \S\,.
\end{align}
Finally, the broken translations produce the displacement operator which must be a top component, since $[G_{\frac12}^\pm,\mathsf{P}_{\uparrow}]=[G_{\frac12}^\pm,\mathsf{P}_{\downarrow}]=[\Gb_{\frac12}^\pm,\mathsf{P}_{\uparrow}]=[\Gb_{\frac12}^\pm,\mathsf{P}_{\downarrow}]=0$.
It is then a purely algebraic exercise to compute the action of the preserved supercharges on these defect operators and one easily realizes that they fit in the two short multiplets shown in figure~\ref{fig:displacement}.
The multiplet on the left is an $(a,c)$  short multiplet, \ie it is annihilated by $G_{-\frac12}^-$ and $\bar{G}_{-\frac12}^+$. Consistently with $su(1,1|1)$ representation theory, the superprimary operator has quantum numbers $L_0=-\frac{\JJ}{2}= \Lb_0=\frac{\bar \JJ}{2}=1$. All operators in this multiplet have charge one under the commutant $u(1)_\CC$. Analogously, the multiplet on the right is a $(c,a)$ multiplet, annihilated by $G_{-\frac12}^+$ and $\bar{G}_{-\frac12}^-$, and it has charge $\CC=-1$ under the commutant. The quantum numbers of the remaining operators can be obtained from those of the supercharges, but for convenience we present the values of $\Dh = L_0 + \Lb_0$ and $r=\frac12\left(\JJ + \bar{\JJ}\right)$. The supersymmetry variations of these supermultiplets are collected in appendix~\ref{app:displ22}.

\subsubsection{Correlation functions}
As we did for the $\mathcal{N}=1$ case, we list the non-vanishing correlation functions involving the stress tensor and the displacement supermultiplets. Since we have already learned that supersymmetry requires the validity of the relation \eqref{relation}, in the following we express all the correlators in terms of $h$. We start again from the one-point function of the stress tensor supermultiplet. The components of the $\mathcal{N}=2$ supercurrent multiplet are summarized in appendix \ref{sec:susyvars}. Together with the stress tensor operator, whose one-point function was given in \eqref{Tonept}, also the scalar superprimary $O_2$ and the $su(2)_R$ current $t_{\mu\mathcal{I}}{}^{\mathcal{J}}$ acquire a non-vanishing one-point function
\begin{align}
  \braket{O_2}_{\S}&=-\frac{3\, h}{2 z \bar z}\,, & \braket{t_{z1}{}^1}_{\S}&=\frac{3\, h}{4 z^2 \bar z}\,, & \braket{t_{\bar z 1}{}^1}_{\S}&=\frac{3\, h}{4 z \bar z^2}\,.
  \label{eq:Neq2onepoints}
\end{align}
where $t_{z\II}{}^\JJ= \frac12(t_{3\II}{}^\JJ+ i t_{4\II}{}^\JJ)$ and $t_{\bar{z}\II}{}^\JJ= \frac12(t_{3\II}{}^\JJ- i t_{4\II}{}^\JJ)$
Moving to defect correlation functions, it is not hard to see that the only non-vanishing correlators are
\be
\begin{aligned}
\label{eq:defecttwopoint}
 \braket{\mathbb{O}_{\uparrow}(w)\mathbb{O}_{\downarrow}(0)}_{\S}&=\frac{-6h}{w^2\bar w^2}\,, \qquad & \braket{\mathbb{D}_{\uparrow}(w)\mathbb{D}_{\downarrow}(0)}_{\S}&=\frac{24h}{w^3\bar w^3}\,, \\
 \braket{\mathbb{\L}^+_{\uparrow}(w)\mathbb{\L}^-_{\downarrow}(0)}_{\S}&=\frac{12h}{w^3\bar w^2}\,, & \braket{\mathbb{\L}_{\uparrow}^-(w)\mathbb{\L}_{\downarrow}^+(0)}_{\S}&=\frac{12h}{w^2\bar w^3} \,.
\end{aligned}
\ee
Finally, there is a long list of bulk to defect correlators. We only spell out those that are relevant for the discussion in section \ref{sec:chiralalgebra}, where we will be interested in a subsector of states that are in cohomology of a particular supercharge. Specifically, we will need correlators involving the $su(2)_R$ current and the displacement superprimary $\mathbb{O}$. These are given by
\be
\begin{aligned}
\label{eq:diplstress}
 \braket{t_{w 2}{}^1(z,w) \mathbb{O}_{\downarrow}(0,0)}_\S&=\frac{3h \bar w}{\pi (w \bar w + z \bar z)^3}\,, &  \braket{t_{\bar w 2}{}^1(z,w) \mathbb{O}_{\downarrow}(0,0)}_\S&=\frac{3h  w}{\pi (w \bar w + z \bar z)^3}\,,\\
  \braket{t_{w 1}{}^2(z,w) \mathbb{O}_{\uparrow}(0,0)}_\S&=\frac{3h \bar w}{\pi (w \bar w + z \bar z)^3}\,, &  \braket{t_{\bar w 1}{}^2(z,w) \mathbb{O}_{\uparrow}(0,0)}_\S&=\frac{3h  w}{\pi (w \bar w + z \bar z)^3}\,,\\
  \braket{t_{\bar{z} 2}{}^1(z,w) \mathbb{O}_{\downarrow}(0,0)}_\S&=-\frac{3h z}{\pi (w \bar w + z \bar z)^3}\,, &  \braket{t_{z 2}{}^1(z,w) \mathbb{O}_{\downarrow}(0,0)}_\S&=\frac{3h  w \bar w}{\pi z (w \bar w + z \bar z)^3}\,,\\
  \braket{t_{z 1}{}^2(z,w) \mathbb{O}_{\uparrow}(0,0)}_\S&=-\frac{3h \bar z}{\pi (w \bar w + z \bar z)^3}\,, &  \braket{t_{\bar{z} 1}{}^2(z,w) \mathbb{O}_{\uparrow}(0,0)}_\S&=\frac{3h  w \bar w}{\pi \bar z (w \bar w + z \bar z)^3}\,,
\end{aligned}
\ee
where $t_{w\II}{}^\JJ= \frac12(t_{1\II}{}^\JJ- i t_{2\II}{}^\JJ)$ and $t_{\bar{w}\II}{}^\JJ= \frac12(t_{1\II}{}^\JJ+ i t_{2\II}{}^\JJ)$.
This concludes our discussion on the kinematics of superconformal defects. We now briefly explore some of the physical consequences of the relation \eqref{relation}.

%% file: sections/3_consequences.tex
%!TEX root = ../susysurfaces.tex
%%%%%%%%%%%%%%%%%%%%%%%%%%%%%%%%%%%%%%%%%%%%%
\section{Physical consequences}
\label{sec:consequences}
%%%%%%%%%%%%%%%%%%%%%%%%%%%%%%%%%%%%%%%%%%%%%

The relation \eqref{relation} provides an interesting identity between apparently independent pieces of defect CFT data. The physical relevance of the operators involved, moreover, leads to a relation between two of the Weyl anomaly coefficients. We also discuss the implications of this relation in different examples, and put forward the proposal of a relation between $C_D$ and $h$ for defects of arbitrary dimension, in $d-$dimensional  SCFTs,  \eqref{genrel}, which could hold for sufficiently supersymmetric defects. Finally, we discuss the implications of the relation \eqref{relation} for the stress tensor defect OPE.

\subsection{Weyl anomaly coefficients}
Even dimensional CFTs are generically affected by Weyl anomalies. The trace of the stress energy tensor, in a generic curved background, acquires a non-vanishing expectation value which can be expressed as a linear combination of geometric structures. The classification of conformal anomalies can be formulated as a cohomology problem: one has to look for solutions to the Wess-Zumino consistency conditions that cannot be expressed as a Weyl variation of a local term. A similar procedure applies to the case of even dimensional defects, where the presence of an induced metric and of the extrinsic curvature leads to a richer range of possibilities \cite{Graham:1999pm}. For the case of a two-dimensional surface, a common basis for the Weyl cohomology is given by \cite{Schwimmer:2008yh}
\begin{align}\label{Weylanom}
\braket{T_{\mu}{}^{\mu}}_{\S}=-\frac{\delta^2(z)}{2\pi} \left(b R_{\S} +d_1 \tilde{K}^i_{ab} \tilde{K}_i^{ab}-d_2 \g^{ab} \g^{cd} W_{acbd}\right)\,,
\end{align}
where $R_{\S}$ is the two-dimensional Ricci scalar, $\tilde{K}^i_{ab}$ is the traceless part of the extrinsic curvature $\tilde{K}^i_{ab}=K^i_{ab}-\frac12 K^i \g_{ab}$, with $K^i=\g^{ab} K^i_{ab}$, and $W_{abcd}$ is the pullback of the bulk Weyl tensor contracted with the inverse of the induced metric $\g^{ab}$. The anomaly coefficients $b$, $d_1$ and $d_2$ appear in several different contexts. The $b$ coefficient, associated to a A-type anomaly, is determined by the expectation value of the spherical defect and it was shown to be monotonically decreasing under defect RG flows \cite{Jensen:2015swa}. This prevents its dependence on defect marginal couplings, although it still allows for a non-trivial dependence on the bulk marginal parameters \cite{Herzog:2019rke,Bianchi:2019umv}. This dependence was shown to be absent in the case of supersymmetric defects preserving at least two supercharges of opposite chirality \cite{Bianchi:2019umv}. The B-type anomaly coefficients $d_1$ and $d_2$, on the other hand, are non-trivial functions of both defect and bulk marginal couplings and they can be mapped to defect CFT data. In four dimensions \cite{Bianchi:2015liz,Lewkowycz:2014jia} they are related to the two-point function of the displacement operator and to  the one-point function of the stress tensor by
\begin{align}\label{anomcoeff}
 d_1&= \frac{\pi^2}{16} C_D\,,  & d_2=3\pi^2 h\,.
\end{align}
This implies, in particular, that $d_1\geqslant 0$. Furthermore, assuming the validity of the averaged null energy condition in the presence of a defect one can prove that $d_2\geqslant 0$ \cite{Jensen:2018rxu}. Crucially, in section \ref{sec:Neq1} we have shown that, for any supersymmetric surface defect
\begin{align}\label{anomalyrel}
 d_1=d_2\,,
\end{align}
which, in particular, implies $d_2 \geqslant0$.
We now consider the implications of this result for some examples of superconformal surface defects.

\subsection{Comparison with holography and higher codimension}\label{holography}
The first holographic computation of the conformal anomaly for a two-dimensional submanifold goes back to the seminal paper \cite{Graham:1999pm} (see \cite{Schwimmer:2008yh} for a reorganization of the result in the basis \eqref{Weylanom}). In that case, the authors find $d_1=d_2$ for holographic theories with an Einstein gravity dual. This is consistent with our result and suggests an extension of the equality $d_1=d_2$ to any superconformal surface defect in dimension higher than four. In other words, if a relation exists between the displacement two-point function and the stress tensor one-point function for superconformal surfaces, consistency with holography requires it to be
\begin{align}\label{conjp2}
 C_D=\frac{q+1}{q-1}\frac{16\pi^{\frac{q-2}{2}}}{\Gamma(\frac{q}{2})} h\,, \qquad \text{for } p=2\,,
\end{align}
where we use $p$ to indicate the defect dimension and $q$ for the codimension. We also used the relation between $d_2$ and $h$ in arbitrary dimension \cite{Jensen:2018rxu}. For the Wilson surface defect in $d=6$ this gives $C_D=\frac{80 \pi h}{3}$, a result that was confirmed by a free theory computation for the theory of a single free tensor multiplet \cite{Gustavsson:2004gj}  and that is not valid for a free non-supersymmetric theory \cite{Henningson:1999xi}. Therefore, we have strong evidence that supersymmetry enforces the relation \eqref{anomalyrel} for any codimension.

After the initial study of \cite{Graham:1999pm,Berenstein:1998ij}, various other holographic computations were performed, both in four and six dimensions \cite{Corrado:1999pi,Buchbinder:2007ar,Gomis:2007fi,Drukker:2008wr,Koh:2008kt,DHoker:2008lup,DHoker:2008rje,Jensen:2013lxa,Gentle:2015jma,Gentle:2015ruo,Estes:2018tnu}. To the best of our knowledge, however, all these results can be related to the value of $b$, \ie to the spherical defect expectation value, or to the value of $d_2$, \ie the stress tensor one-point function. Therefore, the relation \eqref{anomalyrel}, provides a whole new set of predictions for the value of $d_1$, which we briefly summarize.

For the case of the Gukov Witten surface defects \cite{Gukov:2006jk} in $\mathcal{N}=4$ SYM theory, the one-point function of the stress tensor operator was computed in various limits in \cite{Drukker:2008wr}. Consistently with the supersymmetric Ward identities described in section \eqref{sec:Neq2}, the scalar superprimary $O_2$ (in \cite{Drukker:2008wr} it is called $\mathcal{O}_{2,0}$) and the stress tensor one-point function are determined by the same function $d_2$ (or equivalently $h$). The class of defects described in \cite{Gukov:2006jk} are disorder operators characterized by a codimension two singularity for the gauge and scalar fields along the defect profile $\S$. When a $U(N)$ gauge group is broken to a Levi subgroup $L=\prod_{l=1}^{M}U(N_l)$, the defect is labeled by $4M$ parameters $(\a_l,\beta_l,\g_l,\eta_l)$, where $\a_l$ is associated to the gauge field configuration, $\eta_l$ to the $\theta$-angles and $\beta_l+i\gamma_l$ to a complex scalar field (see \cite{Gukov:2006jk} for a detailed description). Prescribing a singular behavior for the complex scalar field breaks the symmetry of rotations in directions orthogonal to the defect, which we are assuming throughout this work, and thus our results do not directly apply. Henceforth we will set $\beta_l=\gamma_l=0$. A semiclassical gauge theory description of these defects is effective in the limit of small 't Hooft coupling $\lambda\ll 1$.
In the opposite regime, \ie $N\gg1$ and $\lambda \gg 1$, the same system admits two different gravitational descriptions. In general, half-BPS surface defects in $\mathcal{N}=4$ are described holographically as a system of intersecting D3 branes \cite{Constable:2002xt}. In the probe approximation, the conformal defect corresponds to $M$ stacks of probe D3 branes in $AdS_5\times S^5$ intersecting the boundary along the defect profile $\S$, where each stack contains $N_l$ coincident D3-branes. Of course, for the probe approximation to be valid, the number of probe D3 branes needs to be small compared to $N$. The marginal parameters of the gauge theory solution are mapped to geometric parameters of the gravity solution. The second strong coupling description consists in a smooth ten dimensional solution of Type IIB supergravity, which is asymptotically $AdS_5\times S^5$ and it captures the complete D3 brane backreaction \cite{Gomis:2007fi}. The stress tensor one-point function has been computed in all these different approximations and it has been reinstated in terms of anomaly coefficients in \cite{Jensen:2018rxu}.
Using the relation \eqref{anomalyrel} we can now complete the list with\footnote{The result of \cite{Gomis:2007fi} also includes a term depending on the $\beta_l$ and $\gamma_l$ parameters that we are setting to zero such that our results can be directly applied. Note that our anomaly coefficients differ from those in \cite{Jensen:2018rxu} by a factor of 12.}
\begin{align}\label{d1N4}
 d_1=\frac14 \left(N^2-\sum_{l=1}^{M} N_l\right)\,.
\end{align}
%for the case where  $\beta_l=\gamma_l=0$ where our results can be directly applied.
Notice that the classical gauge theory computation only captures the term of order $\frac1{\lambda}$, which vanishes when $\beta_l=\gamma_l=0$, while the two holographic descriptions give a result that is consistent with it when the corresponding approximations are taken into account.
The non-trivial agreement between computations in very different regimes of \cite{Gomis:2007fi} hints that the result \eqref{d1N4} may be exact, even though eq.~\eqref{d1N4} was obtained as a large $N$ result. It would be interesting to confirm this expectation through an integrability or a localization computation. Generalizations preserving less supersymmetry were considered in \cite{Koh:2008kt}, but these examples do not preserve orthogonal rotations and therefore we do not consider them here.

\subsection{Supersymmetric R\'enyi entropy}
A physically interesting example of conformal defect is the twist operator \cite{Hung:2014npa,Calabrese:2004eu,Bianchi:2015liz}, an extended probe whose expectation value computes the R\'enyi entropy. The latter can be defined by taking a QFT in flat $d$-dimensional spacetime and considering its density matrix $\rho$, which describes the state of the QFT in a given time slice. Tracing out all the degrees of freedom associated to a region of space $\bar A$, one obtains the reduced density matrix associated to the complementary region $A$
\begin{align}
 \rho_A=\Tr_{\bar A} (\rho) \,.
\end{align}
The R\'enyi entropy is defined as a function of a parameter $n$
\begin{align}\label{eq:defren}
 S_{n}=\frac{1}{1-n}\log \Tr (\rho_A^n)\,,
\end{align}
and the limit $n\to 1$ gives the entanglement entropy between the regions $A$ and $\bar A$. The evaluation of \eqref{eq:defren} in QFT is a notoriously hard task and it is usually tackled by a path integral construction commonly known as the replica trick. For the case of CFTs, however, one can treat the twist operator as a conformal defect \cite{Bianchi:2015liz}. This approach turned out to be particularly useful in the study of the dependence of the R\'enyi entropy on the shape of the entangling surface (the codimension two surface separating the two spacetime regions). In this context, the relation \eqref{anomalyrel} was observed for free theories in \cite{Lewkowycz:2014jia,Lee:2014xwa} and conjectured to hold for any CFT. At the same time various other conjectures on the shape dependence of the R\'enyi entropy were put forward for different geometrical configurations \cite{Mezei:2014zla,Bueno:2015rda,Bueno:2015lza,Bueno:2015qya}. In \cite{Bianchi:2015liz} all these proposals were reinterpreted, in a defect perspective, as a relation between $C_D$ and $h$
\begin{align}\label{conjq2}
 C_D=(p+2)2^{p+2} \frac{\Gamma(\frac{p+3}{2})}{\pi^{\frac{p-1}{2}}} h\,, \qquad  \text{for } q=2\,,
\end{align}
where both $C_D$ and $h$ are now functions of the replica parameter $n$. The proposal was shown to hold in the limit $n\to 1$ \cite{Faulkner:2015csl}, but it failed holographically \cite{Dong:2016wcf,Bianchi:2016xvf}. Interestingly, a supersymmetric generalization of the R\'enyi entropy \eqref{eq:defren} was put forward in \cite{Nishioka:2013haa} (see also \cite{Hama:2014iea,Huang:2014pda,Zhou:2015kaj,Nishioka:2016guu} for higher dimensional generalizations). An important property of the supersymmetric R\'enyi entropy is that it has the same $n\to 1$ limit as the ordinary R\'enyi entropy. Furthermore, for the four-dimensional case, our proof in section \ref{sec:Neq1} obviously applies, leading to the natural expectation that the relation \eqref{conjq2} holds for supersymmetric R\'enyi entropies in any dimension. As a consequence, if supersymmetry enforces a relation between any superconformal defect of codimension 2, for consistency with supersymmetric R\'enyi entropy this relation has to be \eqref{conjq2}. This observation, combined with other empirical data, leads us to formulate a proposal for a general relation between $C_D$ and $h$ in arbitrary dimension, which we describe in the next subsection.

\subsection{A conjecture for the general relation}
The first instance of a conjectured relation between $C_D$ and $h$ appeared in the context of supersymmetric Wilson lines \cite{Lewkowycz:2013laa}, where the displacement two-point function measures the energy emitted by an accelerated particle \cite{Correa:2012at}. Although in a conformal collider setup one would expect the stress tensor one-point function to measure the same energy, it turns out no universal relation can be found between $C_D$ and $h$, and only supersymmetry enforces such a connection \cite{Bianchi:2018zpb}\footnote{The authors of \cite{Lewkowycz:2013laa} were forced to introduce a deterioration of the stress tensor (\ie modify a traceless stress tensor by an automatically conserved term, which spoils its tracelessness) to reproduce a relation between $C_D$ and $h$ that is consistent with holography. A recent discussion on the reasons why the argument of \cite{Lewkowycz:2013laa} does not provide the correct result is given in \cite{Fiol:2015spa}. It would be interesting to try and perform a similar calculation for the case of surface defects.}. Nevertheless, consistency with the holographic predictions allows us to propose that the relation found in \cite{Lewkowycz:2013laa}
\begin{align}\label{conjp1} 
C_D= \frac{q}{q-1}\frac{12\pi^{\frac{q-2}{2}}}{\Gamma(\frac{q}{2})} h\,, \qquad \text{for } p=1\,,
\end{align}
is valid for any superconformal line defect. 

It is now a simple exercise to put together the relations \eqref{conjp2}, \eqref{conjq2} and \eqref{conjp1} to formulate a general relation that is expected to hold for any superconformal defect in any dimension
\begin{align}\label{genrel}
 C_D=\frac{2^{p+1} (q+p-1)(p+2)}{q-1} \frac{\Gamma(\frac{p+1}{2})}{\pi^{\frac{p+1}{2}}} \frac{\pi^{\frac{q}{2}}}{\Gamma(\frac{q}{2})} h\,,
\end{align}
where we assume $q>1$ since the stress tensor one-point function vanishes for $q=1$, consistently with the pole in \eqref{genrel}. Let us stress that, at the moment, we cannot make a statement on the amount of supersymmetry that is needed for this relation to hold, but we claim that, whenever a relation exists it has to take this form. Furthermore, to the best of our knowledge, there is no counterexample to this relation for a defect that preserves a $p$-dimensional superconformal algebra. Notice that, since superconformal algebras exist only for $d\leq6$, only the $p=q=3$ case is not included in the relations \eqref{conjp2}, \eqref{conjq2} or \eqref{conjp1}. It is important to mention that the procedure we used to derive the relation \eqref{relation} in section \ref{sec:Neq1} can be straightforwardly extended to higher dimensions and there is no conceptual obstacle in testing the proposal \eqref{genrel}. We leave this analysis for future work.

\subsection{Stress tensor defect OPE}
As it was already pointed out in \cite{Bianchi:2015liz}, the relation \eqref{conjq2} has intriguing consequences on the stress tensor defect OPE. In light of our proof of the relation \eqref{relation}, we focus on the case of a surface defect in $4d$ and we consider the terms in the stress tensor defect OPE which involve the displacement operator and its conformal descendants. We will show that \eqref{relation} leads to a vanishing coefficient for the most singular terms in a Lorentzian sense, \ie in our language, for $z\to 0$ with fixed $\bar z$.\footnote{Here Lorentzian means that, if we were to insert a defect in Minkowski space, the limit $z\to 0$ at fixed $\bar z$ would correspond to the stress tensor approaching the lightcone of a spacelike defect.}  Matching dimensions and charges under orthogonal rotations it is easy to check that the most singular terms in this limit appear in 
\begin{align}
 T_{w \bar w}(z) &\sim \a \frac{\mathbb{D}_{\downarrow}}{z}\,,  & T_{wz}(z) &\sim \delta \frac{\bar z \pa_w \mathbb{D}_{\downarrow}}{z}\,, \\
 T_{z \bar z}(z) &\sim \zeta \frac{\mathbb{D}_{\downarrow}}{z}\,, & T_{zz}(z) &\sim \zeta  \frac{\bar z \mathbb{D}_{\downarrow}}{z^2}\,.
 \label{TdefOPE2}
\end{align}
Staring at the correlation functions in appendix \ref{sec:allcorr} one immediately realizes that they are not consistent with these defect OPE expansions and therefore we are forced to set $\alpha=\delta=\zeta=0$. As mentioned, it was noted in \cite{Bianchi:2015liz} that this is a consequence of \eqref{relation}. Notice, however, that this does not mean that the stress tensor defect OPE is less singular than one would normally expect. Indeed, other operators may appear that are lighter than the displacement and would lead to more singular terms. Furthermore, it is important to note that in the $T_{zz}(z)$ defect OPE there is a term $\mathbb{D}_{\uparrow}/z$ with a non-vanishing OPE coefficient which would compete with  \eqref{TdefOPE2} in the Euclidean OPE. This is the reason why we need to focus on the Lorentzian OPE limit. Actually, it turns out the contribution to $T_{zz}(z)$ is the only singular term in the stress tensor OPE containing the displacement operator. Its OPE coefficient can be easily computed from the correlators in appendix \ref{sec:allcorr}
\begin{align}
 T_{zz}(z) &\sim \frac{\mathbb{D}_{\uparrow}}{2\pi z}\,, &  T_{\bar z\bar z}(z) &\sim \frac{\mathbb{D}_{\downarrow}}{2\pi \bar z} \,.
\end{align}
All the other terms involving the displacement are non-singular and proportional to a conformal descendant of the displacement operator.

%% file: sections/4_chiralalgebra.tex
%!TEX root = ../susysurfaces.tex
%%%%%%%%%%%%%%%%%%%%%%%%%%%%%%%%%%%%%%%%%%%%%
\section{Chiral algebras of \texorpdfstring{$\NN=(2,2)$}{N=(2,2)} surface defects}
\label{sec:chiralalgebra}
%%%%%%%%%%%%%%%%%%%%%%%%%%%%%%%%%%%%%%%%%%%%%

Any $\NN \geqslant 2$ four-dimensional superconformal field theory possess a subsector isomorphic to a two dimensional chiral algebra \cite{Beem:2013sza}. This subsector is obtained by restricting local operators to lie on a plane, and passing to the cohomology of a nilpotent supercharge, $\qq$, such that the anti-holomorphic dependence is $\qq$-exact, and one obtains a two-dimensional chiral algebra. We will denote the chiral algebra associated to a given SCFT, $\TT$, by $\chi(\TT)$.
An $\NN=(2,2)$ surface defect orthogonal to the plane where we define the chiral algebra, such that it intersects it at a point, preserves the supercharge $\qq$ used for the construction. This defect insertion gives rise, in $\qq$-cohomology, to non-vacuum modules of  $\chi(\TT)$ \cite{defectLCW,Cordova:2017mhb}. The modules introduced by different defects in various SCFTs were studied in \cite{defectLCW,Cordova:2017mhb,Nishinaka:2018zwq} by obtaining the (graded) partition function of the module of the $\chi(\TT)$ introduced by the defect. This is achieved by computing, in four dimensions, the Schur limit of the superconformal index \cite{Gadde:2011uv}, which is an invariant of the SCFT that counts (with signs) certain short multiplets that cannot recombine to form long multiplets. It was shown that this particular limit of the superconformal index matches the (graded) partition function of the chiral algebra \cite{Beem:2013sza,defectLCW,Cordova:2017mhb}, both with and without defects. While the superconformal index provides information on which operators are in $\qq$-cohomology, it suffers from ambiguities and does not always provide enough information to fully identify the modules.
An attempt to obtain directly correlation functions  in Lagrangian $2d$-$4d$ coupled was carried out in \cite{Pan:2017zie} using supersymmetric localization. The authors set up the computation of the correlation function between two defect operators and a bulk operator, however, they were unable to evaluate the expressions and provide results for these correlation functions.

\bigskip

In what  follows we determine which defect operators are in $\qq$-cohomology and we find that the two most universal operators, the defect identity and the displacement multiplet, have a representative in chiral algebra. Other operators, such as defect exactly marginal deformations, can also play a role in chiral algebra.
This provides a new computational tool for defect correlation functions. In particular, the one-point function of the stress tensor, $h$, is related to the dimension in chiral algebra of the defect identity (see footnote \ref{defectid}), $h_\sigma^\chi$, as given in \eqref{eq:hsigma}.
However, not all the defect operators in cohomology can be obtained by the action of chiral algebra generators on the defect identity, namely the defect generically inserts a reducible module over the original chiral algebra. Therefore, if we are handed the module corresponding to a non-trivial defect, it is not simple to identify which chiral algebra operators correspond to the defect operators we want to study, as most gradings of defect operators are not preserved by the construction. From the quantum numbers of defect operators only the commutant $\CC$ (see \eqref{eq:u1s}), and any flavor charges the theory may have, are visible in chiral algebra. With the goal of identifying the defect identity, we determine its chiral algebra properties following from four-dimensional OPE selection rules.
We also propose that all defect operators with charge $\CC$  have chiral algebra dimension $h_\sigma^\chi+\CC$, based on considerations involving correlation functions of defect operators and the superconformal index. Finally, we see how our results are realized in a few examples.

\subsection{Review: Chiral algebras of \texorpdfstring{$4d$ $\NN\geqslant2$}{N>=2} SCFTs}
\label{sec:chiralalgreview}

We start by giving a quick review of the chiral algebra construction without defect insertions, and refer to \cite{Beem:2013sza} for all details. For this construction we restrict operators to lie in the $(x_3,x_4)$ plane, which we parameterize by $z$ and $\zb$ according to \eqref{eq:zzb}. The generators of the $\sl(2) \times \overline{\sl(2)}$ conformal symmetry on the chiral algebra plane are
\be
\begin{alignedat}{4}
&2 \Lchi_{-1}= \PP_{{\bf{1}} \pd} \,, \qquad  2 \Lchi_{+1}& =  \KK^{\pd {\bf{1}}} \,, \qquad 2 \Lchi_0&= \HH + \MM\,, \\
&2 \Lbchi_{-1}= -\PP_{{\bf{2}} \md} \,, \qquad 2 \Lbchi_{+1} &= - \KK^{\md {\bf{2}}} \,, \qquad 2 \Lbchi_0&= \HH - \MM\,,
\end{alignedat}
\ee
where we added the superscript $\chi$ to avoid confusion with the $L_m$ and $\Lb_m$ generators on the defect plane introduced in section~\ref{sec:kin}.
The chiral algebra is obtained by passing to the cohomology of a nilpotent supercharge. There are two such choices, up to an arbitrary phase $\zeta$, that give rise to the same cohomology \cite{Beem:2013sza}:
\be
\begin{alignedat}{3}
&\qq_1 = \QQ^1_{\bf{2}}  + \zeta \St^{2 \md} \,, \qquad \qq_2 &=\SS_1^{\bf{2}} - \frac{1}{\zeta} \Qt_{2 \md} \,,  \\
&\qq_1^\dagger  = \SS_1^{\bf{2}}  +\frac{1}{\zeta} \Qt_{2 \md} \,, \qquad 
\qq_2^\dagger &=   \QQ^1_{\bf{2}}  - \zeta \St^{2 \md}\,.
\label{eq:funnyQ}
\end{alignedat}
\ee
At the origin of the chiral algebra plane, it was shown that the cohomology classes of $\qq_i$ consist of operators satisfying the conditions
\be
\tfrac12 \left(\Delta - (j_1+j_2)\right) - \RR=0\,, \quad r+ (j_1-j_2)=0\,,
\label{eq:Schurops}
\ee
where $\Delta$ is the conformal dimension, $j_1$, $j_2$ are the eigenvalues of $\MM_{\bf{1}}^{\phantom{{\bf{1}}}{\bf{1}}}$ and $\MM^{\pd}_{\phantom{\pd} \pd}$, $\RR$ the cartan of the $su(2)_R$ symmetry and $r$ the $u(1)_r$. These operators are dubbed \emph{Schur operators} as they are the ones that contribute to the Schur limit of the superconformal index that we review in subsection~\ref{sec:index}.

The $\Lchi_{-1,0,1}$ generators of $\sl(2)$ commute with $\qq_i$, and so we are free to translate the operators in the $z$ direction. However, to translate the operators in $\zb$, and have them remain in cohomology, we must perform a twisted translation using the diagonal subalgebra $\widehat{\sl(2)}$ of the $\overline{\sl(2)}\times su(2)_R$
\be
 \Lhchi_{-1}= \Lbchi_{-1} - \zeta \RR_- \,, \qquad \Lhchi_{+1}= \Lbchi_{+1} +\frac{1}{\zeta} \RR_+\,, \qquad \Lhchi_0 = \Lbchi_0 - \RR\,.
 \ee
The twisted $\widehat{\sl(2)}$ is $\qq_i$-exact
\bea
\label{eq:Lhatqexact}
-\{\qq_1, \Qt_{1 \md}\} &= \zeta  \{\qq_2, \QQ^2_{\bf{2}} \}= \Lhchi_{-1}\,, \\
-\frac{1}{\zeta}\{\qq_1, \SS_{2}^{\bf{2}} \} &= - \{\qq_2, \St^{1\md} \}= \Lhchi_{1}\,, \\
\{\qq_1, \qq_1^\dagger \} &= \{\qq_2, \qq_2^\dagger \}= \Lhchi_{0}\,,
\eea
and thus the $\qq_i$-cohomology classes are holomorphic, depending only on $z$.
Operators are then moved to an arbitrary $(z,	\zb)$ position by the \emph{twisted translations}
\be 
\OO(z,\zb) \colonequals e^{z \Lchi_{-1} + \zb \Lhchi_{-1}} \OO(0,0) e^{-z \Lchi_{-1}-\zb \Lhchi_{-1}}  \,,
\label{eq:twistedtrans}
\ee
or equivalently, noting that operators obeying \eqref{eq:Schurops} always transform in non-trivial $su(2)_R$ representations, by
\be 
\OO(z,\zb) \colonequals u_{\mathcal{I}_1}(\zb) \ldots u_{\mathcal{I}_{2R}}(\zb) \OO^{\mathcal{I}_1\ldots \mathcal{I}_{2R}}(z,\zb) \,, \qquad \text{with } \quad
u_\mathcal{I}(\zb) = (1 , - \zeta \zb)\,,
\label{eq:chiraltwist}
\ee
where $\OO$ is in the spin $R$ representation of $su(2)_R$ and $\mathcal{I}_i=1,2$ is an $su(2)_R$ fundamental index. 

The cohomology classes of the twisted translated operators
\be 
\OO(z)\colonequals \left[\OO(z,\zb)\right]_\qq\,,
\ee
depend only on $z$ and have meromorphic OPEs, being those of a two-dimensional chiral algebra. The $\Lchi_0$ weight of a four-dimensional operator in chiral algebra is given by
\be 
\Lchi_0 = \frac{\Delta+(j_1+j_2)}{2} = \Delta-\RR\,.
\label{eq:L0weight}
\ee

\subsubsection*{Stress tensor}

Among the operators in $\qq_i$-cohomology the $su(2)_R$ current, $t_{\mu}^{\mathcal{I}\mathcal{J}}$, will play an important role in the rest of the paper. It gives rise to the chiral algebra stress tensor, and is responsible for the enhancement of geometric $sl(2)$ on the chiral algebra plane to a full Virasoro symmetry.
Explicitly, the chiral algebra stress tensor is obtained by 
\be \label{eq:2dST}
T(z) \colonequals\left[\kappa u_\mathcal{I}(\zb) u_\mathcal{J}(\zb)  t_{{\bf{1}} \pd}^{\mathcal{I} \mathcal{J}} (z,\zb)\right]_\qq=\left[\kappa \left(  t_{{\bf{1}} \pd}^{11} (z,\zb)- 2\zb \zeta  t_{{\bf{1}} \pd}^{12} (z,\zb)+ \zb^2 \zeta^2  t_{{\bf{1}} \pd}^{22}(z,\zb) \right)\right]_\qq\,,
\ee
where $\kappa$ is fixed by demanding the canonical normalization for the two-dimensional stress tensor.
The OPE of the twisted translated $su(2)_R$ current becomes \cite{Beem:2013sza}\footnote{Our conventions for the $su(2)_R$ current and the stress tensor are given in appendix~\ref{sec:susyvars}.}
\be 
T(z) T(0) \sim -\frac{6 c \kappa^2 \zeta^2}{\pi^4} \frac{1}{z^4} + \frac{2 \kappa \zeta}{\pi^2} \frac{T(0)}{z^2} + \qq_i\text{-exact} + \ldots \,,
\ee
thus fixing the normalization to be
\be 
\kappa = \frac{\pi^2}{\zeta}\,.
\label{eq:kappaST}
\ee
We recover the relation between the four-dimensional central charge $c$ -- the two-point function of the stress tensor -- and the two dimensional one\footnote{We take the standard conventions for the central charge in $\NN=2$ SCFTs in which a single free hypermultiplet has $c=\frac{1}{12}$ and a single free vector multiplet has $c=\frac{1}{6}$.}
\be 
\label{eq:c2dc4d}
c_{2d}=-12 c\,.
\ee
The modes of the stress tensor $\LT_{0,\pm1}$ were argued in \cite{Beem:2013sza} to match the global $sl(2)$ modes $\Lchi_{0,\pm1}$, when acting on local operators. Thus, the dimension in chiral algebra, $h^\chi_\OO$, of a bulk operator $\OO$ is given by its eigenvalue under \eqref{eq:L0weight}.

\subsubsection*{Flavor symmetries}

If the four-dimensional theory has a continuos flavor symmetry, \ie a continuos symmetry that commutes with the superconformal algebra,  the general lore states that there will exist a conserved current that generates the symmetry. Conserved flavor currents are a top component of a half-BPS superconformal multiplet  -- $\hat{\BB}_1$ in the classification of \cite{Dolan:2002zh} --  whose superconformal primary is in the cohomology of $\qq_i$. The superprimary corresponds to the moment map operator, a dimension two scalar that is a triplet of $su(2)_R$ and, by belonging to the same multiplet of the current itself, transforms in the adjoint representation of the flavor symmetry. In chiral algebra, flavor symmetries give rise to affine Kac-Moody (AKM) current algebras \cite{Beem:2013sza}, where the current is obtained by the twisted translations of the moment map $\mu$
\be \label{eq:2dflavor}
J^A(z) \colonequals \left[\kappa_J u_\mathcal{I}(\zb) u_\mathcal{J}(\zb)  \mu^{A \, \mathcal{I} \mathcal{J}} (z,\zb)\right]_\qq=\left[\kappa_J \left(  \mu^{A \, 11} (z,\zb)- 2\zb \zeta  \mu^{A \, 12} (z,\zb)+ \zb^2 \zeta^2  \mu^{A \, 22}(z,\zb) \right)\right]_\qq\,.
\ee
Here $A$ is an adjoint index of the flavor symmetry algebra.
The OPE of two moment maps, given in \eqref{eq:momentmapope}, becomes
\be 
J^A(z) J^B(0) \sim  \frac{- k_{4d} \kappa_J^2 \zeta^2 \delta^{AB}}{32 \pi^4 z^2} + \frac{\kappa_J \zeta \, \ii f^{ABC} J^C(0)}{4 \pi^2 \, z}+  \qq \text{-exact}  +\ldots \,,
\label{eq:AKMOPE}
\ee
where $A,B,C$ are again adjoint indices, and $f^{ABC}$ the structure constants of the algebra. After fixing
\be 
\kappa_J= \frac{4 \pi^2}{\zeta}\,,
\label{eq:kappaJ}
\ee
we recognize the OPE of AKM currents with level \cite{Beem:2013sza}\footnote{Note that we work in conventions where the length of the longest root of the flavor algebra is $\sqrt{2}$, which means the level of the current algebra, $k_{2d}$, is equal to the two-point function of the AKM currents.  Our conventions for $k_{4d}$ are the standard for $\NN=2$ SCFTs given for example in \cite{Argyres:2007cn}, where a single free hypermultiplet enjoys an $su(2)$ flavor symmetry with $k_{4d}=1$.  The two-point function of the flavor current is given in 	\eqref{eq:flavorcurrent}.}
\be 
k_{2d} = - \frac{1}{2} k_{4d}\,.
\ee 

\subsection{Chiral algebras with defects}
\label{sec:chiralalgebradefects}

Next we consider introducing an $\NN=(2,2)$ surface defect extended along the $(x_1,x_2)$ directions and intersecting the chiral algebra plane at the origin. The generators of the $4d$ $\NN=2$ SCFT preserved by this defect were discussed in section~\ref{sec:Neq2}, and among them one finds precisely the supercharges used for the cohomological construction of the chiral algebra in \eqref{eq:funnyQ}.\footnote{In this work we restrict to a single defect introduced at the origin of the chiral algebra plane, and do not try to translate the defect.}  Note that when we insert the flat defect at the origin of the chiral algebra plane it will also intersect the chiral algebra plane at infinity.
We now want to consider correlation functions of local operators in the presence of the defect. The local operators can be both defect operators  (inserted at the origin of the chiral algebra plane or at infinity), or bulk operators inserted in an arbitrary position. Let us start by looking at the latter.

\subsubsection*{Bulk operators}

We start by noting that $\Lhchi_{\pm 1}$ are still $\qq$-exact, even though they are given by the action of $\qq$ on a broken supercharge eq.~\eqref{eq:Lhatqexact}, while  $\Lhchi_0$ is $\qq$-exact with respect to preserved supercharges. The full construction briefly reviewed in the previous subsection goes through, with operators in chiral algebra being those in \eqref{eq:twistedtrans}. These operators and their OPEs  (both in four-dimensions and in chiral algebra) are precisely those of the theory without defects, but they are no longer enough to compute correlation functions of bulk operators.
Note that the proof of the independence of chiral algebra correlation functions on marginal deformations, used in \cite{Beem:2013sza,Baggio:2012rr}, does not hold in the presence of the surface defect, as we now have less preserved symmetries. This means that chiral algebra correlation functions, in the presence of the defect, can depend non-trivially on both bulk and defect exactly marginal couplings. In particular, the one point function of the $su(2)_R$ current is generically expected to depend on all couplings.  

\subsubsection*{Defect operators}

We now analyze which defect operators are in $\qq_i$-cohomology, when inserted at the origin (both of defect plane and chiral algebra plane -- these are defect operators and thus cannot be translated in directions orthogonal to defect without translating the defect). For defect operators to be in cohomology they must satisfy the  two conditions given in eq.~\eqref{eq:Schurops}, which we write in terms of defect quantum numbers  (equations \eqref{2dconf} and \eqref{eq:u1s}) as
\be 
L_0 = -\frac12 \JJ \,, \qquad \Lb_0 = \frac12 \bar{\JJ}\,.
\label{eq:Schurdefops}
\ee
Unitarity of the defect theory implies that these are superprimaries of  $(a,c)$ supermultiplets with respect to the two-dimensional $\NN=(2,2)$ defect superalgebra.
The commutant of the defect superalgebra inside the four-dimensional $\NN=2$ algebra, denoted by $\CC$ in \eqref{eq:u1s}, matches $\Lchi_0$ for Schur operators, and this is the only quantum number of defect operators that is visible in cohomology.

Note that the defect also intersects the chiral algebra plane at infinity, and defect operators inserted at this intersection  must satisfy the opposite condition -- $(c,a)$ --  to be in cohomology.\footnote{This also happens for local operators inserted without the defect: at the origin we get $\OO^{1\ldots1}(0)$ and at infinity $\OO^{2\ldots2}(\infty)$, as can be seen by defining the out state from $\OO(z) = u_{\II_1}(\zb) \ldots u_{\II_n}(\zb) \OO^{\II_1 \ldots \II_n}$.} This makes two-point functions of defect operators in chiral algebra non-trivial, with the insertion of conjugate operators at the origin and at infinity.

It was argued in \cite{Cordova:2017mhb,defectLCW,Pan:2017zie} that the cohomological sector of defect operators forms a module over the original chiral algebra without defects, with the chiral algebra generators acting on the cohomology at the origin. In what follows we set out to study the properties of this module. Since the known non-renormalization theorems, that guarantee coupling independence of chiral algebra correlation functions without defects, do not apply, modules can in principle depend on all couplings of the theory. Most work so far has focused on the superconformal index and thus no example of coupling dependence is known to date. While the localization computation  for $2d$-$4d$ coupled systems of \cite{Pan:2017zie} provides a tool for the exact computation of defect correlation functions, their final expression is too hard to evaluate explicitly leaving the question of a possible coupling dependence unanswered. An alternative recipe to obtain localization results for the stress tensor one-point function is through the relation with a deformation in the background geometry and one could hope to extend the derivation of \cite{Bianchi:2019dlw} to the case of surfaces.

\subsection{Notable defect operators in chiral algebra}

We now look at a few noteworthy defect operators that are $(a,c)$ and thus make it to the $\qq_i$-cohomology  at the origin. The conjugate $(c,a)$ operators of the ones discussed here are in cohomology when inserted at infinity, and have opposite charge under $U(1)_{\CC}$.

\subsubsection*{Defect identity}

A trivial example of a defect operator satisfying the conditions \eqref{eq:Schurdefops}  is the defect identity $\hat{\mathbb{1}}$. As such, when inserting a defect orthogonal to the chiral algebra plane, we should think that we are inserting the defect identity. We denote its cohomology class by
\be 
\sigma(0) \colonequals \left[\hat{\mathbb{1}}\right]_\qq\,.
\ee
Since the defect intersect the chiral algebra plane at infinity as well, and the defect  identity is both  $(a,c)$ and $(c,a)$, $\sigma$ is also always inserted at infinity. In what follows we will normalize the defect to have a unit expectation value, such that there is no denominator in \eqref{eq:defcorrfn}.

\subsubsection*{Displacement supermultiplet}

A universal defect operator that must be present in any non-trivial defect is the displacement operator, arising from the breaking of translational invariance \eqref{WardDbarD}. 
For defects that break the $su(2)_R$ symmetry down to a $u(1)$, as the ones we are considering here, the two displacement operators are the top components of the $(a,c)$ and $(c,a)$ superconformal multiplets shown in figure~\ref{fig:displacement}. Both these superconformal multiplets are guaranteed to be present in any non-trivial defect. The superprimary of the former ($\mathbb{O}_{\uparrow}$) is a Schur operator, thus visible in chiral algebra at the origin with $\CC=1$, while the superprimary of the latter ($\mathbb{O}_{\downarrow}$)  is in cohomology when inserted infinity, and has $\CC=-1$.

\subsubsection*{Marginal deformations}

Exactly marginal deformations of the defect $\NN=(2,2)$ theory can be accommodated as top components of either $(c,a)$, $(a,c)$, $(a,a)$ or $(c,c)$ multiplets with $\CC=0$ and $L_0=\Lb_0=\frac12$. In particular this means that the defect conformal manifold always has even real dimension, as we need a pair of conjugate multiplets.
As pointed out above, correlation functions in chiral algebra can be non-trivial functions on this conformal manifold. Moreover, for marginal deformations arising from an $(a,c)$ defect supermultiplet, the superprimary of the multiplet makes it to the cohomology at the origin \eqref{eq:Schurdefops}. It will give rise, in chiral algebra, to an operator with $\CC=0$, that appears indistinguishable from the defect identity. The conjugate multiplet $(c,a)$ will appear in cohomology at infinity.

\subsubsection*{Broken flavor symmetries}

Whenever a defect breaks a flavor symmetry we have the following Ward identity
\be 
\partial_\mu J^{\mu A} = - \delta _\DD^{2}(x) \mathbb{J}^A (x)\,,
\ee
where $A$ runs over the generators of the flavor symmetry that were broken. This implies there is a scalar defect operator $\mathbb{J}$ of dimension two for each broken generator of the flavor symmetry. The flavor current is a top component of the $\hat{\BB}_1$ multiplet, and thus $\mathbb{J}$ is a defect top component as well. Just like the exactly marginal deformations, this top component can be accommodated in multiplets that are either $(c,a)$, $(a,c)$, $(a,a)$ or $(c,c)$.  Note that when $\mathbb{J}$ is uncharged under the preserved flavor symmetries, it corresponds to an exactly marginal operator. 

\subsubsection*{Broken extra supersymmetry}
\label{sec:extrasusy}

Bulk theories with supersymmetry algebras larger than $\NN=2$ will have extra supercurrents, as well as a larger $R-$symmetry.
From an $\NN=2$ point of view the extra $R-$symmetry appears as a flavor symmetry, namely $u(1)_f$ ($su(2)_f$) for theories with $\NN=3$  ($\NN=4$) supersymmetry. There will be a $\hat{\BB}_1$ multiplet for this ``flavor'' symmetry, and if the symmetry is broken by the defect then all the considerations above apply.

Furthermore, we are guaranteed there will exist additional four-dimensional superconformal multiplets, containing the extra supercurrents and extra $R-$symmetry currents -- $\DD_{\frac12,(0,0)}$ and $\bar{\DD}_{\frac12,(0,0)}$ in the classification of \cite{Dolan:2002zh}. Each of these multiplets contains a Schur operator, and thus has a representative in the bulk chiral algebra.
If the defect breaks some of the extra supercharges, the non-conservation of the supercurrent in a $\DD_{\frac12,(0,0)}$  ($\bar{\DD}_{\frac12,(0,0)}$) multiplet gives rise to two top components of two multiplets. In this case, the extra $R-$symmetry currents, transforming as an $su(2)_R$ doublet, with $u(1)_r$ charge $+1$ ($-1$ respectively), and charged under the ``flavor'' symmetry, will also be broken, giving rise to defect operators in the aforementioned defect multiplets.
The superconformal primaries of these multiplets have $\hat{\Delta}=\frac{3}{2}$, $\hat{\ell}=0$, and $\CC=\pm \frac12$, and they can be accommodated in $(c,c)$, $(c,a)$, $(a,c)$ or $(a,a)$ multiplets. Whenever they belong to $(a,c)$ multiplets the superconformal primary is a Schur operator, with the corresponding value of $\CC$, and seen in chiral algebra.

\subsection{Correlation functions and defect operator dimensions in chiral algebra}
\label{sec:corrfunct}

We can now study in chiral algebra correlation functions involving any number of bulk Schur operators and two defect operators (one at origin and one at infinity),
\be 
\langle \hat{\OO}_0(\infty) \OO_1(z_1) \ldots \OO_n(z_n) \hat{\OO}_{n+1}(0)\rangle_\S\,.
\ee
Here $\hat{\OO}_i$ denote defect Schur operators and $\OO_i$ twisted translated bulk Schur operators, which depend only on $z_i$.
Note that even if the defect operators are trivial, \ie the defect identity, they still give rise in chiral algebra to a non-vacuum operator, thus the above is always a $(n+2)$-point function in chiral algebra (provided none of the $\OO_i$ are the identity), even if it is a lower point function in four dimensions.

In chiral algebra, a Schur bulk operator will give rise to a two-dimensional operator with weight given by \eqref{eq:L0weight}, as can be checked by showing that the OPE of the two-dimensional stress tensor with a Schur operator reproduces precisely the action of $\Lchi_0$ \cite{Beem:2013sza}.
To answer the same question for defect operators we must consider their OPE with the stress tensor. We will do so for the two universal  supermultiplets present in any defect -- the defect identity and the displacement supermultiplet. For a generic Schur operator we will just constrain the form of correlation functions involving one bulk operator and two defect operators. Altogether these results lead us to the following proposal for the dimension in chiral algebra of an $(a,c)$ defect operator $\hat{\OO}$ 
\be
h^\chi_{\hat{\OO} }= h^\chi_\sigma + \CC_{\hat{\OO}}\,,
\label{eq:hSchurdefect}
\ee
where $h^\chi_\sigma$ is the dimension of the defect identity.
In particular this relation means that the monodromy as a bulk operator, $\OO_2$, goes around the defect follows simply from its dimension $h^\chi_{\OO_2}$ in chiral algebra, and the values of $\CC$
\be \label{eq:3pt}
\langle [\hat{\OO}_1 (\infty)]_\qq \, \OO (z)\, [\hat{\OO}_2(0)]_\qq \rangle = \frac{\lambda}{z^{h^\chi_{\OO} + h^\chi_{\hat{\OO}_2}-h^\chi_{\hat{\OO}_1}}} =  \frac{\lambda}{z^{h^\chi_{\OO} + \CC_{\hat{\OO}_2}-\CC_{\hat{\OO}_1}}} \,,
\ee
thus allowing $h^\chi_\sigma$ to be any real number without introducing branch cuts in the correlators.

\subsubsection{One-point function of the \texorpdfstring{$su(2)_R$}{su(2)R} current}

Denoting the operator that the defect identity inserts in the chiral algebra plane by $\sigma$, the stress tensor one-point function gives rise to the following three-point function in chiral algebra
\be 
\langle \sigma(\infty) T(z) \sigma(0) \rangle  =  - 2 \kappa \zb \zeta \langle t_{{\bf{1}} \pd 1}^{\phantom{{\bf{1}} \pd} 1}(z,\zb) \rangle_{\Sigma}  =-\frac{3\pi^2 h}{ z^2}\,,
\label{eq:chiralTonepoint}
\ee
where we used \eqref{eq:Neq2onepoints}.
Comparing with the expected result for the two-dimensional three-point function
\be 
\langle \sigma(\infty) T(z) \sigma(0) \rangle= \lim\limits_{z_3 \to \infty} z_3^{2 h^\chi_\sigma} \langle \sigma(z_3) T(z) \sigma(0) \rangle = \frac{h^\chi_\sigma}{z^2}\,,
\ee
we find the dimension of the defect identity in chiral algebra
\be 
h^\chi_{\sigma} = - 3\pi^2 h  = - d_2 = -\frac{\pi^2 C_D}{16} \leqslant 0 \,.
\label{eq:hsigma}
\ee
where $h$ and $d_2$ are defined in \eqref{Tonept} and \eqref{Weylanom}.
Note that the defect identity always gives rise to a negative dimension operator in chiral algebra, due to the relation \eqref{relation} and positivity (unitarity) of the displacement two-point function  $C_D$.

\subsubsection{Displacement supermultiplet correlation functions}

The displacement supermultiplet is shown in figure~\ref{fig:displacement}. The superprimary of the $(a,c)$ multiplet, $\mathbb{O}_{\uparrow}$, is in cohomology when inserted at the origin, while the superprimary of the $(c,a)$ multiplet, $\mathbb{O}_{\downarrow}$, is in cohomology at infinity. The former has $\CC=1$ and the latter has $\CC=-1$. From the defect two-point function \eqref{eq:defecttwopoint} of these operators we obtain the following chiral algebra two-point function
\be
\langle \left[\mathbb{O}_{\downarrow}(\infty)\right]_\qq \left[\mathbb{O}_{\uparrow}(0)\right]_\qq \rangle = - 6 h\,,
\label{eq:OO}
\ee
and from the bulk to defect two-point functions in eq.~\eqref{eq:diplstress} we get 
\begin{align}
\label{eq:OTs}
 \langle \left[\mathbb{O}_{\downarrow}(\infty)\right]_\qq T(z) \sigma(0) \rangle = - \frac{ 6 h \pi}{\zeta z} \,, \qquad
  \langle \sigma(\infty) T(z) \left[\mathbb{O}_{\uparrow} (0)\right]_\qq\rangle = - \frac{6h \zeta \pi}{z^3}\,.
\end{align}
where we used eqs.~\eqref{eq:2dST}~and~\eqref{eq:kappaST}.\footnote{Note that there is an explicit $\zeta$ appearing but it can be absorbed in the normalization of the chiral algebra operator arising from $\mathbb{O}_{\uparrow}$, which we took to be trivial.}
From \eqref{eq:OTs} together with \eqref{eq:3pt}, we find that the dimension of $\mathbb{O}_{\uparrow}$ in chiral algebra is
\be
h^\chi_{\mathbb{O}_{\uparrow}} = h^\chi_\sigma + 1\,,
\ee
compatible with \eqref{eq:hSchurdefect}. Then we can predict, from a chiral algebra computation, the value of the following twisted correlator
\be 
\langle \left[\mathbb{O}_{\downarrow}(\infty)\right]_\qq  T (z) \left[\mathbb{O}_{\uparrow} (0)\right]_\qq  \rangle = \frac{h^\chi_\sigma + 1}{z^2}\,,
\ee
which may also follow from four-dimensional superconformal Ward identities.\footnote{Note that even though the  correlation function of two defect and one bulk operator has a cross-ratio, after passing to the $\qq_i$ cohomology it becomes a chiral three-point function. In particular, the four-dimensional correlator of two defect operators at positions $x_{1,2}$ and a bulk operator at position $x_3$ depends on a single conformally invariant crossration given by
\be 
\xi = \frac{|x_{3}^\perp|^2 |x_{12}^{\parallel}|^2}{(|x_{13}^\parallel|^2 + |x_3^\perp|^2)(|x_{23}^\parallel|^2 + |x_3^\perp|^2)}\,,
\ee
where $x_{ij}=x_i-x_j$ and $\parallel$ ($\perp$) denotes the distance parallel (orthogonal) to the defect. When the bulk operator is restricted to lie on the chiral algebra plane, and the defect operators are placed at the origin and at infinity we have $\xi=1$, and thus the chiral algebra captures the value of the three-point function for $\xi=1$.}

\subsubsection{Correlation functions of Schur operators}

Let us now consider a generic three-point function of a bulk Schur operator, $\OO$, restricted to the chiral algebra plane, a defect $(c,a)$ Schur operator $\hat{\OO}_1$ placed at infinity, and a defect $(a,c)$ Schur operator $\hat{\OO}_2$ placed at the origin.
The Schur conditions fix the quantum numbers of the defect operators to be
\be 
\Dh_1= -s_1 - 2 R_1 \geqslant 0 \,, \qquad \Dh_2= s_2 +2 R_2 \geqslant 0\,, \qquad \hat{\ell}_1=r_1 \, \qquad \hat{\ell}_2 = - r_2\,, 
\label{eq:defectSchurqn}
\ee
where $\Dh=L_0 + \Lb_0$ and is positive due to defect unitarity, and $s$ and $\hat{\ell}$ are the eigenvalues of $\MM$ and $\MM_\parallel$ respectively. The bulk operator transforms in a non-trivial representation of $su(2)_R$ of spin $R$, obeying \eqref{eq:Schurops}. The chiral algebra operator 
$\OO(z) = \left[\OO(z,\zb) \right]_\qq$ is made from the twisted translations in \eqref{eq:chiraltwist}, and thus involves summing over all  $su(2)_R$ components.
Considering the component with Cartan eigenvalue $R^0$, and imposing the symmetries of the problem we find
\be 
\zb^{R-R^0} \langle [\hat{\mathcal{O}}_1(\infty)]_\qq  \, \mathcal{O}_{R^0}(z,\zb) \, [\hat{\mathcal{O}}_2(0)]_\qq \,\rangle=\frac{\lambda_{12\OO}}{z^{h^\chi_{\mathcal{O}}+\CC_1+\CC_2 }} = \frac{\lambda_{12\mathcal{O}}}{z^{h^\chi_{\OO} + h^\chi_{\hat{\OO}_2}-h^\chi_{\hat{\OO}_1}}}\,,
\label{eq:threepointchiral}
\ee
where we must have $R_0+R_1+R_2=0$ and $r+r_1+r_2=0$. The three-point function of the twisted translated bulk operator will be a sum of terms like \eqref{eq:threepointchiral}, ranging over all the values of $R^0$ in the spin $R$ representation, and with suitable coefficients.
In chiral algebra, $\OO$ will have dimension $h^\chi_\OO$ given by \eqref{eq:L0weight}. 
Eq.~\eqref{eq:threepointchiral} shows that if two defect Schur operators have non-zero three-point function with some bulk operator, then their dimensions in chiral algebra differ from their value of $\CC$ by the same constant shift, compatible with \eqref{eq:hSchurdefect}.\footnote{Note that the conjugate operators appear at infinity and thus their dimensions are $h^\chi_{\hat{\OO}} = h^\chi_\sigma - \CC_\OO$, since they must have the same dimension in chiral algebra, and opposite value of $\CC$. } Taking $\hat{\OO}_1$ to be the defect identity we see that all operators that appear in the defect OPE of bulk Schur operators have dimensions given by \eqref{eq:hSchurdefect}. Note that there can be defect Schur operators that do not appear in the defect OPE of any bulk Schur operator.

\paragraph{Chiral algebra OPE from sending $\OO$ to $\hat{\OO}_2$:}
Sending $z\to0$ in chiral algebra amounts to taking the chiral algebra OPE between $\OO$  and $[\hat{\OO}_2]_\qq$, and extracting the operator conjugate to $[\hat{\OO}_1]_\qq$.
Note that unless $\hat{\OO}_2$ is the defect identity, the limit $z\to 0$ is not controlled by the defect OPE of the bulk operator, since there is a defect operator inserted. However, in chiral algebra both this limit and the defect OPE appear on the same footing, as an OPE between two chiral algebra operators. While the strength of the singularity of the defect OPE is controlled by the dimension of the bulk operator, when $\hat{\OO}_2$ is not the defect identity we see from \eqref{eq:threepointchiral} that the OPE can be arbitrarily singular -- the singularity is controlled by $\CC_2$ which is not bounded from above by unitarity.
When $\hat{\OO}_1$ is the defect identity, \ie when $\hat{\OO}_2$ appears in the defect OPE of $\OO$, unitarity requires the defect operator at the origin to obey
\be 
\CC_2 \geqslant -R_0\,.
\label{eq:nondefectOPE}
\ee
The inequality is saturated only if $\hat{\OO}_2$ has zero dimension, which we only allow for $R_0=0$, since there should be a single defect identity and it should be uncharged under $\CC$ to preserve all the symmetries.

\paragraph{Chiral algebra OPE from the defect OPE of $\OO$:}
Taking instead $\hat{\OO}_2$ to be the defect identity we can probe which operators are allowed to appear in the defect OPE of the bulk Schur operator $\OO$. In this case unitarity of the defect operator at infinity requires
\be 
\CC_1 \leqslant  R_0\,.
\label{eq:defectOPEC1bound}
\ee
This bounds the strength of the singularity of the defect OPE of a Schur operator in the spin $R$ representation of $su(2)_R$ by $z^{-h^\chi_\OO-R}$.
The operators $\hat{\OO}_1$ that can have non-trivial two-point function with the bulk Schur operator $\OO$ have dimensions given by \eqref{eq:defectSchurqn} with $R_1=-R_0$ and $r_1=-r$, and subject to \eqref{eq:defectOPEC1bound}.
These operators are the conjugates of the operators $\hat{\OO}$ that appear in the defect OPE of $\OO$, which must then satisfy
\be 
\OO(z) \sigma(0) \sim  [\hat{\OO}(0)]_\qq \,, \quad \text{with } \hat{\Delta}_{\hat{\OO}}=R_0 + \CC_{\hat{\OO}}\,, \quad R_0=-R, \dots,R\,, \quad \text{ and where } \CC_{\hat{\OO}} \geqslant -R_0\,,
\label{eq:defectopeC}
\ee
giving us OPE selection rules for $\OO$.
The OPE of the twisted translated Schur operator is obtained by summing the components with different $R_0$ according to \eqref{eq:chiraltwist}.
 In what follows we will spell out a few of these selection rules in detail for relevant bulk operators such as the $su(2)_R$ current and the moment maps. 

\subsection{OPE selection rules and properties of defect chiral algebras}
\label{sec:OPEselectionrules}

We will use the preserved symmetries to obtain defect OPE selection rules for different bulk Schur operators. This tells us which $(a,c)$ defect Schur operators can appear in the defect OPE of a given Schur bulk operator, which translates in chiral algebra to the OPE of the twisted translated bulk operator with $\sigma$.  Let us stress that there can be Schur defect operators that do not appear in the defect OPE of twisted-translated bulk Schur operators (see also the discussion in section \ref{sec:vortexfreehyper}). We will derive necessary but not sufficient conditions for an operator to appear, and in particular we will not impose particular shortening conditions the bulk operators may obey. As the rules we obtain are already quite restrictive, and enough for the purposes of this work, we leave obtaining the complete selection rules for the full superconformal multiplets for future work.

\subsubsection{\texorpdfstring{$su(2)_R$}{su(2)R} current defect OPE}
Using the conserved symmetries we find the following OPE between the components of the $su(2)_R$ current and the defect located at the origin
\begin{alignat}{2}
t^{11}_{{\bf{1}} \pd} &\sim \vert_{(a,c)} \hat{\OO}_{\hat{\Delta}= \CC+1, \hat{\ell}=0, s=\CC-1}\,, 	\qquad &\CC \geqslant -1 \,, \nn\\
\label{eq:SU2RdefOPE}
t^{12}_{{\bf{1}} \pd}  &\sim \vert_{(a,c)} \hat{\OO}_{\hat{\Delta}= \CC, s= \CC, \hat{\ell}=0} \,, 	\qquad &\CC \geqslant 0 \,,\\
t^{22}_{{\bf{1}} \pd}  &\sim \vert_{(a,c)}  \hat{\OO}_{\hat{\Delta}= \CC-1, s= \CC+1, \hat{\ell}=0} \,,	\qquad &\CC \geqslant 1  \,, \nn
\end{alignat}
where we listed only operators that are in the chiral algebra, and thus are $(a,c)$.
Here $\hat{\ell} = L_0 - \Lb_0$, $s$ is the eigenvalue of $\MM$, and the condition on $\CC$ comes from imposing unitarity of the defect operators, \ie  $\hat{\Delta} = L_0 + \Lb_0 \geqslant 0$. 
From the four-dimension selection rules \eqref{eq:SU2RdefOPE} we obtain the chiral algebra OPE using \eqref{eq:2dST}
\be 
\begin{split}
T(z) \sigma(0) \sim &\frac{0}{z^3} + \kappa \frac{\sum_i b^{11}_{i, \Dh=-s=1}\left[\hat{\OO}_{i, \Dh=-s=1}(0)\right]_\qq -2 \zeta b^{12}_{\hat{\mathbb{1}}}  \sigma(0)}{z^2} \\
&+ \kappa \frac{ -2\sum_i  \zeta b^{12}_{i,\Dh=s=1}  \left[\hat{\OO}_{i, \Dh=1, s=1}(0)\right]_\qq +   b^{11}_{\mathbb{O}_{\uparrow}}  \left[\mathbb{O}_{\uparrow}(0)\right]_\qq }{z} + \ldots\,.
\label{eq:TDOPE}
\end{split}
\ee
where the $b$ coefficients are four-dimensional bulk to defect couplings, their superscript labels the R-symmetry components in \eqref{eq:SU2RdefOPE} and the sums run over possible degenaracies. Although operators with the same value of $\mathcal{C}$ are indistinguishable in chiral algebra, for this analysis we find it useful to keep track of their four-dimensional origin. In \eqref{eq:TDOPE} we excluded the most singular term, a defect operator with $\CC=-1$, scaling dimension zero, and charged under transverse spin, \ie a charged defect identity, since we do not expect such an operator in a neutral defect that preserves $u(1)_\CC$. The absence of $z^{-3}$ and higher terms means $\LT_{+n} |\sigma\rangle=0$, for $n>0$. Operators with $\CC=0$ and $\hat{\Delta}>0$ can contribute to the $z^{-2}$ pole of the OPE, implying the action of $\LT_0$ on $\sigma$ is not necessarily diagonal. These operators correspond to superprimaries of multiplets whose top component are exactly marginal defect operators. Note that we have excluded half-integer powers of $\CC$ since we expect the stress tensor to be single valued around the defect.

The OPE coefficients of the defect identity, $\sigma$, and the displacement supermultiplet, $\mathbb{O}_{\uparrow}$, can be computed from $4d$ correlation functions, since  defect operators of different dimensions must be orthogonal. From \eqref{eq:chiralTonepoint} we find
\be 
 b^{12}_{\hat{\mathbb{1}}}  =\frac{3}{2}h = -\frac{h_\sigma}{2\pi^2}  \,.
 \ee
Then we have that $\LT_{-1} |\sigma \rangle$, that is $\partial \sigma$,  is a linear combination of the displacement supermultiplet and another type of multiplet:
\be
\partial\sigma(0) =  -2  \kappa \sum_i \zeta b^{12}_{i,\Dh=s=1}  \left[\hat{\OO}_{i, \Dh=1, s=1}(0)\right]_\qq +   \kappa b^{11}_{\mathbb{O}_{\uparrow}} \left[\mathbb{O}_{\uparrow}(0)\right]_\qq \,,
\label{eq:partialSigmadef}
\ee
with $i$ running over possible degenerate operators.
Computing the $4d$ two-point function \eqref{eq:OTs} from the chiral algebra OPE \eqref{eq:TDOPE} and using the normalization of $\Od_{\uparrow}$ given in \eqref{eq:OO} we also obtain
\be 
b^{11}_{\Od_{\uparrow}}  = \frac{1}{\pi}\,.
\ee
Then the chiral algebra two-point function
\be 
\langle \partial\sigma(\infty) \partial \sigma(0) \rangle  = \langle \sigma | 2 \LT_0 |\sigma \rangle = \langle \sigma | (2 h^\chi_\sigma |\sigma \rangle + |\mathrm{extra} \rangle)= 2 h^\chi_\sigma\,,
\ee
should match the computation in $4d$ following \eqref{eq:partialSigmadef}. After plugging the right coefficient for $\Od_{\uparrow}$ we find  that $b^{12}_{i,\Dh=s=1}  =0$.
\medskip

All in all, we obtain the following properties for the defect identity $|\sigma\rangle$
\begin{align}
\LT_{n} |\sigma \rangle &= 0\,, \;n>0\,,  & \LT_0 |\sigma \rangle &= h^\chi_\sigma |\sigma \rangle  + b^{11}_{ \Dh=-s=1} |\hat{\OO}_{\Dh=-s=1}\rangle\,, & \LT_{-1} |\sigma \rangle &= \frac{\kappa}{\pi} |\Od_{\uparrow} \rangle\,,
\end{align}
allowing for logarithmic representations, where additional operators with the same dimension as the defect identity in chiral algebra appear under the action of $\LT_0$. These other operators indicate the presence of \emph{exactly marginal} operators, so we expect them to be present only in theories with a defect conformal manifold. Distinguishing these operators from $|\sigma\rangle$ may be hard in chiral algebra, since they have the same properties.

\subsubsection{Preserved flavor symmetries}

Suppose now the bulk theory has a \emph{non-abelian} flavor symmetry, that is not broken by the defect, then selection rules fix the OPE of the AKM current \eqref{eq:2dflavor} as
\be 
J^A(z) \sigma(0) \sim \frac{0}{z^2} + \kappa_J \frac{\sum_i b^{11}_{i, J \Dh=-s=1}\left[\hat{\OO}_{i, \Dh=-s=1}^A (0)\right]_\qq }{z} + \ldots\,,
\ee
where again the first term is absent since it would correspond to a defect identity charged under transverse spin. In writing the above we assumed the current to be single valued around the defect otherwise the powers would not be integers.
This implies
\be 
J_{n}^A |\sigma \rangle =0\,, \; n \geqslant 1\,, \qquad J_0^A |\sigma \rangle = \sum_i |\hat{\OO}_i^A \rangle\,,
\ee
where once again the multiplet containing $\hat{\OO}^A$ accommodates a  dimension two scalar, neutral under all the $u(1)$s preserved by the defect, but now in the adjoint of the flavor group.

If the symmetry is instead \emph{abelian} the moment map as well as the associated current are clearly uncharged under the preserved symmetry and they can get a one-point function (see the discussion around eq.~\eqref{u1curr1pt} for the possibility of abelian preserved currents acquiring a one-point function) consistently with supersymmetric Ward identities. In particular, for the abelian AKM associated to the four-dimensional moment map we have
\be 
J(z) \sigma(0) \sim \frac{0}{z^2} + \kappa_J \frac{\sum_i  b^{11}_{i, J \Dh=-s=1}\left[\hat{\OO}_{i, \Dh=-s=1}(0)\right]_\qq - 2 \zeta b^{12}_{J \hat{\mathbb{1}}}  \sigma(0)  }{z} + \ldots\,,
\ee
for a single-valued current, and thus
\be 
J_{n} |\sigma \rangle =0\,, \; n \geqslant 1\,, \qquad J_0 |\sigma \rangle = \sum_i |\hat{\OO}_ i \rangle- 2 \kappa_J \zeta b^{12}_{J \hat{\mathbb{1}}} |\sigma \rangle\,,
\ee
where the multiplet $\hat{\OO}$ accommodates an  \emph{exactly marginal operator}, but now also the defect identity can appear under the action of $J_0$. We will see an example of this when computing the one-point function of a $u(1)_f$ flavor current for a monodromy defect in section~\ref{sec:monodromy}.

\subsubsection{Broken flavor symmetries}

If a non-abelian flavor symmetry is broken  the moment maps can acquire a one-point function and we find 
\be 
J^A(z) \sigma(0) \sim \frac{0}{z^2} + \kappa_J \frac{\sum_i b^{11 A}_{i, J \Dh=-s=1} \left[\hat{\OO}_{i, \Dh=-s=1}(0)\right]_\qq  - 2 \zeta b^{12A}_{J \hat{\mathbb{1}}}  \sigma(0) }{z} + \ldots\,,
\ee
where $ b^{12}_{J \hat{\mathbb{1}}}  $ can be related to the one-point function of the moment map, and $A$ runs over the broken flavor currents. Note that if the Cartan of the flavor symmetry is preserved then the one-point function must vanish. Here we have  assume again that the moment map is single valued around the defect, which does not hold generically, \eg, in the monodromy defect considered in section~\ref{sec:monodromy}. However, the modes $J^A_{n \geqslant 1}$ will always annihilate the defect identity, as the strength of the singularity is always less than two.

\bigskip

All in all, the defect identity is annihilated by the positive modes of the stress tensor and, if the bulk theory has a flavor symmetry, also annihilated by modes with $n \geqslant 1$ of the flavor current, irrespectively of the preservation or single-valuedness of the current. It should also be uncharged under any preserved flavor symmetries.

\subsection{Superconformal Index}
\label{sec:index}

As shown in \cite{Beem:2013sza}, and briefly reviewed in appendix~\ref{app:index}, the graded partition function of the two-dimensional chiral algebra matches the Schur limit of the superconformal index
\be
\II (q) = \Tr_{\mathcal{H}}\left( (-1)^{2(j_1-j_2)} q^{\Lchi_0} \right) =  \Tr_{\mathcal{H}} \left( (-1)^F q^{\CC} \right)\,,
\label{eq:SchurIndex}
\ee
where we used the two Cartans of the four-dimensional superconformal theory preserved by the chiral algebra, $F=2(j_1-j_2)$ and $\Lchi_0$, which for Schur operators matches $\CC$. This fact has also recently been proven using localization in \cite{Pan:2019bor}.
In \eqref{eq:SchurIndex} the trace is taken over the Hilbert space of the theory $\mathcal{H}$ in radial quantization, and the index counts (with signs) operators that satisfy \eqref{eq:Schurops}, \ie that are in the cohomology of  $\qq$ and thus make it to the chiral algebra.
The superconformal index can also be enriched by the presence of defects, by doing radial quantization about a point in the defect, now counting (with signs) the spectrum of defect operators. The Schur limit of the superconformal index  in the presence of an $\NN=(2,2)$ surface defect once again counts operators that have the right quantum numbers to be in chiral algebra, \ie that are $(a,c)$ defect operators.\footnote{These are precisely the same operators counted by the Schur index computed in \cite{Cordova:2017mhb}, even though they are referred to as  $(c,c)$ there.}
It was argued in \cite{defectLCW,Cordova:2017mhb} that the Schur index should then match the  character of the module introduced by the defect.
Note that $\Lchi_0$ does not match the action of the zero mode of the stress tensor $\LT_0$ on a defect operator. For instance, the defect identity has $\CC=0$, while it was argued to have a dimension in chiral algebra, $h^\chi_\sigma$, given by the one-point function of the stress tensor \eqref{eq:hsigma}. This means that for the index to match the graded partition function we must have
\be 
\LT_0= \Lchi_0 + h^\chi_\sigma= \CC  + h^\chi_\sigma\,,
\ee
holding for all $(a,c)$ defect operators, such that the partition function and index match, up to an overall power of $q^{h^\chi_\sigma}$. This matches our proposal given in eq.~\eqref{eq:hSchurdefect}. It also allows for $h^\chi_\sigma$ to have a dependence on both bulk and defect marginal couplings, while the index is invariant under all continuous parameters.

It was shown in \cite{Beem:2017ooy} that the character of the vacuum module, or equivalently the superconformal index of the theory without defects,  obeys certain linear modular differential equations (see also \cite{Arakawa:2016hkg}). The solutions of these equations form a vector-valued modular form, and thus one expects there to exist modules over the original chiral algebra whose characters appear under the modular transformations of the vacuum module.\footnote{The modular properties are of the chiral algebra partition function which is a trace of $q^{\LT_0-c_{2d}/24}$. This means that the characters that come out of this computation always have the prefactor $q^{\LT_0-c_{2d}/24}$ and are not normalized as the superconformal index where operators neutral under all cartans contribute as $1$. }  The dimension of these modules arising from interpreting the functional form of the character as that of a highest weight module, are given for several $\NN=2$ SCFTs in \cite{Beem:2017ooy}. 
We would then like to understand how these modules fit with those inserted by defects, particularly in light of the properties of the module inserted by the defect identity established in the previous subsection.
The $q \to 1$ behavior of the Schur index, \ie the vacuum module character, was related to the $a$ and $c$ anomaly coefficients of the four-dimensional theory in \cite{DiPietro:2014bca,Buican:2015ina,  Ardehali:2015bla}. Using the S-modular transformation the $q \to 1$ behavior of the vacuum character maps to the small $q$ behavior of the characters into which the vacuum character transforms under the S transformation. These characters are either solutions of the aforementioned modular linear differential equation  or of its conjugate, and the $q\to0$ limit is controlled by the character with the lowest dimensional state.\footnote{Here and in the following we ignore the subtleties of the case of $\frac{1}{2}-\mathbb{Z}$ graded chiral algebras where one needs to consider the conjugate differential equation -- we refer to \cite{Beem:2017ooy} for the precise treatment of these cases.} Thus one can relate the lowest dimension among the characters appearing under the modular transformation, $h^\chi_{min}$ to the $a$ and $c$  anomalies \cite{Beem:2017ooy,Cecotti:2015lab}, and bound it by the Hofman-Maldacena bounds \cite{Hofman:2008ar}:
\be\label{hmin}
h^\chi_{min} = 2 a - \frac{5}{2}c\,, \qquad -\frac{3}{2}c \leqslant h^\chi_{min} \leqslant 0\,.
\ee
As we have seen in \eqref{eq:hsigma} the defect identity always inserts a module of negative dimension, thus being compatible with $h^\chi_{min}$ being identified with a defect insertion. 
However, the dimensions identified from the functional form of the characters, of which $h^\chi_{min}$ is the lowest, do not exhaust all modules of the VOA.
An example is the case of Gukov Witten defects in $\mathcal{N}=4$ SYM, which have been reviewed in section~\ref{holography}. From \eqref{d1N4}  we get that the defect identity produces in chiral algebra an operator with dimension\footnote{As discussed in section~\ref{holography} this was obtained at large $N$ but since there are hints it holds at finite $N$ we will assume it here.}
\be 
h^\chi_\sigma =-  \frac{1}{4} \left(N^2 - \sum\limits_{l=1}^M N_l^2 \right)\,.
\label{eq:hsigmaGW}
\ee
It is curious to note that for $N=2,\ldots,7$ the partitions of $N$ of length two (\ie $M=2$) produce exactly the dimensions of the modules that are solutions to the modular linear differential equations of \cite{Beem:2017ooy}, given in table~5 of that reference. These dimensions appear with large degeneracies, which are presumably distinguished by considering the differential equations graded by the Cartan of the  $su(2)_f$  flavor symmetry, that $\NN=4$ SYM has when viewed as an $\NN=2$ theory, which is preserved by the defect \cite{Peelaerswip}.\footnote{We thank W.~Peelaers for many discussions on these solutions and on the defects they could correspond to. While the unflavored solution of the differential equation is logarithmic \cite{Beem:2017ooy}, it has been checked that for gauge group $SU(2)$ this is not the case for the solution graded by the Cartan of the flavor symmetry \cite{Peelaerswip,Beem:2017ooy}.}
The dimensions of all defects with $M\neq 2$ do not appear in the lists of \cite{Beem:2017ooy} and their dimension is generically lower than \eqref{hmin} (for $\mathcal{N}=4$ SYM with gauge group $SU(N)$ $h^\chi_{min}=-\frac{N^2-1}{8}$). As described in the example of a monodromy defect in section~\ref{sec:monodromy}, different modules, of distinct dimensions, can have the same functional form for their characters (and thus for the superconformal index). This provides a possible resolution for accommodating the missing modules in this case. We leave understanding how these modules are accommodated, and what is the significance of the module corresponding to $h^\chi_{min}$ for future work.

The superconformal index in the presence of surface defects has been computed in different examples by a variety of approaches. For defects admiting a Lagrangian description in terms of a $2d$-$4d$ coupled system the index was computed in \cite{Gadde:2013ftv,Nakayama:2011pa,Cordova:2017ohl}. In \cite{Cordova:2017ohl,Cordova:2017mhb,Fluder:2019dpf} the index was computed via a conjectural formula  in terms of the $2d$-$4d$ BPS spectrum in the Coulomb branch of the theory, being thus applicable to non-Lagrangian theories as well. Finally, for vortex defects a prescrisption to compute index was given in \cite{Gaiotto:2012xa} and used in a variety of different theories \cite{Alday:2013kda,Bullimore:2014nla,Cordova:2017mhb,Nishinaka:2018zwq,Watanabe:2019ssf}. Since some of the examples we will consider below are precisely vortex defects we will give a brief summary of how they are obtained.

\subsubsection*{Vortex defects}

Vortex defects admit a uniform construction using renormalization group flows along Higgs branches \cite{Gaiotto:2012xa}.\footnote{Some of these vortex defects have been given a microscopic description in terms of $2d$-$4d$ coupled systems, see \eg, \cite{Gadde:2013ftv, Bullimore:2014nla}.}
To consider a vortex defect in a specific SCFT $\TT$ one starts by embedding $\TT$ in an ultraviolet theory, $\TT_{UV}$, that flows to $\TT$ upon Higgsing a $u(1)_f$ flavor symmetry. By giving a constant expectation value to the Higgs branch operator, we find the original theory $\TT$ in the infrared, together with free hypermultiplets \cite{Gaiotto:2012xa}.
Instead, by turning on a position dependent expectation value for the Higgs branch operator that triggers the flow, we find $\TT$ with a surface defect inserted. 
This construction motivated the prescription of \cite{Gaiotto:2012xa},  whereby the Schur index of vortex defects in $\TT$ is computed from that of $\TT_{UV}$ by taking a certain residue in the fugacity associated to the $u(1)_f$, and stripping off the index of the decoupled free hypermultiplets that arise in the infrared.
This prescription has been used to compute the Schur index of different vortex defects in \cite{Cordova:2017mhb,Nishinaka:2018zwq,Watanabe:2019ssf}, which were then matched to the characters of non-vacuum modules of $\chi(\TT)$.
Note that a particular theory $\TT$ can be embedded in different $\TT_{UV}$ theories and the vortex defects created from different ultraviolet theories can be distinct.

In \cite{Cordova:2017mhb} a proposal was put forward for creating vortex defects in chiral algebra.
The Higgsing of a flavor symmetry $G$, by giving a nilpotent expectation value to the moment map operator of that flavor symmetry, has been given a conjectural image in chiral algebra in \cite{Beem:2013sza,Beem:2014rza}.
The nilpotent expectation value corresponds to giving an expectation value to the raising component of an $su(2)$ embedded in $G$. In the above discussion, the Higgsed $u(1)_f$ corresponds to the Cartan of $su(2)$ under the embedding. The chiral algebra procedure that implements this Higgsing, taking  $\chi(\TT_{UV})$ to $\chi(\TT)$, is a quantum Drinfeld-Sokolov (qDS) reduction of the flavor symmetry of the chiral algebra, with respect to the aforementioned embedding.
To produce a vortex defect instead, \ie Higgsing with a non-constant expectation value,  the authors of \cite{Cordova:2017mhb} propose that one should first perform a spectral flow \cite{Schwimmer:1986mf,Lerche:1989uy} for the $u(1)_f$ in $\chi(\TT_{UV})$, followed by a qDS reduction, with respect to the $su(2)$ embedding used. Thus the module introduced by the vortex defect is conjectured to be the qDS reduction of a spectral flow of the vacuum module of $\chi(\TT_{UV})$.

\subsubsection*{Spectral flow}

The spectral flow \cite{Schwimmer:1986mf,Lerche:1989uy} of a $\widehat{u(1)}_f$ flavor symmetry with current $J$, by $\alpha$ units, acts as an outer automorphism of the current algebra, such that commutation relations are unchanged.
In section~\ref{sec:examples} we will need the action of the spectral flow for the Cartan of an $\widehat{su(2)}_f$ flavor symmetry. We will take the $\widehat{su(2)}_f$ generators in the spin basis, denoting the Cartan by $J$, under which the remaining two generators have charge $\pm1$.\footnote{Note that in \eqref{eq:AKMOPE} we took the length of the longest root of the flavor algebra to be $\sqrt{2}$, which means that $J_n=\frac{1}{\sqrt{2}} J^3_n $.} Then the spectral flow acts as follows in the chiral algebra generators (see for example  \cite{Lesage:2002ch})
\be 
 J_n \rightarrow J_n + \alpha  \frac{k}{2} \delta_{n,0}\,, \quad L_n \rightarrow L_n + \alpha J_n + \frac{k}{4} \alpha^2 \delta_{n,0}\,, \quad \OO_{r} \to \OO_{r+ \alpha q_\OO}\,,
 \label{eq:spectralflow}
\ee
where $\OO$ is any operator charged under $u(1)_f$, with charge $q_\OO$, and with uncharged operators unaffected.
The commutation relations of the chiral algebra are unaffected by the spectral flow, but the spectral-flowed vacuum $|\alpha \rangle$ now obeys
\be 
\begin{split}
&J_{n \geqslant0} |\alpha \rangle = - \alpha \frac{k}{2} \delta_{n,0} |\alpha \rangle\,,  \quad
L_{n\geqslant0} |\alpha\rangle = \frac{\alpha^2 k}{4}\delta_{n,0} |\alpha\rangle\,, \quad L_{-1} |\alpha \rangle = - \alpha J_{-1} |\alpha \rangle\,,\\
& \OO_{r+ \alpha q_\OO > - h^\chi_\OO+ \alpha q_\OO}  |\alpha \rangle =0\,,  \quad \text{ for }\OO_r \text{ of dimension }h_\OO \text{ and charge } q_\OO\,.
\end{split}
\label{eq:alphaprops}
\ee
Note that the spectral-flowed vacuum is  still a Virasoro and AKM primary, but now it has non-zero charge and dimension.
However, for charged operators $\OO$, of charge $q_\OO$ and dimension $h^\chi_\OO$, it is no longer the case that modes of weight larger than $-h^\chi_\OO$ annihilate the spectral-flowed vacuum.
Indeed, from \eqref{eq:alphaprops} we see that an operator of charge $q_\OO$ has the following expansion
\be
\OO_{q_\OO} (z) |\alpha \rangle = \sum\limits_{m\leqslant -h^\chi_\OO} \frac{1}{z^{m + h^\chi_\OO + q_\OO \alpha} } (\OO_{q_\OO})_{m + q_\OO \alpha}| \alpha \rangle\,,
\label{eq:defectOPEmonodromy}
\ee
thus getting a pole of order  $q_\OO \alpha$.

\subsection{Chiral algebras of defect SCFTs}
\label{sec:examples}
We now look at a few examples to see how the above results are realized.

\subsubsection*{Monodromy defect for the free hyper}
\label{sec:monodromy}

Monodromy defects are codimension two defects at the end of a topological domain wall that implements an action of the flavor group.\footnote{A similar construction was used in \cite{Dedushenko:2019yiw} to define an R-symmetry monodromy defect.} 
Introducing a monodromy defect for a $u(1)_f$ flavor symmetry makes operators of charge $q$ under that symmetry pick up a phase of
\be 
\OO_q (e^{2 \pi \ii} z) = e^{- 2 \pi \ii \alpha q} \OO_q(z)\,,
\label{eq:monodromy}
\ee
when going around the monodromy defect.
The free hypermultiplet has an $su(2)_f$ flavor symmetry, with the hypermultiplet scalars transforming as a doublet. We want to introduce a monodromy defect for the the Cartan of $su(2)_f$, in the spin basis. Under this $u(1)_f$ the scalars in the hypermultiplet have a charge $\pm \frac12$, and thus pick up a phase between $0$ and $2\pi$ according to \eqref{eq:monodromy}, with $0 \leqslant \alpha < 2$.  
In \cite{Cordova:2017mhb} it was proposed that the introduction of a monodromy defect would correspond, in chiral algebra, to the spectral flow \eqref{eq:spectralflow} with respect to the $u(1)_f$ flavor symmetry.
Under the spectral flow the vacuum is mapped to $|\alpha\rangle $, and obeys the conditions spelled out in eq.~\eqref{eq:alphaprops}, which imply it is  a Virasoro and AKM primary, consistently with the conclusions of section~\ref{sec:OPEselectionrules} on the properties of the defect identity. The form of \eqref{eq:defectOPEmonodromy} is also compatible with the monodromy condition \eqref{eq:monodromy} when interpreted as a defect OPE, and after identifying $\alpha$ in both equations. In the following, we will carefully analyze how to create a monodromy defect compatible with the preserved $\mathcal{N}=(2,2)$ supersymmetry and discuss how the spectral flow results are reproduced for $0 \leqslant \alpha < 2$. Note that according to our discussion the spectral flow with $\alpha \geqslant 2$ is not interpreted as a monodromy defect, since the monodromy defect only exists for $\alpha <2$. 

We start by checking that the most singular term in the defect OPE, produced by the monodromy defect under this identification, is consistent with the OPE selection rules. 
The free hypermultiplet gives rise in chiral algebra to a pair of free sympletic bosons \cite{Beem:2013sza}, that is a $\beta \gamma$ system with weights $h^\chi_\beta=h^\chi_\gamma=\frac12$, which has central charge $c_{2d}=-1$.
The $su(2)_f$ flavor symmetry of the free hypermultiplets gets enhanced in chiral algebra to an AKM current algebra $\widehat{su(2)}_{-\frac12}$, generated by the following currents in the spin basis:
\be 
J =-\frac{1}{2} (\beta \gamma)\,, \qquad J^+ = \frac{1}{2} (\beta \beta) \,, \qquad J^- =  - \frac{1}{2}(\gamma \gamma)\,.
\ee
The defect OPE of the free hypermultiplets ($\hat{\BB}_{\frac12}$ multiplets in the classification of \cite{Dolan:2002zh}) is constrained by the selection rules given in \eqref{eq:defectopeC}, to which we must supplement the monodromy condition \eqref{eq:monodromy}. This condition translates into the requirement that the defect operators, $\hat{\OO}$, appearing in the defect OPE obey
\be 
\CC_{\hat{\OO}} = \frac{1}{2} - q \alpha +\mathbb{Z}\,,
\label{eq:Cinmonodromy}
\ee
where the $u(1)_f$ charge is $q =\frac12$ ($q=-\frac12$) for $\beta$ ($\gamma$).
We can thus write the  $\beta(z)$ defect OPE as
\be 
\beta(z) \sigma(0) \sim \sum\limits_{\CC > -\frac12}  \frac{b_{\hat{\OO}} \, \hat{\OO}_{\Dh = \frac12 +  \CC }(0)}{z^{\frac12-\CC }}   \,,
\label{eq:betasigmaOPEfreehyper}
\ee
with $\CC$ subject to \eqref{eq:Cinmonodromy}. Here we also imposed that the dimensions of defect operators appearing in a free scalar defect OPE are constrained as shown in \cite{Billo:2016cpy} by the equation of motion of a free scalar.\footnote{The strict inequality follows from demanding that the only operator with $\Dh=0$ is the defect identity which has $s=0$, where $s$ is the eigenvalue of $\MM$.}
$\gamma(z)$ will have an analogous OPE.
Recalling that $\beta$ and $\gamma$ have charges $\pm\frac12$, the singularity of order $q \alpha < 1$  predicted by eq.~\eqref{eq:defectOPEmonodromy}  matches precisely the most singular term allowed in the above selection rule.  Similarly, since all bulk operators in chiral algebra are made out of $\gamma$ and $\beta$, selection rules will be consistent with an  operator $\OO_q$, of charge $q$ having a pole of order $q \alpha < 2 q$, as predicted by \eqref{eq:defectOPEmonodromy}.

However, the identification of the defect identity with the spectral flowed vacuum,
\be 
|\sigma \rangle = |\alpha \rangle\,,
\ee
allows us to infer dynamical information about the defect theory.
From eq.~\eqref{eq:defectOPEmonodromy} it is clear that when $\alpha >0$, $\gamma(z)$ has a regular OPE while  $\beta(z)$ does not. This is a dynamical statement about the values of $b_{\hat{\OO}}$ appearing in the OPE \eqref{eq:betasigmaOPEfreehyper}.
More importantly, the scaling weight of $|\alpha \rangle$ given in \eqref{eq:alphaprops} gives the one point function of the stress tensor in the presence of a monodromy defect
\be
h= \frac{\alpha^2 }{24 \pi^2}\,,
\label{eq:hmonodromy}
\ee
where we used that the $\widehat{u(1)}_f$ is  the Cartan of the $\widehat{su(2)}_f$ flavor symmetry of the free hypermultiplets, and has level $k_{2d}=-\frac12$.
Similarly we can compute the one-point function of the flavor current supermultiplet. Since the Cartans of the $su(2)_f$  flavor symmetry and the $su(2)_R$ are preserved by the defect, only the neutral component of the moment map will acquire a one-point function. Starting from \eqref{eq:2dflavor}, this gives rise in chiral algebra to the three-point function
\be 
\langle  \sigma(\infty)  J^3 (z) \sigma(0) \rangle = - 2  \kappa_J  \zeta \zb \langle \mu^{3\, 12}(z,\zb) \rangle_\S\,,\
\ee
where $J^3$ is related to the spin basis we took for the spectral flow by $J  \sqrt{2} =  J^3$.
Identifying the defect identity with the spectral-flowed vacuum we find
\be
\langle  \sigma(\infty) J^3(z) \sigma(0) \rangle = \frac{\alpha }{2 \sqrt{2}} \frac{1}{z} \,, \quad \Rightarrow \quad \langle \mu^{3\, 12}(z,\zb) \rangle_\S = -\frac{\alpha }{16 \sqrt{2} \pi^2 }\frac{1}{z \zb }\,.
\label{eq:oneptmumonodromy}
\ee 

Let us now check these predictions, and thus the identification of the monodromy defect chiral algebra with the spectral flow in the range $0 \leqslant \a <2$, by computing these correlation functions explicitly, using the fact that the bulk theory is free.
The free hypermultiplet contains two complex scalars, which we will denote by $\QQ$ and $\tilde{\QQ}$, and which are highest weights of $su(2)_R$, and rotated under the $su(2)_f$ flavor symmetry. They can be grouped in the following doublet of $su(2)_R$ and $su(2)_f$
\be
\QQ^ {\II \hat \II} \colonequals \left(\begin{array}{cc} \QQ & \tilde{\QQ} \\ \tilde{\QQ}^* &  -\QQ^*  \end{array}\right)\,,
\ee
where $\II$ ($\hat{\II}$) is an $su(2)_R$ ($su(2)_f$) fundamental index. The chiral algebra fields are obtained as
\be
\beta(z) \colonequals \left[\QQ^{\hat{\II}=1}(z,\zb)\right]_\qq\,, 
\qquad
\gamma(z) \colonequals \left[\QQ^{\hat{\II}=2}(z,\zb)\right]_\qq\,.
\label{eq:betagamma}
\ee
We start from a trivial defect, where defect operators are obtained by evaluating bulk operators at the location of the defect, keeping in mind that derivatives in directions orthogonal to the defect give rise to new defect primaries.
The trivial defect OPE of a free scalar is simply given by a Taylor expansion,
\be
\QQ(z,\bar z) \sim \QQ(0,0)+\sum_{n>0} \left(\frac{1}{n!} z^n  \pa_z^n  \QQ(0,0)+\frac{1}{n!}\zb^n \pa_{\zb}^n \QQ(0,0) \right) \,,
\label{eq:trivialdef}
\ee
where we took into account the equations of motion to write only a sum over defect primaries. The OPE of the remaining free scalars will be the same.
We now introduce the monodromy as a deformation of \eqref{eq:trivialdef} according to \eqref{eq:monodromy}, assuming $0 \leqslant \alpha <2$,
\be
\begin{split}
\QQ(z,\zb) & \sim  \sum_{\substack{n > \frac12\alpha-1\\n  \in \mathbb{N}_0}}b_n z^{n-\frac12 \alpha}  \hat{\OO}_{\Dh=1+n-\frac12 \alpha, s=n-\frac12 \alpha}+\sum_{n \geqslant 0} b'_n\zb^{n+\frac12 \alpha} \hat{\OO}_{\Dh=1+n+\frac12 \alpha, s=-n-\frac12\alpha } \,,\\
\tilde{\QQ}(z,\zb) &\sim \sum_{n \geqslant 0}  \tilde{b}_n z^{n+\frac12 \alpha}  \hat{\OO}_{\Dh=1+n+\frac12 \alpha, s=n+\frac12 \alpha}+\sum_{\substack{n > \frac12\alpha-1 \\n  \in \mathbb{N}_0} }\tilde{b}'_n\zb^{n-\frac12 \alpha} \hat{\OO}_{\Dh=1+n-\frac12 \alpha, s=-n+\frac12\alpha }  \,,
\end{split}
\label{eq:monodromyOPE}
\ee
where we omit the $\RR$ and $r$ quantum numbers of defect operators since they are the same for all operators as for the bulk operator. The range of $n$ of the sums is further constrained with respect to \eqref{eq:trivialdef} by imposing $\Dh >0$.
In deforming \eqref{eq:trivialdef} there is an ambiguity of where to include $\QQ(0,0)$, that affects all of the $n=0$ terms of the sums in \eqref{eq:monodromyOPE}. Below we will fix this ambiguity imposing that the defect defined by \eqref{eq:monodromyOPE} is compatible with supersymmetry.
The defect OPEs of the conjugate scalars are obtained trivially from \eqref{eq:monodromyOPE}. From \eqref{eq:betagamma} we obtain the chiral algebra OPEs
\be 
\beta(z) |\sigma \rangle \sim  \sum_{n > \frac12\alpha-1} b_n \frac{ \hat{\OO}_{\Dh=1+n-\frac12 \alpha, s=n-\frac12 \alpha} }{z^{-n+\frac12 \alpha} } |\sigma \rangle \,, \qquad
\gamma(z) |\sigma \rangle \sim \sum\limits_{n \geqslant 0} \tilde{b}_n\frac{ \hat{\OO}_{\Dh=1+n+\frac12 \alpha, s=n+\frac12 \alpha}}{z^{-n-\frac12\alpha}} |\sigma \rangle \,, 
\ee
where we kept only $(a,c)$ operators on the right hand side (notice that in this very simple case the defect OPEs of $\tilde Q^*$ and $Q^*$ do not contribute in chiral algebra).
We recover the spectral flow result \eqref{eq:defectOPEmonodromy} if $b_0$ is not zero. We now show that $b_0=0$ is incompatible with having an $\NN=(2,2)$ supersymmetric defect in the free hypermultiplet theory.
Following \cite{Gaiotto:2013nva} we can compute the two-point functions of the free scalars, by solving the equation of motion imposing the correct monodromy.\footnote{We thank M.~Meineri for discussions on this defect.}
Since the Cartans of both $su(2)$ are preserved, we have two non-trivial two-point functions to compute, $\langle \QQ(x_1) \QQ^*(x_2) \rangle$ and $\langle \tilde{\QQ}(x_1) \tilde{\QQ}^*(x_2)\rangle$. The computation proceeds exactly as in \cite{Gaiotto:2013nva}, except that we impose the monodromy condition \eqref{eq:monodromy}, and that we allow for all operator dimensions present in~\eqref{eq:monodromyOPE}, namely
\be 
\hat{\Delta}= |s| + 1\,, \quad  s =  - \frac{\alpha}{2} +  \mathbb{Z} \,; \qquad \hat{\Delta} = 1 + s\,, \quad s= - \frac{\alpha}{2}\,,
\label{eq:QQterms}
\ee
for $\langle\QQ (x_1) \QQ^* (x_2)\rangle $, and
 \be 
\hat{\Delta}= |s| + 1\,, \quad  s =  \frac{\alpha}{2} +  \mathbb{Z} \,; \qquad \hat{\Delta} = 1 - s\,,\quad s=  \frac{\alpha}{2}\,,
\label{eq:QQtterms}
\ee
for  $\langle\tilde{\QQ} (x_1) \tilde{\QQ}^* (x_2) \rangle $. The aforementioned ambiguity, in the split of the $s=0$ operator in the trivial defect ($\alpha=0$), to a defect operator of dimension $\hat{\Delta}=1 \pm \frac12\alpha$,  is parametrized by giving a defect OPE coefficient $b_{\QQ}^2$ ($b_{\tilde{\QQ}}^2$) to the  operator with $\hat{\Delta}=1 - \frac12\alpha$,  in the   $\langle\QQ (x_1) \QQ^* (x_2)\rangle $ ($\langle\tilde{\QQ} (x_1) \tilde{\QQ}^* (x_2) \rangle $) two-point function.  Then, the operator with  $\hat{\Delta}=1 + \frac12\alpha$ gets an OPE coefficient of $1-b_{\QQ}^2$ ($1-b_{\tilde{\QQ}}^2$). The former operators do not appear in computation of \cite{Gaiotto:2013nva}, and we add them by hand.\footnote{Note that this split explicitly breaks the continuity if we were to consider $\alpha=2$. Taking $\alpha=2$ in \eqref{eq:monodromyOPE} we see that the condition $n > \frac12 \alpha + 1$ removes the $b_0$ and $\tilde{b}'_0$ terms from the sum, and we recover the $\alpha=0$ OPE, as expected from \eqref{eq:monodromy}. If  $b_0=\tilde{b}'_0=0$  the spectrum would be continuous at  $\alpha=2$, and all quantities computed with the $\alpha <2$ assumption evaluated at $\alpha=2$ would yield the $\alpha=0$ result.  Taking these to be non-zero our expressions are valid strictly for $\alpha <2$ and the trivial defect is not recovered upon setting $\alpha=2$.}
However, the two-point functions constructed, and quoted in eq.~\eqref{eq:two-pointmonodromies},  are crossing symmetric, and are obtained assuming $0 \leqslant \alpha < 2$. 
Setting both  $b_{\QQ}^2$ and $b_{\tilde{\QQ}}^2$ to zero is not compatible with supersymmetry, as it would make the defect OPE of the scalars non-singular, and in turn setting the one-point function of the stress tensor superprimary to  zero, which is not possible for a non-trivial defect in an $\NN=2$ SCFT. In free theory, we can construct the superprimary of the stress tensor multiplet, the $su(2)_R$ current, and the moment map, out of $\QQ$ and $\tilde{\QQ}$, as they do not contain fermions. Their expressions with our normalizations are quoted in \eqref{eq:O2tmufromQ}. We can thus compute their one-point functions, from the two-point functions of the complex scalars, by taking the coincident limit. We impose supersymmetry by demanding that the one-point functions of the first two operators are related as given in eq.~\eqref{eq:Neq2onepoints}, which fixes $b_{\QQ}^2=1$ and $b_{\tilde{\QQ}}^2 = 0$.\footnote{If we had assumed $-2<\alpha <0$ the roles of $\QQ$ and $\tilde{\QQ}$ would be interchanged.} Here we assumed the simplest possible defect, with $b^2_{\QQ/\tilde{\QQ}}$ independent of $\alpha$. After fixing these coefficients we get a value for $h$, as well as for the one-point function of the moment map, that precisely match \eqref{eq:hmonodromy} and \eqref{eq:oneptmumonodromy}, thus confirming the identification of the monodromy defect with the spectral flow.

Finally, we note that the functional form of the spectral flowed graded partition function for $\alpha >1$ is identical to that with spectral flow parameter $\alpha -2$, and in particular a naive series expansion in $q$ would not feature the correct dimension for the spectral flowed vacuum. This follows from the radius of convergence of the $q$ expansion being altered by the spectral flow as discussed at length in \cite{Lesage:2002ch}. This makes it hard to read off from the superconfonformal index the module introduced by the defect. Note that evaluating \eqref{hmin} for the free hyper one finds $h^\chi_{min}=-\frac{1}{8}$, which corresponds precisely to the dimension of the spectral flowed vacuum with $\alpha=1$.

\subsubsection*{Vortex defect for the free hypermultiplet}
\label{sec:vortexfreehyper}

Even though the bulk theory is free --  a single free hypermultiplet -- vortex defects are strongly interacting and little is known about their dynamics. The Schur index in the presence of a vortex defect was computed in \cite{Cordova:2017mhb}. There, the authors started from the so-called $(A_1,D_4)$ Argyres-Douglas theory, or $H_2$ theory, which has an $su(3)$ flavor symmetry. To go to the free hypermultiplet theory, we consider an embedding of  $su(2)$ in $su(3)$, and give a constant expectation value to the moment map component corresponding to the raising operator of $su(2)$ under the embedding.\footnote{In an enlarged class $\SS$ of type $A_1$, the $(A_1,D_4)$ theory is described by a sphere with a single regular maximal puncture, and a single irregular puncture. The flavor symmetry of the theory is $su(3)$, even though in class $\SS$ only an $su(2)$ flavor symmetry, associated with the regular puncture, and a $u(1)$ associated to the irregular one are visible. The Higgsing corresponds to closing the regular puncture, and one ends up with only the irregular puncture, finding the $(A_1,A_1)$ theory, \ie a single free hypermultiplet.} The infrared theory is then a single free hypermultiplet, that has an $su(2)$ flavor symmetry.
If instead one turns on a position dependent expectation value, one finds a vortex defect in the free hyper theory, which breaks the flavor symmetry of the free hyper down to $u(1)$. The canonical surface defect, \ie with vortex number one, has a well-understood spectrum of $2d$-$4d$ BPS particles \cite{Gaiotto:2011tf}, allowing for the computation of the index by the Coulomb branch formula \cite{Cordova:2017mhb}. 
For defects $\mathbb{S}_r$, with higher vortex numbers $r$, the index was instead computed with the Higgsing prescription. Up to an overall power of $q$ this result matches the character of the qDS reduction of the spectral flowed vacuum module of $\chi((A_1,D_4))$, \ie of the $\widehat{su(3)}_{-\frac32}$ AKM current  algebra.\footnote{Similarly for $r=1$ the index in \eqref{eq:vortexinfreehyperindex} matches the Coulomb branch computation up to an overall power of $q$. For the purposes of identifying $h^\chi_\sigma$ the normalization resulting from the chiral algebra computation is the relevant one.} The spectral flow by $r$ units is done with respect to the Cartan of an  $\widehat{su(2)}_{-\frac32}$ subalgebra of   $\widehat{su(3)}_{-\frac32}$, on which one then performs a qDS reduction. The result is  \cite{Cordova:2017mhb}
\be 
\II_{\mathbb{S}_r} = (-1)^r \sum\limits_{n=-\infty}^{+\infty} \frac{x^n q^{-\frac{n}{2}}}{(q)_\infty} \sum\limits_{k=0}^{+\infty} \frac{(1- q^{(2k+1+|n|)(r+1)})q^{\frac{1}{2}(2k + |n| -\frac{r}{2})(2k + |n| + 1 - \frac{r}{2}})}{(q)_\infty}\,,
\label{eq:vortexinfreehyperindex}
\ee
where $(q)_\infty = (q;q)_\infty$ is the q-Pochhammer symbol. The variable $x$ keeps track of the $u(1)$ flavor symmetry preserved by the defect, \ie the Cartan of the $su(2)$ flavor symmetry of the bulk chiral algebra, a $\beta \gamma$ system.
The representation theory of free hypermultiplet chiral algebra is very rich and has been studied in detail in \eg, \cite{Lesage:2002ch,Lesage:2003kn}. Here we only make use of some of the features relevant for accommodating the modules introduced by the vortex defect and its operators.\footnote{As pointed out in \cite{Lesage:2002ch} some care is needed when interpreting characters obtained from spectral flows, due to the fact that the region of convergence of the character when written as a power series in $q$ is modified by the flow, even if the functional form of the character appears to fall back on the original module, giving rise to fake periodicities, just like the ones described in the monodromy defect example above. To have a one-to-one map between irreducible modules and characters one needs to view the characters as distributions \cite{Creutzig:2012sd}. We thank L.~Eberhardt for showing us this reference.}
From \eqref{eq:vortexinfreehyperindex} we see that for $ r \neq 1$ the index displays the existence of defect operators whose dimensions are unbounded from below. In particular, while for each definite $u(1)$ charge the weights of the operators are bounded from below, there can be operators with arbitrarily large positive, or negative, $u(1)$ charge and correspondingly arbitrarily negative weight. 
Modules displaying these properties are present in the $\beta\gamma$ chiral algebra. They are obtained from the vacuum module by the spectral flow, and they were dubbed ``deeper twists'' in \cite{Lesage:2002ch}.  The dimension of the spectral flowed vacuum is $-\alpha^2/8$, and it can be made arbitrarily negative (see \eqref{eq:spectralflow}). Furthermore spectral flowed modules with $\alpha \geqslant 2 $ can have a spectrum of dimensions unbounded from below, as positive modes of $\beta$ and $\gamma$ (and the charged flavor currents made from them)  start having a non-zero action  \eqref{eq:spectralflow}. This action matches the type of structure seen in the index of more negative dimensions having larger charges.
Note that this, in turn, implies that the OPE of $\beta$ with the spectral flowed vacuum has a singularity larger than one. This singularity is incompatible with the defect OPE of $\beta$ -- see the selection rules given in eq.~\eqref{eq:defectopeC}.  Thus, in these cases the spectral flowed vacuum cannot be identified with $\left[\hat{\mathbb{1}}\right]_\qq$. The chiral algebra operator corresponding to a spectral flowed vacuum must arise from a non-trivial defect operator, such that the chiral algebra OPE can be made arbitrarily negative as discussed around \eqref{eq:nondefectOPE}.
As such, the defect introduces at least two distinct modules, a ``deeper twist'' as well as a module that can accommodate the defect identity.
Finally, note that due to the aforementioned selection rules the $\beta$ and $\gamma$ descendants of the defect identity will never have dimension smaller than that of the defect identity. This implies the defect operators with lower dimensions must belong to different modules, and not obtained from the defect OPE of $\beta$ and $\gamma$. Since the bulk theory
is made out of $\beta$ and $\gamma$ composites we conclude these defect operators do not appear in the
defect OPE of Schur operators.

\subsubsection*{Vortex defects for the $(A_1,A_2)$ Argyres-Douglas theory}
With the purpose of computing $h$ from chiral algebra, the non-trivial step consists of identifying the defect identity in chiral algebra. In \cite{Cordova:2017mhb} the Schur index in the presence of vortex defects for the $(A_1, A_{2n})$ Argyres-Douglas SCFTs were also obtained. All these theories have minimal models as their chiral algebras, making easier the task of identifying the defect identity, which should be annihilated by the positive modes of the stress tensor, according to the results of section~\ref{sec:OPEselectionrules}.
Let us focus on the Lee-Yang minimal model, which is the chiral algebra of the $(A_1,A_2)$ Argyres-Douglas theory. The chiral algebra is simply Virasoro with central charge $c_{2d}= - \frac{22}{5}$, and all bulk operators can then be made out of normal ordered products of the stress tensor and its derivatives.
The only non-vacuum module corresponds to a highest weight representation of dimension $-\frac{1}{5}$. Since the defect identity gives rise to a negative dimensional operator in chiral algebra this is the only module that can accommodate it. Non-trivial defects in the $(A_1,A_2)$ theory will then have 
\be 
h=\frac{1}{15 \pi^2}\,.
\ee
For Argyres-Douglas SCFTs with higher values of $n$ there are more modules that can accommodate the defect identity, and thus identifying $\sigma$, and obtaining $h$, requires analyzing the chiral algebra in detail.

%% file: sections/acknowledgments.tex
%!TEX root = ../susysurfaces.tex

\acknowledgments

We are especially grateful to Marco Meineri for collaboration in earlier stages of this work and many useful suggestions.
We have greatly benefited from discussions with
Philip Argyres,
Chris Beem,
Marco Bill\`o,
Nikolay Bobev,
Adam Chalabi,
Lorenz Eberhardt,
Francesco Galvagno,
Alberto Lerda,
Mario Martone,
Andy O'Bannon,
Wolfger Peelaers,
Leonardo Rastelli,
Balt van Rees,
Brandon Robinson
and
Volker Schomerus. 
The work of LB is supported by the European Union’s Horizon 2020 research and innovation programme under the Marie Sklodowska-Curie grant agreement No 749909. 
This work benefited from the 2019 Pollica summer workshop, which was supported in part by the Simons Foundation (Simons Collaboration on the Non-perturbative Bootstrap) and in part by the INFN.
Part of this work was carried out at  Perimeter Institute for Theoretical Physics during the ``Bootstrap 2019'' and the ``Boundaries and Defects in Quantum Field Theory'' conferences. Research at Perimeter Institute is supported by the Government of Canada through Industry Canada and by the Province of Ontario through the Ministry of Economic Development \& Innovation. 
ML also acknowledges FCT-Portugal under grant PTDC/MAT-OUT/28784/2017.

%Further, in order to keep our records of such manuscripts up to date, I ask that you inform us by sending an email to: pipublications@perimeterinstitute.ca. If you are posting the paper to the arXiv, please send us the assigned number. If you do not post to the arXiv, please be certain to send us the journal reference details.

%% file: sections/A_conventions.tex
%!TEX root = ../susysurfaces.tex
%%%%%%%%%%%%%%%%%%%%%%%%%%%%%%%%%%%%%%%%%%%%%
\section{Conventions and superconformal algebras}
\label{sec:conventions}
%%%%%%%%%%%%%%%%%%%%%%%%%%%%%%%%%%%%%%%%%%%%%

In this appendix we summarize the conventions used throughout the paper and collect the different superconformal algebras and supersymmetry variations used.

\subsubsection*{Conventions}
We lift and lower $su(2)$ indices as $\phi^a = \epsilon^{a b} \phi_b$, $\phi_a = \epsilon_{ab} \phi^b$, where we take $\epsilon_{12}=1$, $\epsilon^{12}=-1$. Our sigma matrix are taken to be
\be
 \sigma^{\mu}_{\aa \bbd} =(\sigma^a,\ii\mathbb{1})\,, \qquad  (\bar \sigma^{\mu})^{ \aad \bb} =(\sigma^a,-\ii \mathbb{1})\,,
\ee
where $\sigma^a$ are the Pauli matrices, and $\alpha= {\bf{1}}, \,{\bf{2}}$, $\dot{\alpha}= \dot{{\bf{1}}},\, \dot{{\bf{2}}}$. With the exception of $K^{\aad \aa}$ all fields go from vector to spinor indices by $\OO_{\aa \aad} = \sigma^\mu_{\aa \aad} P_\mu$, and then spinor indices are raised and lowered with epsilon tensors as described above. We will use the notation
$X_z  = \frac12 \sigma^{\mu}_{\bf{1} \dot{\bf{1}}} X_\mu$, $X_{\zb}  = -\frac12 \sigma^{\mu}_{\bf{2} \dot{\bf{2}}} X_\mu$, $X_w  = \frac12 \sigma^{\mu}_{\bf{1} \dot{\bf{2}}} X_\mu$ and $X_{\bar{w}}  = \frac12 \sigma^{\mu}_{\bf{2} \dot{\bf{1}}} X_\mu$.

\subsubsection*{$4d$ conformal algebra}

The commutation relations for the $4d$  conformal algebra are given by
\begingroup
\allowdisplaybreaks[1]
\be
\begin{alignedat}{4}
&[\MM_{\aa}^{~\bb},\MM_{\gamma}^{\phantom{\gamma}\delta}]	&~=~&	\delta_{\gamma}^{~\bb}\MM_{\aa}^{~\delta}-\delta_{\aa}^{~\delta}\MM_{\gamma}^{~\bb}~,\\
&[\MM^{\aad}_{~\bbd},\MM^{\ggd}_{~\ddd}]			&~=~&	\delta^{\aad}_{~\ddd}\MM^{\ggd}_{~\bbd}-\delta^{\ggd}_{~\bbd}\MM^{\aad}_{~\ddd}~,\\
&[\MM_{\aa}^{~\bb},\PP_{\gamma\ggd}]				&~=~&	\delta_{\gamma}^{~\bb}\PP_{\aa\ggd}-\tfrac12\delta_{\aa}^{\phantom{\aa}\bb}\PP_{\gamma\ggd}~,\\
&[\MM^{\aad}_{~\bbd},\PP_{\gamma\ggd}]				&~=~&	\delta^{\aad}_{~\ggd}\PP_{\gamma\bbd}-\tfrac12\delta^{\aad}_{\phantom{\aad}\bbd}\PP_{\gamma\ggd}~,\\
&[\MM_{\aa}^{~\bb},\KK^{\ggd\gamma}]				&~=~&	-\delta_{\aa}^{~\gamma}\KK^{\ggd\bb}+\tfrac12\delta_{\aa}^{\phantom{\aa}\bb}\KK^{\ggd\gamma}~,\\
&[\MM^{\aad}_{~\bbd},\KK^{\ggd\gamma}]				&~=~&	-\delta^{\ggd}_{~\bbd}\KK^{\aad\gamma}+\tfrac12\delta^{\aad}_{\phantom{\aad}\bbd}\KK^{\ggd\gamma}~,\\
&[\HH,\PP_{\aa\aad}]							&~=~&	\PP_{\aa\aad}~,\\
&[\HH,\KK^{\aad\aa}]							&~=~&	- \KK^{\aad\aa}~,\\
&[\KK^{\aad\aa},\PP_{\bb\bbd}]					&~=~&	4 \delta_{\bb}^{\phantom{\bb}\aa}\delta^{\aad}_{\phantom{\aad}\bbd}\HH+4\delta_{\bb}^{\phantom{\bb}\aa}\MM^{\aad}_{\phantom{\aad}\bbd}+4\delta^{\aad}_{\phantom{\aad}\bbd}\MM_{\bb}^{\phantom{\bb}\aa}~,
\end{alignedat}
\label{eq:confalg}
\ee
\endgroup
where we took $P_{\aa \aad} = \sigma^\mu_{\aa \aad} P_\mu$, $K^{ \aad \aa} = \bar{\sigma}_\mu^{\aad \aa} K_\mu$  and
\be 
\MM_{\bb}^{\phantom{\bb} \aa}=-\tfrac14 \bar{\sigma}^{\mu \aad \a} \sigma_{\nu \bb \aad} M_{\mu \nu}\,, \qquad
\MM^{\aad}_{\phantom{\aad} \bbd}=- \tfrac14  \bar{\sigma}^{\mu \aad \a} \sigma_{\nu \aa \bbd} M_{\mu \nu}\,.
\ee

\subsection{Four-dimensional \texorpdfstring{$\NN=1$}{N=1} superconformal algebra}
\label{sec:Neq1algebra}

The $4d$ $\NN=1$ superconformal algebra supplements the generators whose commutation relations are given in \eqref{eq:confalg} by four Poincar\'{e} supercharges ($\QQ_{\aa}$, $\Qt_{\ad}$), four conformal supercharges ($\SS^{\aa}$, $\St^\aa$), and a $u(1)_{\hat{r}}$ $R-$symmetry $\hat{r}$ under which the supercharges are charged as follows
\be
 [\hat{r},\QQ_{\aa}]= \frac{1}{2} \QQ_{\aa}\,, \qquad  [\hat{r},\Qt_{\ad}]=- \frac{1}{2} \Qt_{\ad}\,, \qquad
  [\hat{r},\SS^{\aa}]= -\frac{1}{2} \SS^{\aa}\,, \qquad  [\hat{r},\St^{\ad}]= \frac{1}{2} \St^{\ad}\,.
\ee
The remaining commutation relations of the supercharges among themselves and with the generators of the conformal algebra can be read from \eqref{eq:Neq2Qcomm} and \eqref{eq:Neq2bosQcomm} where one should set $\II,\JJ=1$ and $\RR^{1}_{\phantom{1}1} = \frac32 \hat{r}$. 

\subsubsection*{Stress tensor multiplet}
\label{sec:susyvarsNeq1}

The supersymmetry variations of the $\NN=1$ stress tensor multiplet are given by
\be
\begin{aligned}
&\delta j_\mu= \frac{1}{2}\left(J_{\mu }^\aa \xi_\aa - \tilde{J}_{\mu\aad }\bar\xi^{\aad}\right)\,,\\
&\delta J^{\mu }_\aa=\frac{1}{2}\sigma_{\nu\alpha\aad }\bar{\xi}^{\aad }T^{\mu\nu}
+\frac{1}{4}\left({\sigma^{\mu\nu}}{\sigma^{\lambda}}-3\sigma^{\lambda}{\bar{\sigma}}^{\mu\nu}\right)_{\aa\aad}\bar{\xi}^{\aad } \partial_\nu j_\lambda\,,\\
&\delta \tilde{J}_{\mu \ad }=\frac{1}{2}\sigma^{\nu}_{\aa \aad }\xi^{\aa}T_{\mu\nu} 
-\frac{1}{4}\left({\bar{\sigma}^{\mu\nu}}{\bar{\sigma}^{\lambda}}-3\bar{\sigma}^{\lambda}{\sigma}^{\mu\nu}\right)^{\bbd \aa}\epsilon_{\aad \bbd} \xi_\aa \partial_\nu j_\lambda\,,\\
&\delta T_{\mu\nu}=- \frac{1}{2} \xi^\aa {{\sigma^{\mu\lambda}}_\aa}^\bb \partial_\lambda J_\bb^{\nu}-\frac{1}{2}\bar{\xi}_{\aad}{\bar{\sigma}^{\mu\lambda\aad}}_{\phantom{\mu\lambda\aad}\bbd}\partial_\lambda\tilde J^{\nu\bbd }+\mu\leftrightarrow\nu\,,
\end{aligned}
\end{equation}
where $j_\mu$ is the $u(1)_r$ current, $J^\mu_\aa$ and $\tilde{J}_{\mu \ad}$ the supersymmetry currents, and $T_{\mu \nu}$ the stress tensor which we take to be canonically normalized according to \eqref{eq:STtwopoint}.
Here $\delta$ is defined by
\be
\delta \OO = \left[\delta, \OO \right]= \left[\xi^\a Q_\alpha + \bar{\xi}^{\ad} \Qt_{\ad} , \OO \right]\,,
\ee
and the coefficients in the supersymmetry variations can be fixed by imposing the algebra
\bea
(\delta_1 \delta_2 - \delta_2 \delta_1 )\OO & = \left[\left[\delta_1, \delta_2\right],\OO\right] = - \left(\xi_{1}^\aa \bar{\xi}_2^{\aad} - \xi_{2}^\aa \bar{\xi}_1^{\aad} \right)  \left[\{ Q_\aa , \tilde{Q}_{\bbd} \} , \OO \right]\nn \\
&= - \frac{1}{2}  \left(\xi_{1}^\aa \bar{\xi}_2^{\aad} - \xi_{2}^\aa \bar{\xi}_1^{\aad} \right) (\sigma_\mu)_{\aa \aad} \partial_\mu \OO \,.
\eea

\subsection{Two-dimensional \texorpdfstring{$\NN=(2,0)$}{N=(2,0)} superconformal algebra}
\label{sec:2d20algebra}

For the a surface defect preserving $\NN=(2,0)$ supersymmetry inside a $4d$ $\NN=1$ SCFT the bosonic generators of the conformal algebra on the defect \eqref{2dconf} are supplemented by the supercharges given in \eqref{eq:N20charges} and the $R-$symmetry  generator given by $J=3 r- \MM$.  They obey the following algebra
\be
\begin{aligned}
\{G^+_r, G^-_s\}&= L_{r+s} + \frac{r-s}{2} J\,,   &\left[L_m, G^\pm_r \right] = \left(\frac{m}{2}- r\right) G^\pm_{m+r}\,, \qquad
 &\left[J, G^\pm_r \right] = \pm  G^\pm_r\,, \\
\left[ L_m,L_n \right] &= (m-n)L_{m+n}\,, \qquad &[\Lb_m,\Lb_n]= (m-n)\Lb_{m+n}\,, \qquad &
\end{aligned}
\ee
where $ r,s=\pm \frac12$ and $m,n,=0,\pm 1$, since we consider the global superalgebra, with no Virasoro enhancement, due to the absence of a defect stress tensor. The commutant of this superalgebra inside the $4d$ $\NN=1$ superconformal algebra is $Z= -r + \MM$. A short (anti)chiral multiplet is annihilated ($G^-_{-\frac12}$) $G^+_{-\frac12}$  and obeys ($L_0 = -\frac{1}{2} J$) 
$L_0 = \frac{1}{2} J$.

\subsection{Four-dimensional \texorpdfstring{$\NN=2$}{N=2} superconformal algebra}
\label{sec:superalgebra}

We collect here the commutation relations of the four-dimensional $\NN=2$ superconformal algebra used in sections~\ref{sec:Neq2}~and~\ref{sec:chiralalgebra}.
The bosonic subalgebra consists of the conformal algebra displayed in eq.~\eqref{eq:confalg}, together with the $su(2)_R \oplus u(1)_r$ $R-$symmetry. We define the $R-$symmetry generators as $\RR^\II_{\phantom{1}\JJ}$ as
\be
\RR^1_{\phantom{1}2}=\RR^+~,\qquad\RR^2_{\phantom{2}1}=\RR^-~,\qquad\RR^1_{\phantom{1}1}=\frac12 r+\RR~,\qquad\RR^2_{\phantom{1}2}=\frac12 r-\RR~,
\ee
where we follow the conventions of \cite{Dolan:2002zh} for the $u(1)_r$ charge $r$, and where the $su(2)_R$ generators obey the standard algebra
\be
[\RR^+,\RR^-]=2\RR~,\qquad [\RR,\RR^{\pm}]=\pm\RR^{\pm}\,.
\ee
Then, the $R-$symmetry generators $\RR^\II_{\phantom{1}\JJ}$ obey the commutation relations
\be
[\RR^\II_{\phantom{\II}\JJ},\RR^{\mathcal{K}}_{\phantom{\KK}\LL}]=\delta^\mathcal{K}_{\phantom{\KK}\JJ}\RR^\II_{\phantom{\II}\LL}-\delta^\II_{\phantom{\II}\LL}\RR^\mathcal{K}_{\phantom{\KK}\JJ}~.
\ee

The eight conformal and eight superconformal supercharges have the following non-zero commutation relations
\be
\begin{alignedat}{2}
&\{Q_{\aa}^\II,\,\tilde{Q}_{\JJ\aad}\}  			&~=~~&\tfrac12	\delta^\II_{\phantom{\II}\JJ} \PP_{\aa\aad}~,\\
&\{\tilde{S}^{\II\aad},\,S_{\JJ}^{\phantom{\aa}\aa}\} &~=~~&	\tfrac12 \delta^\II_{\phantom{\II}\JJ} \KK^{\aad\aa}~,\\
&\{Q_{\aa}^\II,\,S^{\phantom{\aa}\bb}_\JJ\}     	&~=~~&	\tfrac12 \delta^\II_{\phantom{\II}\JJ}\delta_{\aa}^{\phantom{\aa}\bb}\HH   + \delta^\II_{\phantom{\II}\JJ} \MM_{\aa}^{\phantom{\aa}\bb}-\delta_\aa^{\phantom{\aa}\bb} \RR^\II_{\phantom{\II}\JJ}~,\\
&\{\tilde{S}^{\II\aad},\,\tilde{Q}_{\JJ\bbd}\}		&~=~~&	\tfrac12 \delta^\II_{\phantom{\II}\JJ}\delta^{\aad}_{\phantom{\aad}\bbd}\HH + \delta^\II_{\phantom{\II}\JJ} \MM^{\aad}_{\phantom{\aad}\bbd}+\delta^{\aad}_{\phantom{\aad}\bbd} \RR^\II_{\phantom{\II}\JJ}~,
\end{alignedat}
\label{eq:Neq2Qcomm}
\ee
and the commutators of the supercharges with the bosonic symmetry generators are
\begingroup
\allowdisplaybreaks[1]
\be
\label{eq:Neq2bosQcomm}
\begin{alignedat}{2}
&[\MM_{\aa}^{~\bb},Q_{\gamma}^\II]	=	\delta_{\gamma}^{~\bb} Q_{\aa}^\II -\tfrac12\delta_{\aa}^{\phantom{\aa}\bb} Q_{\gamma}^\II~,	\qquad
&[&\MM^{\aad}_{~\bbd},\tilde{Q}_{\II \ddd}]			= \delta^{\aad}_{~\ddd}\tilde{Q}_{\II \bbd} -\tfrac12\delta^{\aad}_{\phantom{\aad}\bbd}\tilde{Q}_{\II \ddd}~,\\
&[\MM_{\aa}^{~\bb},S_{\II}^{\phantom{\aa}\gamma}]				=	-\delta_{\aa}^{~\gamma}S_{\II}^{\phantom{\aa}\bb}+\tfrac12\delta_{\aa}^{\phantom{\aa}\bb} S_{\II}^{\phantom{\aa}\gamma}~,
\qquad &[&\MM^{\aad}_{~\bbd},\tilde{S}^{\II\ggd}]				=	-\delta^{\ggd}_{~\bbd}\tilde{S}^{\II\aad}+\tfrac12\delta^{\aad}_{\phantom{\aad}\bbd}\tilde{S}^{\II\ggd}~,\\
&[\HH,Q_{\aa}^\II]							=	\tfrac12 Q_{\aa}^\II~,\qquad
&[&\HH,\tilde{Q}_{\II \aad}]							= 	\tfrac12 \tilde{Q}_{\II \aad}~,\\
&[\HH, S_{\II}^{\phantom{\aa}\aa}]							=	-\tfrac12  S_{\II}^{\phantom{\aa}\aa}~,
&[&\HH, \tilde{S}^{\II\aad} ]							=	-\tfrac12 \tilde{S}^{\II\aad} ~,\\
&[\RR^\II_{\phantom{\II}\JJ},Q_{\aa}^\mathcal{K}]=	\delta_{\JJ}^{~\mathcal{K}} Q_{\aa}^\II -\frac{1}{4} \delta_{\JJ}^{\II} Q_{\aa}^\mathcal{K}~,
&[&\RR^\II_{\phantom{\II}\JJ},\tilde{Q}_{\mathcal{K} \aad}]	=	-\delta_{\mathcal{K}}^{~\II} \tilde{Q}_{\JJ \aad} +\frac{1}{4} \delta_{\JJ}^{\II} \tilde{Q}_{ \mathcal{K} \aad}~,\\
&[\KK^{\aad\aa},Q_{\bb}^\II]					=	2 \delta_{\bb}^{\phantom{\bb}\aa}\tilde{S}^{\II\aad}~,
&[&\KK^{\aad\aa},\tilde{Q}_{\II \bbd}]					=	2 \delta_{\bbd}^{\phantom{\bbd}\aad} S_{\II}^{\phantom{\aa}\aa}~,\\
&[\PP_{\aa\aad},S_{\II}^{\phantom{\aa}\bb}]					=	- 2 \delta_{\aa}^{\phantom{\aa}\bb}\tilde{Q}_{\II \aad}~,
&[&\PP_{\aa\aad},\tilde{S}^{\II\bbd} ]					=	-2 \delta_{\aad}^{\phantom{\aad}\bbd} Q_{\aa}^\II~,
\end{alignedat}
\ee
\endgroup
where the commutators of $\RR^\II_{\phantom{\II}\JJ}$ with $\SS$ and $\tilde{\SS}$  are omitted since they are identical to those of the $\QQ$ and $\tilde{\QQ}$ generators with the same index structure.

\subsubsection*{Stress tensor supermultiplet}
\label{sec:susyvars}

The stress tensor belongs to the  $\hat{\CC}_{0,(0,0)}$ superconformal multiplet in the notation of \cite{Dolan:2002zh}, and its supersymmetry variations were obtained in \cite{Fisher:1982fu}. Here we collect the variations, after correcting a few typos, and normalizing canonically the stress tensor ($T_{\mu \nu}$), the supersymmetry currents ( $J_{\aa}^{\mu \II}$ and $\bar{J}_{\mu \aad, \II}$,), the $u(1)_r$ current ($j_\mu$) and the $su(2)_R$ current (${t_{\mu \II}}^\JJ$).
For reference the canonically normalized stress tensor has a two-point function given by (see \eg, \cite{Osborn:1993cr})
\be
\langle T_{\mu \nu}(x) T_{\rho \sigma}(0) \rangle = \frac{40c}{\pi^4 x^8} \mathcal{I}_{\mu \nu, \rho \sigma}(x)\,,\\
\label{eq:STtwopoint}
\ee
where 
\be 
\mathcal{I}_{\mu \nu, \rho \sigma}(x) = \frac12\left(I_{\mu \rho}(x) I_{\nu \rho}(x) + I_{\mu \rho}(x) I_{\nu \sigma}(x)\right) - \frac14 \delta_{\mu \nu}\delta_{\rho\sigma}\,, \qquad I_{\mu \nu}(x) = \delta_{\mu \nu} - 2 \frac{x_\mu x_\nu}{x^2}\,,
\ee
and where $c$ is the usual central charge normalized such that a free $\NN=2$ hypermultiplet has $c=\tfrac{1}{12}$. 
Similarly, the $su(2)_R$ current has two-point function\footnote{This differs from the $su(2)_R$ current defined in \cite{Beem:2013sza} by $t_{here}^{IJ}= 2 \ii J_{there}^{IJ}$.}
\be 
\langle t_\mu^{\II \JJ} (x)t_\nu^{\mathcal{K}\LL}(0) \rangle =  -\frac{3 c}{ \pi^4} \frac{I_{\mu \nu}}{x^6} \epsilon^{\mathcal{K} (\II} \epsilon^{\JJ) \LL}\,,
\label{eq:su2rOPE}
\ee
where supersymmetry fixes its two-point function in terms of $c$, see \eg, \cite{Shapere:2008zf}. Here the brackets mean indices are symmetrized and we always take symmetrizations with strength one. The three-point function of conserved currents is given, for example, in \cite{Osborn:1993cr}, where Ward identities are used to fix the coefficient of the three-point function that survives after the chiral algebra twist of eq.~\eqref{eq:chiraltwist} in terms of the two-point function.

Defining 
\be
\delta \OO = \left[\delta, \OO \right]= \left[\xi_\II^\a Q_\alpha^\II + \bar{\xi}^{\ad \II} \Qt_{\ad \II} , \OO \right]\,,
\label{eq:delta}
\ee
the supersymmetry variations of the stress tensor multiplet read
\begingroup
\allowdisplaybreaks[1]
\begin{align}
& \delta O_2 =\bar\chi_{\aad \II}\bar\xi^{\aad \II}+\xi^\aa_\II\chi_\aa^\II\,, \nonumber \\
&\delta\chi_\alpha^\II={H_\aa}^\bb \xi^\II_\bb + \frac{1}{2} \sigma^\mu_{\aa \aad }j_\mu\bar{\xi}^{\aad \II}-\frac{1}{2}t_{\mu \JJ}^{\phantom{\mu \JJ} \II}{\sigma^\mu}_{\aa \aad}\bar{\xi}^{\aad \JJ} +\frac14\; \sigma^{\mu}_{\aa\aad}\partial_\mu O_2\bar\xi^{\aad \II}\,, \nonumber  \\
&\delta{\bar{\chi}}_{\aad \II}={\bar{H}_{\aad}}^{\phantom{\aad}\bbd} \bar{\xi}_{\II \bbd}  + \frac{1}{2} \sigma^\mu_{\aa \aad }j_\mu \xi_{\II}^{\aa} - \frac{1}{2}{t_{\mu \II}}^{\JJ} {\sigma^\mu}_{\aa \aad}\xi_{\JJ}^{\aa} - \frac14 \; \sigma^{\mu}_{\aa\aad}\partial_\mu O_2 \xi_{\II}^{\aa}\,, \nonumber \\
&\delta{H_\alpha}^\beta=-\frac{1}{4} \left(J_{\mu \II}^\bb \sigma^\mu_{\aa \aad}\bar{\xi}^{\aad \II}+\bar{\xi}_{\aad \II}{\bar\sigma_\mu}^{\aad \bb}J_{\aa}^{\mu \II}\right) + \frac{1}{6} \left(\bar{\xi}_{\aad  \II }{\bar{\sigma}}^{\mu\aad \bb}\partial_{\mu}\chi_\alpha^\II+\partial_\mu\chi_\II^\bb \sigma^\mu_{\aa \aad}\bar{\xi}^{\aad \II}\right)\,,\nonumber  \\
& \delta {\bar{H}_\aad}^{\phantom{\aad}\bbd} = \frac{1}{4} \left(  \bar{J}_{\mu \aad}^\II \xi_{\II \aa} \bar{\sigma}_\mu^{\bbd \aa} + \sigma^\mu_{\aa \aad} \xi^{\JJ \aa} \bar{J}_{\mu \JJ}^\bbd \right) + \frac{1}{6}\left( \xi_ \II^\aa \partial_\mu \bar{\chi}_{\aad}^\II  \epsilon^{\ggd \bbd} \sigma^\mu_{\aa \ggd} - \partial_\mu \bar{\chi}^{\bbd \II} \xi_{\II \beta} \bar{\sigma}_\mu^{\ddd \bb}\epsilon_{\ddd \aad}\right)\,, \nonumber \\
&\delta j_\mu= \frac{1}{2}\left(J_{\mu \II}^\aa \xi^\II_\aa - \bar{J}_{\mu\aad \II}\bar\xi^{\aad \II }\right)-\frac{2}{3}\left(\xi_\II^\aa {\sigma_{\mu\nu\aa}}^\bb\partial^\nu\chi_\bb^\II+\bar{\xi}_{\dot{\alpha}\II}\bar{\sigma}_{\mu\nu\;\bbd}^{\phantom{\bbd} \aad}\partial^\nu\bar{\chi}^{\bbd \II}\right)\,,\nonumber \\
\label{eq:susyvars}
&\delta {t_{\mu \II}}^\JJ=-\left(J_{\mu \II}^\aa \xi_\aa^\JJ+\bar{\xi}_{\aad \II}{\bar J}_\mu^{\aad \JJ}-\frac{1}{2}\delta_\II^\JJ\left(J_{\mu \KK}^\aa \xi_\aa^\KK +\bar\xi_{\aad \KK}\bar J^{\aad \KK}_\mu\right) \right)\\
&\phantom{\delta {t_{\mu \II}}^\JJ=}+\frac{1}{3}\left(\xi^\aa_\II {\sigma_{\mu\nu\alpha}}^\beta\partial^\nu\chi_\bb^\JJ-\partial^\nu\chi^\aa_\II{\sigma_{\mu\nu\aa}}^\bb \xi^\JJ_\bb +\partial^\nu{\bar{\chi}}_{\aad \II}\bar{\sigma}_{\mu\nu\;\bbd}^{\phantom{\bbd} \aad}\bar{\xi}^{\bbd \JJ}-\bar{\xi}_{\aad \II}\bar{\sigma}_{\mu\nu\;\bbd}^{\phantom{\bbd} \aad}\partial^\nu\bar{\chi}^{\bbd \JJ}\right)\,,\nonumber \\
&\delta J^{\mu \II}_\aa=\frac{1}{2}\sigma_{\nu\alpha\aad }\bar{\xi}^{\aad \II}T^{\mu\nu} -\left(\partial_\nu {H_\aa}^\bb{{\sigma^{\mu\nu}}_\bb}^\gamma+\frac{1}{3}{{\sigma^{\mu\nu}}_\aa}^\bb\partial_\nu{H_\bb}^\gamma\right)\xi^\II_\gamma\nonumber \\
&\phantom{\delta J^{\mu i}_\alpha=}+\frac{1}{12}\left({\sigma^{\mu\nu}}{\sigma^{\lambda}}-3\sigma^{\lambda}{\bar{\sigma}}^{\mu\nu}\right)_{\aa\aad}\bar{\xi}^{\aad \JJ}\left(\delta^\II_\JJ\partial_\nu j_\lambda+2\partial_\nu {t_{\lambda \JJ}}^\II \right)\,,\nonumber \\
&\delta \bar{J}_{\mu \ad \II}=\frac{1}{2}\sigma^{\nu}_{\aa \aad }\xi^{\aa}_{\II}T_{\mu\nu} - \left(\partial_\nu \bar{H}_{\aad \bbd} {{{\bar{\sigma}}_{\mu\nu}}{}^{\bbd}}_{\ggd} +\frac{1}{3} {{\bar{\sigma}}_{\mu\nu}}{}^{\ddd}_{\bbd} \partial_\nu{\bar{H}^\bbd}_{\phantom{\bbd}\ggd} \epsilon_{\aad \ddd}\right)\bar{\xi}_\II^\ggd\nonumber \\
&\phantom{\delta J^{\mu i}_\alpha=}+\frac{1}{12}\left({\bar{\sigma}^{\mu\nu}}{\bar{\sigma}^{\lambda}}-3\bar{\sigma}^{\lambda}{\sigma}^{\mu\nu}\right)^{\bbd \aa}\epsilon_{\aad \bbd} \xi^{\JJ}_\aa \left(-\epsilon_{\II\JJ} \partial_\nu j_\lambda+2\partial_\nu {t_{\lambda \JJ \II}} \right)\,,\nonumber \\
&\delta T_{\mu\nu}=- \frac{1}{2} \xi^\aa_\II {{\sigma^{\mu\lambda}}_\aa}^\bb \partial_\lambda J_\bb^{\nu \II}-\frac{1}{2}\bar{\xi}_{\aad \II}{\bar{\sigma}^{\mu\lambda\aad}}_{\phantom{\mu\lambda\aad}\bbd}\partial_\lambda\bar J^{\nu\bbd \II}+\mu\leftrightarrow\nu\,. \nonumber 
\end{align}
\endgroup
The coefficients of all the variations can be checked by imposing that for all operators $\OO$ we have that
\bea
(\delta_1 \delta_2 - \delta_2 \delta_1 )\OO & = \left[\left[\delta_1, \delta_2\right],\OO\right] = - \left(\xi_{1, \II}^\aa \bar{\xi}_2^{\aad, \JJ} - \xi_{2, \II}^\aa \bar{\xi}_1^{\aad, \JJ} \right)  \left[\{ Q_\aa^\II , \tilde{Q}_{\bbd \JJ} \} , \OO \right]\nn \\
&= - \frac{1}{2}  \left(\xi_{1, \II}^\aa \bar{\xi}_2^{\aad, \II} - \xi_{2, \II}^\aa \bar{\xi}_1^{\aad, \II} \right) (\sigma_\mu)_{\aa \aad} \partial_\mu \OO \,,
\eea
as follows from the $\NN=2$ superalgebra.

\subsubsection*{Flavor currents supermultiplet}

The conserved currents for a global symmetry of an $\NN=2$ SCFT are one of the top components of the half-BPS $\hat{\BB}_1$ multiplet in the classification of \cite {Dolan:2002zh}.
We take the flavor current OPE following the conventions of \cite{Argyres:2007cn}, which has two-point function
\be
 \langle J^A_\mu(x)   J^B_\nu ( 0)  \rangle =  \frac{3 k_{4d}}{4 \pi^4} \delta^{AB} \frac{I_{\mu \nu}}{x^6} \,,
 \label{eq:flavorcurrent}
\ee
where $A,B,C$ are adjoint flavor indices, and we are using normalizations such that long roots of a Lie algebra have length $\sqrt{2}$ as in \cite{Argyres:2007cn}.
In the same conventions, using the supersymmetric Ward identities of \cite{Dolan:2001tt}, the OPE of superprimary of the  $\hat{\BB}_1$ multiplet, the moment map of the flavor symmetry, reads\footnote{Note that the conventions here differ from those of \cite{Beem:2013sza} by $\mu_{here} = \ii/\sqrt{2} \mu_{there}$.}
\be
\mu^{A\, \II\JJ}(x)\mu^{B\, \KK\LL}(0)\sim
 \frac { k_{4d} }{ 32 \pi^4 } \frac{\epsilon^{\mathcal{K}(\II}\epsilon^{\JJ)\LL} \delta^{AB}}{x^4}
-\frac{ 1}{ 4 \pi^2 }   \frac{\ii f^{ABC}\mu^{C\,(\II(\KK}\epsilon^{\LL)\JJ)}}{x^2}+\cdots\,.
\label{eq:momentmapope}
\ee

\subsection{Two-dimensional \texorpdfstring{$\NN=(2,2)$}{N=(2,2)} superconformal algebra}
\label{app:defectalgebra}

In sections \ref{sec:Neq2} and \ref{sec:chiralalgebra} we consider a surface defect in a $4d$ $\NN=2$ SCFT that preserves a $\NN=(2,2)$ superconformal algebra. In this appendix we collect the commutation relations of the algebra and the supersymmetry variations of the displacement supermultiplet. 
The generators of the two-dimensional superconformal algebra are given in eqs.~\eqref{2dconf},~\eqref{eq:defQ},~\eqref{eq:defS} and~\eqref{eq:u1s}.
The $\NN=(2,2)$ two-dimensional superconformal algebra reads
\be
\begin{alignedat}{3}
\{G^+_r, G^-_s\}&= L_{r+s} + \frac{r-s}{2} \JJ\,, \qquad
&\{\Gb^+_r, \Gb^-_s\}&= \Lb_{r+s} + \frac{r-s}{2} \bar{\JJ}_{r+s}\,,\\
\left[ L_m,L_n \right] &= (m-n)L_{m+n}\,, &[\Lb_m,\Lb_n]&= (m-n)\Lb_{m+n}\,, \\
\left[L_m, G^\pm_r \right] &= \left(\frac{m}{2}- r\right) G^\pm_{m+r}\,, &\left[\Lb_m, \Gb^\pm_r \right] &= \left(\frac{m}{2}- r\right) \Gb^\pm_{m+r}\,,\\
\left[\JJ, G^\pm_r \right] &= \pm  G^\pm_r\,,  &\left[\bar{\JJ}, \Gb^\pm_r \right] &= \pm \Gb^\pm_r\,,
\end{alignedat}
\ee
where $ r,s=\pm \frac12$ and $m,n,=0,\pm 1$ since there is no defect stress tensor, and thus we consider the global superalgebra with no Virasoro enhancement.
A  short multiplet of the left-moving part of the $2d$ superconformal algebra obeying $L_0 = \frac{1}{2} \JJ$ is annihilated by $G^+_{-\frac12}$ and is called chiral, while an antichiral operator obeys $L_0 = -\frac{1}{2} \JJ$ and is annihilated by $G^-_{-\frac12}$. A similar definition holds with adding bars in the generators for the right-moving part of the algebra.

\subsubsection*{The displacement supermultiplet for a  \texorpdfstring{$2d$ $\NN=(2,2)$}{2d N=(2,2)} defect}
\label{app:displ22}

We take the supersymmetry variations of the $4d$ $\NN=2$ superconformal algebra \eqref{eq:delta} and set to zero the parameters corresponding to the non-preserved supercharges, getting the preserved variations
\be 
\delta_p \OO= \left[ \xi_1^{\bf{2}} \Gb^+_{-\tfrac12} + \xi_ 2^{\bf{1}} G^+_{-\tfrac12} + \bar{\xi}^{\pd 1} \Gb^-_{-\tfrac12} +  \bar{\xi}^{\md 2} G^-_{-\tfrac12} , \OO \right]\,.
\ee
Then the supersymmetry variations of the supermultiplet containing the displacement operator shown in figure \ref{fig:displacement} are as follows:
\be 
\begin{alignedat}{2}
\delta_p \Od_{\uparrow} &=- \xi_ 2^{\bf{1}}  \Ld^+_{\uparrow} + \bar{\xi}^{\pd 1} \Ld_{\uparrow}^- \,, \qquad  &\delta_p \Od_{\downarrow} &= \xi_ 1^{\bf{2}}  \Ld_{\downarrow}^+ - \bar{\xi}^{\md 2} \Ld_{\downarrow}^- \,, \\
\delta_p \Ld_{\uparrow}^+ &=  \bar{\xi}^{\pd 1} \Dd_{\uparrow} - \bar{\xi}^{\md 2} \partial_{w} \Od_{\uparrow} \, \qquad &\delta_p \Ld_{\downarrow}^+ &=  \bar{\xi}^{\md 2} \Dd_{\downarrow} + \bar{\xi}^{\pd 1} \partial_{\wb} \Od_{\downarrow} \,,\\
\delta_p \Ld_{\uparrow}^-  &=  \xi_ 2^{\bf{1}} \Dd_{\uparrow} + \xi_1^{\bf{2}} \partial_{\wb}\Od_{\uparrow} \,,\qquad &\delta_p \Ld_{\downarrow}^-  &= \xi_ 1^{\bf{2}} \Dd_{\downarrow} - \xi_2^{\bf{1}} \partial_{w}\Od_{\downarrow} \,,\\
\delta_p \Dd_{\uparrow} &=  \xi_1^{\bf{2}}  \partial_{\wb} \Ld_{\uparrow}^+ +  \bar{\xi}^{\md 2} \partial_{w}\Ld_{\uparrow}^-\,, \qquad 
&\delta_p \Dd_{\downarrow} &=  \xi_2^{\bf{1}}  \partial_{w} \Ld_ {\downarrow}^+ + \bar{\xi}^{\pd 1} \partial_{\wb}\Ld_{\downarrow}^-\,.
\end{alignedat}
\label{eq:displvar}
\ee

%% file: sections/B_index.tex
%!TEX root = ../susysurfaces.tex
%%%%%%%%%%%%%%%%%%%%%%%%%%%%%%%%%%%%%%%%%%%%%
\section{Superconformal index}
\label{app:index}
%%%%%%%%%%%%%%%%%%%%%%%%%%%%%%%%%%%%%%%%%%%%%

The superconformal index \cite{Romelsberger:2005eg,Kinney:2005ej} is an important invariant of $4d$ superconformal field theories that encodes protected information about the spectrum of the theory.  It counts (with signs) all protected multiplets that cannot recombine to form long multiplets, and is invariant under exactly marginal deformations of the SCFT.
Here we briefly review the superconformal index of an $\NN=2$ SCFT and refer to the review \cite{Rastelli:2014jja} for all details.\footnote{We follow the conventions of \cite{Beem:2013sza,Rastelli:2014jja}, which differ slightly from those \cite{Gadde:2011uv,Gaiotto:2012xa}.}
We compute the $\NN=2$ superconformal index with respect to the $\Qt_{2 \dot{\mathbf{2}}}$ supercharge, as the trace over the Hilbert space of the SCFT in radial quantization
\be 
\II(p,t,q) = \Tr_{\mathcal{H}}(-1)^F  t^{R+r} p^{\tfrac{\Delta-2j_1-2R-r}{2}} q^{\tfrac{\Delta+2j_1 -2R-r}{2}}e^{-\beta \left(\Delta - 2 j_2+r-2R\right)}\,,
\label{eq:fullindex}
\ee
where $F=2(j_1-j_2)$ is the fermion number. For theories with additional symmetries the index can be further refined by additional fugacities conjugate to Cartans of the symmetry that commute with the ones already introduced and with $\tilde{Q}_{2 \md}$ . The superconformal index defined like this is independent of $\beta$,  receiving only contributions from operators with 
\be 
\Delta = 2j_2 -r+2R\,,
\label{eq:indexops}
\ee
and is independent of any continuous parameters of the theory - see \cite{Rastelli:2014jja}. The index is then the most general invariant that counts (with signs) the short multiplets of the theory that cannot recombine to form long multiplets. 

The Schur limit of the superconformal index is obtained by setting $t=q$ \cite{Gadde:2011uv},
\be 
\II(q)=\Tr_{\mathcal{H}}(-1)^F q^{\Delta -R } p^{\tfrac{\Delta-2j_1-2R-r}{2}} e^{-\beta \left(\Delta - 2 j_2+r-2R\right)}\,.
\ee
and it becomes independent of $p$, receiving only contributions from operators satisfying also $\Delta-2j_1-2R-r=0$, which we already used to simplify the exponent of $q$. Together with \eqref{eq:indexops} we obtain precisely the conditions necessary for operators to contribute to the chiral algebra \eqref{eq:Schurops}. From the four-dimensional Cartans of the $\NN=2$ SCFT the chiral algebra preserves $\Lchi_0$ and $r=j_2-j_1$, and thus one can define its graded partition function as\footnote{Here we omitted an overall power of $q^{-c_{2d}/24}$ which must be included in the partition function for it to have the modular properties described in \cite{Beem:2017ooy}. See also \cite{Bobev:2015kza} for a discussion of this prefactor when relating the index to the partition function on $S^1 \times S^3$.}
\be 
Z(q,x) = \Tr\left( q^{\LT_0} x^{2(j_2-j_1)}  \right) = \Tr\left( q^{\Delta - R} x^{F}\right)\,,
\label{eq:partfunct}
\ee
where $\LT_0$ is the zero mode of the $2d$ stress tensor, which matches $\Lchi_0$ when acting on local operators.
For $x=-1$ matches precisely the definition of the Schur limit of the superconformal index.

The same set of operators is also counted by the Macdonald limit of the index, obtained by setting $p=0$ in \eqref{eq:fullindex}, meaning that we compute
\be 
\II(t,q) = \Tr_{\mathcal{H}_M} (-1)^F  t^{R+r} q^{j_1+j_2-r} \,,
\ee
where $\mathcal{H}_M$ denotes a restriction of the Hilbert space to operators having $\Delta-2j_1-2R-r=0$. It thus counts the same operators as those contributing to the chiral algebra, but refines the counting by the additional fugacity $t$. Recovering this information from the chiral algebra itself is an open question, as the $R$ grading of the four-dimensional SCFT is lost when passing to the chiral algebra, and so the refinment is by a Cartan not preserved by the chiral algebra. See, however,  \cite{Song:2016yfd,Bonetti:2018fqz,Beem:2019tfp,Beem:2019snk,Xie:2019zlb} for proposals on recovering the full Macdonald index from the chiral algebra.

\subsubsection*{Superconformal index with defects}

The index defined above counts local operators in four-dimensional $\NN=2$ SCFTs. To count operators living on the $\NN=(2,2)$ surface defect we define the index instead by doing radial quantization centered on a point on the defect
\be 
\II(p,y,q) = \Tr_{\mathcal{H}_{\deft}}(-1)^{F}  t^{R+r} p^{\Lb_0-\frac12 \bar{\JJ}} q^\CC y^{\bar{\JJ}}  e^{-\beta \left(2L_0+\JJ\right)}\,.
\label{eq:fulldefectindex}
\ee
The above is precisely the same formula as \eqref{eq:fullindex}, except that the trace is now over the Hilbert space of the defect theory, and where we introduced the two-dimensional quantum numbers and re-defined fugacities as $t=q y$.
Written in this way it becomes clear the index simply corresponds to an elliptic genus for the $\NN=(2,2)$ two-dimensional theory \cite{Witten:1986bf} on the defect, refined by a flavor fugacity $q$ that keeps track of the $u(1)_\CC$ global symmetry of the defect theory \cite{Gadde:2013wq}. Recall that the index is computed with respect to the $G^-_{-\tfrac12}=\Qt_{2 \dot{\mathbf{2}}}$ supercharge, and so it will count operators that are anti-chiral on the left, \ie obeying $2L_0 = -\JJ$.

The Schur limit of the index becomes $y=1$, with the index once more becoming independent of $p$, and thus receiving contributions only from operators that are also chiral on the right, \ie with $2\Lb_0= \bar{\JJ}$.  All in all it counts $(a,c)$ defect operators, graded by their $u(1)_\CC$ flavor charge, as written in \eqref{eq:SchurIndex}. As shown in \ref{sec:chiralalgebradefects} $(a,c)$ operators are precisely those that contribute to the chiral algebra, and it was argued in \cite{Cordova:2017mhb} that it again matches, up to an overall power of $q$ since the index is normalized such that operators with $\CC=0$ contribute as $q^0$, the graded partition function of the chiral algebra, now in the presence of the defect. The graded chiral algebra partition function is given by \eqref{eq:partfunct} with $x=-1$, noting that now $\LT_0$ differs from $\Lchi_0$ by the dimension of the defect identity in chiral algebra, $h_\sigma$, producing an overall power of $q^{h_\sigma}$.

Finally, the Macdonald limit of the index, \ie setting $p=0$, becomes a trace over the restricted Hilbert space of operators that are chiral on the right, thus counting the same as the Schur index but keeping track of the $\bar{\JJ}$ quantum number of operators as well. As such, it can distinguish some of the operators that appear degenerate in the chiral algebra. Recall that this refinement involves refining the index by a Cartan that is not preserved by the chiral algebra, and thus its interpretation in chiral algebra is not clear. In \cite{Watanabe:2019ssf} the conjectured prescription of \cite{Song:2016yfd} was used to attempt to recover the Macdonald index from the chiral algebra, but the authors found disagreements with the expressions for the superconformal indices in some examples.

%% file: sections/C_correlators.tex
%!TEX root &= ../susysurfaces.tex
%%%%%%%%%%%%%%%%%%%%%%%%%%%%%%%%%%%%%%%%%%%%%
\section{Stress tensor displacement correlator for \texorpdfstring{$\mathcal{N}=(2,0)$ surface in $\mathcal{N}=1$}{N=(2,0) surface in N=1}}
\label{sec:allcorr}
%%%%%%%%%%%%%%%%%%%%%%%%%%%%%%%%%%%%%%%%%%%%%
In this appendix we spell out all the bulk to defect correlators of stress tensor and displacement supermultiplet. In the following the defect operator is always taken to be in the origin.
\begingroup
\allowdisplaybreaks[1]
\begin{align*}
\langle j_z \mathbb{D}_{\uparrow}\rangle=\frac12 \langle J_{z 1} \mathbb{\L}_{\uparrow}^-\rangle &= \frac{3 h \bar{z}^2}{\pi  (w
   \bar{w}+z \bar{z})^4}\,, & 
   \langle j_{\bar{w}} \mathbb{D}_{\uparrow}\rangle =\frac12 \langle J_{z 2} \mathbb{\L}_{\uparrow}^-\rangle&= -\frac{3 h w \bar{z}}{\pi  (w
   \bar{w}+z \bar{z})^4}\,,\\
   \langle j_w\mathbb{D}_{\uparrow}\rangle=\frac12 \langle J_{w 1} \mathbb{\L}_{\uparrow}^-\rangle&= -\frac{3 h \bar{w} \bar{z}}{\pi  (w \bar{w}+z
   \bar{z})^4}\,,&
   \langle j_{\bar{z}}\mathbb{D}_{\uparrow}\rangle=\frac12\langle J_{w 2} \mathbb{\L}_{\uparrow}^-\rangle &= -\frac{3 h w \bar{w}}{\pi  (w \bar{w}+z
   \bar{z})^4} \,,\\
   \langle j_z\mathbb{D}_{\downarrow}\rangle=\frac12\langle \tilde{J}_{z 2}
   \mathbb{\L}_{\downarrow}^+\rangle &= \frac{3 h w \bar{w}}{\pi  (w \bar{w}+z
   \bar{z})^4}\,,&
   \langle j_w \mathbb{D}_{\downarrow}\rangle=\langle \tilde{J}_{w
   2} \mathbb{\L}_{\downarrow}^+\rangle &= \frac{3 h
   z \bar{w}}{\pi  (w \bar{w}+z \bar{z})^4}\,,\\
   \langle j_{\bar{w}}
   \mathbb{D}_{\downarrow}\rangle=\frac12 \langle \tilde{J}_{\bar{w} 2}
   \mathbb{\L}_{\downarrow}^+\rangle &= \frac{3 h w z}{\pi  (w \bar{w}+z
   \bar{z})^4}\,,&
   \langle j_{\bar{z}} \mathbb{D}_{\downarrow}\rangle=\frac12 \langle
   \tilde{J}_{\bar{z} 2} \mathbb{\L}_{\downarrow}^+\rangle &=
   -\frac{3 h z^2}{\pi  (w \bar{w}+z \bar{z})^4}\,,\\
   \langle J_{\bar{z} 2} \mathbb{\L}_{\uparrow}^-\rangle=-\langle \tilde{J}_{z 1}
   \mathbb{\L}_{\downarrow}^+\rangle &=
   \frac{6 h w^2 \bar{w}}{\pi  \bar{z} (w \bar{w}+z \bar{z})^4}\,,&\langle
   J_{\bar{w} 2} \mathbb{\L}_{\uparrow}^-\rangle=-\langle \tilde{J}_{\bar{w} 1}
   \mathbb{\L}_{\downarrow}^+\rangle &= \frac{6 h w^2}{\pi  (w
   \bar{w}+z \bar{z})^4}\,,\\
   \langle T_{z z} \mathbb{D}_{\uparrow}\rangle &= \frac{12 h \bar{z}^3}{\pi  (w \bar{w}+z\bar{z})^5}\,,&
   \langle T_{\bar{z} \bar{z}} \mathbb{D}_{\uparrow}\rangle &= \frac{12 h w^2 \bar{w}^2}{\pi \bar{z} (w \bar{w}+z \bar{z})^5}\,,\\
   \langle T_{w w}\mathbb{D}_{\uparrow}\rangle &= \frac{12 h \bar{w}^2 \bar{z}}{\pi  (w \bar{w}+z\bar{z})^5}\,,&
   \langle T_{\bar{w} \bar{w}} \mathbb{D}_{\uparrow}\rangle &= \frac{12 h w^2 \bar{z}}{\pi  (w \bar{w}+z \bar{z})^5}\,,\\
   \langle T_{z w} \mathbb{D}_{\uparrow}\rangle &= -\frac{12 h \bar{w} \bar{z}^2}{\pi  (w \bar{w}+z \bar{z})^5}\,,&
   \langle T_{z \bar{w}} \mathbb{D}_{\uparrow}\rangle &= -\frac{12 h w \bar{z}^2}{\pi  (w \bar{w}+z \bar{z})^5} {\addtocounter{equation}{1}\tag{\theequation}} \\
   \langle T_{z \bar{z}} \mathbb{D}_{\uparrow}\rangle &= -\frac{12 h w \bar{w} \bar{z}}{\pi  (w \bar{w}+z \bar{z})^5}\,,&
   \langle T_{w \bar{w}} \mathbb{D}_{\uparrow}\rangle &=\frac{12 h w \bar{w} \bar{z}}{\pi  (w \bar{w}+z \bar{z})^5}\,,\\
   \langle T_{w \bar{z}} \mathbb{D}_{\uparrow}\rangle &= \frac{12 h w \bar{w}^2}{\pi  (w\bar{w}+z \bar{z})^5}\,,&
   \langle T_{\bar{w} \bar{z}} \mathbb{D}_{\uparrow}\rangle &=\frac{12 h w^2 \bar{w}}{\pi  (w \bar{w}+z \bar{z})^5}\,,\\
   \langle T_{z z} \mathbb{D}_{\downarrow}\rangle &= \frac{12 h w^2 \bar{w}^2}{\pi  z (w \bar{w}+z \bar{z})^5}\,,& 
    \langle T_{\bar{z} \bar{z}} \mathbb{D}_{\downarrow}\rangle &= \frac{12 h z^3}{\pi  (w \bar{w}+z \bar{z})^5}\,,\\
   \langle T_{w w} \mathbb{D}_{\downarrow}\rangle &= \frac{12 h z \bar{w}^2}{\pi  (w \bar{w}+z \bar{z})^5}\,,&
   \langle T_{\bar{w} \bar{w}} \mathbb{D}_{\downarrow}\rangle &= \frac{12 h w^2 z}{\pi  (w \bar{w}+z \bar{z})^5}\,,\\
   \langle T_{z w} \mathbb{D}_{\downarrow}\rangle &= \frac{12 h w \bar{w}^2}{\pi  (w \bar{w}+z \bar{z})^5}\,,&
   \langle T_{z \bar{w}} \mathbb{D}_{\downarrow}\rangle &= \frac{12 h w^2 \bar{w}}{\pi  (w\bar{w}+z \bar{z})^5}\,,\\
   \langle T_{z \bar{z}}\mathbb{D}_{\downarrow}\rangle &= -\frac{12 h w z \bar{w}}{\pi  (w \bar{w}+z \bar{z})^5}\,,&
   \langle T_{w \bar{w}} \mathbb{D}_{\downarrow}\rangle &= \frac{12 h w z \bar{w}}{\pi  (w \bar{w}+z \bar{z})^5}\,,\\
   \langle T_{w \bar{z}} \mathbb{D}_{\downarrow}\rangle &= -\frac{12 h z^2 \bar{w}}{\pi  (w \bar{w}+z \bar{z})^5}\,,&
   \langle T_{\bar{w} \bar{z}} \mathbb{D}_{\downarrow}\rangle &= -\frac{12 h w z^2}{\pi  (w \bar{w}+z \bar{z})^5}\,.
\end{align*}
\endgroup

%% file: sections/D_monodromy.tex
%!TEX root &= ../susysurfaces.tex
%%%%%%%%%%%%%%%%%%%%%%%%%%%%%%%%%%%%%%%%%%%%%
\section{Monodromy defect }
\label{app:monodromy}
%%%%%%%%%%%%%%%%%%%%%%%%%%%%%%%%%%%%%%%%%%%%%
In this appendix we collect the results for a monodromy defect in the free hypermultiplet theory described in section~\ref{sec:monodromy}. 
Imposing \eqref{eq:monodromy}, the two-point functions of the free hypermultiplet scalars read
\be 
\begin{split}
\langle \QQ \QQ^* \rangle &=  a^{\alpha} e^{-i \frac12 \alpha\theta_{12}}\left(\frac{1}{-1+a^2 e^{-i \theta_{12}}}+a^{-2 \alpha} \left(b^2_{\QQ}+\frac{1}{-1+a^2 e^{i \theta_{12}}}\right)-b^2_{\QQ}+1\right)\,,\\
\langle \tilde{\QQ} \tilde{\QQ}^* \rangle &=a^{\a} \left(\left(b_{\tilde{\QQ}}^2 \left(a^{-2 \a}-1\right)+1\right) e^{i \a \theta_{12}}-\frac{e^{i \frac12 \a \theta_{12}}}{1-a^2 e^{i \theta_{12}}}+\frac{a^{-2 \a} e^{i (\frac12 \a+1) \theta_{12}}}{a^2 -e^{i \theta_{12}}}\right)\,,
\end{split}
\label{eq:two-pointmonodromies}
\ee
with
\be 
a= \frac{1}{2} \left(\sqrt{\frac{r_1}{r_2}+\frac{r_2}{r_1}-2}+\sqrt{\frac{(r_1+r_2)^2}{r_1 r_2}}\right)\,,
\ee
and where we placed the two bulk operators on the same plane. Here $r_1$ and $r_2$ denote the distance of each of the operators to the defect, and $\theta_{12}$ the angular separation between the two operators.
These two-point functions are normalized such that far away from the defect the bulk scalars have unit two-point function. As such we have that
\be 
\begin{split}
O_2 = \frac{1}{4 \pi ^2}\left( \QQ \QQ^* + \tilde{\QQ} \tilde{\QQ}^*\right)\,,& \qquad
t^{12}_{i} = \frac{1}{8 \pi ^2}\left( \partial_{i}\QQ \QQ^*-\QQ \partial_{i}\QQ^* + \partial_{i}\tilde{\QQ} \tilde{\QQ}^*-\tilde{\QQ} \partial_{i}\tilde{\QQ}^* \right)\,,\\
\mu^{3\, 12} &= \frac{1}{8\sqrt{2} \pi^2}\left( \QQ \QQ^* - \tilde{\QQ} \tilde{\QQ}^*\right)\,,
\end{split}
\label{eq:O2tmufromQ}
\ee
and the respective one-point functions can be computed by taking the coincident limit of \eqref{eq:two-pointmonodromies}. The results  match precisely the prediction from the spectral flow quoted in \eqref{eq:hmonodromy} and \eqref{eq:oneptmumonodromy}.
Notice that to compute other one-point functions one would need the fermion propagators as well.\footnote{Getting the one-point functions of the $su(2)_R$ current, or of the stress tensor, from the bulk conformal block expansion of the two-point functions \eqref{eq:two-pointmonodromies} is not straightforward as one needs to disentangle different operators. For example the $su(2)_R$ and $su(2)_f$ currents appear degenerate.}